\documentclass{article}

\usepackage[colorlinks = true,
            linkcolor = blue,
            urlcolor  = blue,
            citecolor = blue,
            anchorcolor = blue]{hyperref}
            
\usepackage{graphicx} 

\usepackage{alex_typical}
\usepackage{shorthands}
\usepackage[margin=1in]{geometry}

\definechangesauthor[name={Chris Kang}, color=red]{CK}
\definechangesauthor[name={Ben Ide}, color=blue]{BI}
\definechangesauthor[name={Adam Bene Watts}, color=purple]{ABW}

\usetikzlibrary{external}
\tikzset{
    pctrl color/.store in=\pctrlcolor,
    pctrl color=black,
    pctrlstyle/.style={
        line width=1.25pt,
        fill=\pctrlcolor,
        inner sep=3pt,
        shape = circle,
        path picture={
            
            \draw[white, line width=0.6pt]
                ([yshift=2pt]path picture bounding box.center) -- ([yshift=-2pt]path picture bounding box.center)
                ([xshift=2pt]path picture bounding box.center) -- ([xshift=-2pt]path picture bounding box.center);
            \draw[white, line width=0.6pt]
                (path picture bounding box.center) circle [radius=2pt];
        },
    }
}
\DeclareExpandableDocumentCommand{\pctrl}{O{}{m}}{%
  |[pctrlstyle,#1]| {}%
}

\newcommand{\mygate}{%
  \gategroup[
    wires=2,
    steps=3,
    style={
      draw,
      fill=white,
      rounded corners=1pt,
      inner xsep=2pt,
      inner ysep=2pt
    },
    label style={
      label position=center,
      anchor=mid,
      yshift=-1.6ex
    }
  ]{\texttt{AND}}%
}

\newcommand{\andtwo}{%
  \gategroup[
    wires=3,
    steps=3,
    style={
      draw,
      fill=white,
      rounded corners=1pt,
      inner xsep=2pt,
      inner ysep=2pt
    },
    label style={
      label position=center,
      anchor=mid,
      yshift=-1.6ex
    }
  ]{\texttt{AND}}%
}

\newcommand{\quantikzcz}{%
  \gategroup[
    wires=2,
    steps=3,
    style={
      draw,
      fill=white,
      rounded corners=1pt,
      inner xsep=2pt,
      inner ysep=2pt
    },
    label style={
      label position=center,
      anchor=mid,
      yshift=-1.6ex
    }
  ]{\texttt{CZ}}%
}

\newcommand{\phasepoly}{%
  \gategroup[
    wires=2,
    steps=3,
    style={
      draw,
      fill=white,
      rounded corners=1pt,
      inner xsep=2pt,
      inner ysep=2pt
    },
    label style={
      label position=center,
      anchor=mid,
      yshift=-1.6ex
    }
  ]{\texttt{PHASE}}%
}

\usepackage{pifont}

\usepackage{tikz}
\usepackage{tikzit}

\tikzstyle{boundary vertex}=[inner sep=0mm, minimum size=1mm, shape=circle, draw=black, fill=black]
\tikzstyle{grey_dot}=[fill={rgb,255: red,191; green,191; blue,191}, draw=black, shape=circle, minimum size=1.5mm, inner sep=0mm]
\tikzstyle{blue_dot}=[fill={rgb,255: red,202; green,251; blue,255}, draw=black, shape=circle, minimum size=1.5mm, inner sep=0mm]
\tikzstyle{white_dot}=[fill=white, draw=black, shape=circle, minimum size=1.5mm, inner sep=0mm]
\tikzstyle{red_dot}=[fill=red, draw=black, shape=circle, minimum size=1.5mm, inner sep=0mm]
\tikzstyle{box}=[fill=white, draw=black, shape=rectangle]
\tikzstyle{scalable_dot}=[fill=white, draw=black, shape=circle, minimum size=6.5mm, inner sep=0mm]
\tikzstyle{scalable_box}=[fill=white, draw=black, shape=rectangle, minimum height=6.5mm, minimum width=6.5mm]
\tikzstyle{green_box}=[fill=green, draw=black, shape=rectangle]
\tikzstyle{blue_box}=[fill=blue, draw=black, shape=rectangle]

\tikzstyle{arrow}=[->]
\tikzstyle{red_arrow}=[->, draw=red]
\tikzstyle{cyan_arrow}=[->, draw=cyan]
\tikzstyle{red_dash}=[-, dashed, draw=red]
\tikzstyle{grey dash}=[-, fill=none, draw={rgb,255: red,191; green,191; blue,191}, dashed]
\tikzstyle{dashed arrow}=[->, dashed]
\tikzstyle{blue_edge}=[-, draw=cyan]
\tikzstyle{red_edge}=[-, draw=red]
\tikzstyle{green_edge}=[-, fill=none, draw=green]
\tikzstyle{green_dash}=[-, draw=green, dashed]

\usepackage{mdframed}
\usepackage{xcolor}

\mdfdefinestyle{aidisclosure}{
    linecolor=gray!55,
    linewidth=0.6pt,
    backgroundcolor=white,
    roundcorner=0pt,
    innertopmargin=10pt,
    innerbottommargin=10pt,
    innerleftmargin=12pt,
    innerrightmargin=12pt,
    skipabove=12pt,
    skipbelow=12pt
}

\title{On Reaction Depth and 
Fast Fault-Tolerant Boolean Logic
}

\author{Christopher Kang$^{1,2}$\footnote{\href{mailto:chris.kang.math@gmail.com}{chris.kang.math@gmail.com}}, Alexander Cowtan$^1$\footnote{\href{mailto:alexander.cowtan@xanadu.ai}{alexander.cowtan@xanadu.ai}}, 
Adam Bene Watts$^{1,3}$, Guillaume Dauphinais$^1$, Benjamin Ide$^1$}
\date{%
    $^1$Xanadu, Toronto, ON M5G 2C8, Canada\\%
    $^2$Department of Computer Science, University of Chicago, Chicago, IL 60637, USA \\%
    $^3$Department of Mathematics and Statistics and Institute for Quantum Science and Technology, University of Calgary \\%
}

\begin{document}

\maketitle

\begin{abstract}
Reducing runtimes of quantum applications is critical for practical viability. On quantum computers, the runtime can depend both on the speed of quantum operations and on how often the program must wait for the classical coprocessor to determine adaptive operations. Some recent approaches to reducing this overhead have focused on minimizing the ``reaction depth'' of a computation, informally defined as the number of times the classical coprocessor must be consulted while running a quantum program.

In this work, we formalize the notion of reaction depth. We find several subtleties in this theory. For example, we distinguish the reaction depth of a circuit from that of a circuit run, and show that the magic depth of a unitary can be higher than the reaction depth once compiled into Clifford computation and magic state injection. 

We then focus on the specific case of coherent Boolean logic, a common and expensive component in quantum algorithms. We use a natural gateset for fast operations with Low-Density Parity Check (LDPC) codes, which consists of homogeneous $Z$- and $X$-type Pauli Product Measurements (PPMs), alongside a supply of Toffoli states. We prove that, using this gateset, any Boolean arithmetic circuit can be implemented on a quantum computer with constant quantum depth and constant reaction depth, at the cost of exponential space and leaving some qubits ``dirty''. Using the same gateset we find two efficient \gateautoand circuits that require only a Toffoli state and homogeneous PPMs. We then give efficient implementations of the quantum ripple-carry adder and quantum schoolbook multiplier, which require a low number of heterogeneous PPMs for uncomputation. We find implementations competitive with the state-of-the-art despite the restricted gateset.


\end{abstract}


\newpage
\tableofcontents

\newpage
\section{Introduction}

Ensuring fast runtimes for quantum applications is critical~\cite{mcardle2025fast}: logical clock cycles of fault-tolerant quantum computers are expected to operate approximately at 1kHz-1MHz~\cite{webber2022impact,acharya2024quantumerrorcorrectionsurface,aghaee2025scaling,ransford2026helios,bluvstein2026quera}, around $10^3$-$10^6 \times$ slower than classical computers. Furthermore, quantum algorithms often require sampling overheads of thousands or millions of shots~\cite{mcardle2025fast,fomichev2025fast}. These significant time overheads impugn the utility of quantum applications.

At small scales, logical operations are primarily constrained by the speed of quantum gates and measurements~\cite{ye2026real,perlin2026fault}. As architectures become larger and more complex, however, the speed can also become constrained by the classical coprocessor required for error correction. For example, architectures using quantum Low-Density Parity Check (LDPC) codes~\cite{breuckmann2021ldpc,panteleev2021degenerate,panteleev2022asymptotically,bravyi2024high,he2025extractors} now enjoy a variety of schemes for parallel and fast logic~\cite{cowtan2025fast, cowtan2025parallel, gu2026qgpu, zhang2024time, baspin2025fast,xu2025batched,golowich2025constant,tan2025single,zheng2025high,hillmann2024single, zhou2024algorithmic, huang2023homomorphic, xu2025fast, li2026transversal,cowtan2026nearoptimal,baspin20262fast2surgery} to meet the demands of large quantum algorithms.\footnote{In particular, on high-rate Calderbank-Shor-Steane (CSS) LDPC codes, fast and fault-tolerant logical operations are typically substantially easier to design when they preserve the CSS code structure: namely, when the operations are composed of homogeneous $Z$- and $X$-type Pauli Product Measurements~\cite{cowtan2025fast,baspin2025fast,baspin20262fast2surgery}.} These schemes can
incur significant classical overheads from decoding.
In certain settings, such classical overheads can require stalls in the quantum processor so that the classical co-processor can resolve logical measurement outcomes and determine adaptive quantum operations~\cite{Terhal2015, skoric2023parallel, barber2025realtime, chamberland2023techniques, battistel2023realtime}.\footnote{Note that it is the \textit{logical} program which stalls; syndrome extraction continues on the physical qubits while the logical qubits idle.} 
Classical bottlenecks are most relevant when implementing universal logic via gate teleportation, when feedforward corrections are non-Pauli and therefore are not easily tracked in frame~\cite{litinski2019game,chamberland2018fault}.

At one extreme, we can perform \textit{reaction-limited} computation, where runtimes are limited primarily by the reaction time of the classical coprocessor and the \textit{reaction depth}, the number of times the classical coprocessor must be consulted to run the logical circuit~\cite{fowler2013timeoptimal, gidney2025how}.

Some prior research in algorithms and compilation has implicitly reduced reaction depth via the proxy of magic depth~\cite{selinger2013quantum}. Coherent Boolean logic is particularly important to compile efficiently~\cite{khattar2026leveraging}, as it forms critical and expensive subroutines in many large-scale quantum algorithms~\cite{low2026denser, babbush2026securing, gidney2021factoring, gidney2025how,cain2026shor,motlagh2025quantum,gidney2018halving,babbush2018encoding,dalzell2025quantum,vonburg2021quantum,lee2021even,mukhopadhyay2024quantum,dalzell2024shortcut}. 


In this work, we have two contributions:
\begin{enumerate}
    \item We develop a general theory of the reaction depth of quantum circuits. We find several subtleties in this theory; for example, the magic depth of a unitary can be higher than the reaction depth once compiled into Clifford computation and magic state injection, see Appendix~\ref{app:reaction_depth}. Additionally, for the reaction depth to be efficiently computable in general, it must only be an upper bound on the number of times the classical coprocessor may need to be interacted with in a given shot of the logical circuit. We formalize reaction depth, such that it can be calculated automatically using common graph traversal algorithms.
    \item We then focus on Boolean arithmetic. We prove that any Boolean logic circuit possesses an implementation with constant quantum depth and reaction depth, leaving some qubits in a ``dirty'' state. \footnote{A \textit{dirty} state is an arbitrary, possibly entangled and non-Pauli state, which must be reset to $\ket{0}$ to be \textit{clean}.} This implementation uses only Toffoli resource states, along with $Z$- and $X$-type Pauli product measurements (PPMs), a natural gateset for architectures based on high-rate LDPC codes. Along the way, we find efficient implementations of auto-corrected \gateand gates (\gateautoand), and compile adders and multipliers into efficient quantum circuits which use gates amenable to application on high-rate quantum LDPC codes.
\end{enumerate}

Our compilation of Boolean logic explicitly considers reaction depth. Furthermore, it is tailored to qLDPC codes; existing studies on reaction-limited coherent Boolean logic are typically predicated on the assumption that the underlying code is the 2D surface code~\cite{low2026denser, gidney2025how, gidney2019flexible, gidney2021factoring, gidney2025how, fowler2019low}. Thus, they assume gates which may be harder in high-rate codes, and are frequently compiled into topological diagrams which only have a meaningful interpretation for the 2D surface code. 

\subsection{Summary of results} 

While the reaction depth of quantum programs has been studied before~\cite{gidney2019flexible,litinski2022active,gidney2025how,low2026denser,fowler2013timeoptimal}, it has not previously been formalized mathematically. Our first contribution is to ground the reaction depth of quantum programs in a formal framework.
We give a definition of the reaction depth of a circuit in Section~\ref{sec:reaction_limited_defs}, which is summarized informally below.

Consider a quantum circuit with measurements and classically controlled gates contingent on the results of those measurements. These can have a causal dependence, as some gates must be performed before the next measurement can occur, leading to paths of classical corrections through the circuit.

\begin{define}[Reaction depth, informal]
    Let $\mathcal{C}$ be a quantum circuit. The reaction depth $r(\mathcal{C})$ of $\mathcal{C}$ is the longest path of classical corrections in $\mathcal{C}$.
\end{define}

We can consider the reaction depth to be an upper bound on the number of times the classical co-processor must be interacted with sequentially to execute the circuit in general. Similarly, we define the reaction depth $r_b(\mathcal{C})$ of a circuit \textit{run} with measurement outcomes $b$, which is the number of times the classical co-processor must be interacted with in a given circuit execution. Because the results of quantum measurements are probabilistic, and there may be classical branching as a consequence, the reaction depth of a circuit is not the same as the reaction depth of a circuit run, which can be lower, i.e. $r_b(\mathcal{C}) \leq r(\mathcal{C})$. Indeed, the reaction depth of a given circuit execution is not generally known at compile time.

We also formalize the notion of \textit{reaction-limited computations}. This term has been colloquially defined as circuits with runtime limited by the reaction speed of the classical control system and measurements. We refine the definition slightly to include an explicit criterion for reaction-limited computations.
\begin{define}[Reaction-limited]\label{def:reaction_limited_computation}
    Suppose a quantum architecture $\mathcal{A}$ has classical reaction time $\tau$ and a quantum circuit $\mathcal{C}$ has reaction depth $r(\mathcal{C})$ and runtime $t$. Then, $\mathcal{C}$ is \textit{reaction-limited} if and only if
    \begin{align*}
        t \sim r(\mathcal{C}) \cdot \tau.
    \end{align*}
\end{define}

In Section~\ref{sec:boolean_results} we turn to coherent Boolean circuits and compiling fast Boolean logic. We focus on the gateset $\{\mathtt{PPM}_Z, \mathtt{PPM}_X, \toffstate\}$ consisting of homogeneous $Z$- and $X$-type PPMs, that is measurements of Pauli products containing only either $\{Z, I\}$ terms or $\{X, I\}$ terms respectively, supplemented by $CCX\ket{++0}$ ($\ket{CCX}$) states, leading to our primary result.

\begin{theorem}[Informal version of~\cref{thm:arbitrary_boolean}]\label{thm:const_reaction_depth}
    Let $\mathtt{B}$ be a Boolean function with $n$ input variables. Then $\mathtt{B}$ admits an implementation as a quantum circuit $\mathcal{C}$ using the gateset $\{\mathtt{PPM}_Z, \mathtt{PPM}_X, \toffstate\}$ in constant quantum depth and reaction depth, using $\mathcal{O}(2^n)$ ancilla qubits, which are left ``dirty'', i.e. in a garbage state.
\end{theorem}


Refs.~\cite{gidney2025how, low2026denser} previously demonstrated that QROM circuits could be implemented with constant reaction depth; we show that a restrictive set of PPMs suffices to perform these circuits with constant quantum and reaction depth.


\begin{table}[]
    \centering
    \begin{tabular}{|l|c|c|c|c|}
    \hline
        & \textbf{Space} & \textbf{Depth} & \textbf{cDepth} & \textbf{PPMs}\\
    \hline
        AutoCCZ \cite[Figure 4]{gidney2019flexible} & $\geq 9$ & $6$ & $1$ & non-homogeneous \\
        Semi-auto CCZ \cite[Figure 8]{gidney2025how} & $6$ & $5$ & $1$ & non-homogeneous \\
        AV \gateand \cite[Figures 14b, 15a]{litinski2022active} & $7$ & $4$ & $4$ & non-homogeneous \\
    \hline
        Z-type \gateautoand, this work (\cref{lem:parity_autoand}, \href{https://algassert.com/quirk#circuit=%7B%22cols%22%3A%5B%5B%22Y%5Et%22%2C%22X%5Et%22%5D%2C%5B1%2C1%2C1%2C%22%E2%80%A2%22%2C%22%E2%80%A2%22%2C%22X%22%5D%2C%5B1%2C1%2C%22zpar%22%2C1%2C1%2C%22zpar%22%2C%22zpar%22%2C1%2C1%2C1%2C1%2C1%2C%22X%22%5D%2C%5B1%2C1%2C%22zpar%22%2C%22zpar%22%2C1%2C%22zpar%22%2C1%2C%22zpar%22%2C1%2C1%2C1%2C1%2C1%2C%22X%22%5D%2C%5B1%2C1%2C%22zpar%22%2C1%2C%22zpar%22%2C%22zpar%22%2C1%2C1%2C%22zpar%22%2C1%2C1%2C1%2C1%2C1%2C%22X%22%5D%2C%5B1%2C1%2C%22zpar%22%2C%22zpar%22%2C%22zpar%22%2C%22zpar%22%2C1%2C1%2C1%2C%22zpar%22%2C1%2C1%2C1%2C1%2C1%2C%22X%22%5D%2C%5B1%2C1%2C%22%E2%80%A6%22%5D%2C%5B1%2C%22zpar%22%2C1%2C1%2C%22zpar%22%2C1%2C1%2C1%2C1%2C1%2C1%2C%22X%22%5D%2C%5B%22zpar%22%2C1%2C1%2C%22zpar%22%2C1%2C1%2C1%2C1%2C1%2C1%2C%22X%22%5D%2C%5B1%2C1%2C1%2C1%2C1%2C1%2C1%2C1%2C1%2C1%2C%22Measure%22%2C%22Measure%22%5D%2C%5B1%2C1%2C1%2C1%2C1%2C1%2C%22H%22%2C1%2C1%2C1%2C%22%E2%97%A6%22%2C%22%E2%97%A6%22%5D%2C%5B1%2C1%2C1%2C1%2C1%2C1%2C1%2C%22H%22%2C1%2C1%2C%22%E2%97%A6%22%2C%22%E2%80%A2%22%5D%2C%5B1%2C1%2C1%2C1%2C1%2C1%2C1%2C1%2C%22H%22%2C1%2C%22%E2%80%A2%22%2C%22%E2%97%A6%22%5D%2C%5B1%2C1%2C1%2C1%2C1%2C1%2C1%2C1%2C1%2C%22H%22%2C%22%E2%80%A2%22%2C%22%E2%80%A2%22%5D%2C%5B1%2C1%2C1%2C%22H%22%2C%22H%22%2C%22H%22%2C%22H%22%2C%22H%22%2C%22H%22%2C%22H%22%5D%2C%5B1%2C1%2C1%2C%22Measure%22%2C%22Measure%22%2C%22Measure%22%2C%22Measure%22%2C%22Measure%22%2C%22Measure%22%2C%22Measure%22%2C1%2C1%2C%22Measure%22%2C%22Measure%22%2C%22Measure%22%2C%22Measure%22%5D%2C%5B1%2C1%2C%22X%22%2C1%2C1%2C1%2C%22zpar%22%2C1%2C1%2C1%2C%22%E2%97%A6%22%2C%22%E2%97%A6%22%2C%22zpar%22%5D%2C%5B1%2C1%2C%22X%22%2C1%2C1%2C1%2C1%2C%22zpar%22%2C1%2C1%2C%22%E2%97%A6%22%2C%22%E2%80%A2%22%2C1%2C%22zpar%22%5D%2C%5B1%2C1%2C%22X%22%2C1%2C1%2C1%2C1%2C1%2C%22zpar%22%2C1%2C%22%E2%80%A2%22%2C%22%E2%97%A6%22%2C1%2C1%2C%22zpar%22%5D%2C%5B1%2C1%2C%22X%22%2C1%2C1%2C1%2C1%2C1%2C1%2C%22zpar%22%2C%22%E2%80%A2%22%2C%22%E2%80%A2%22%2C1%2C1%2C1%2C%22zpar%22%5D%2C%5B1%2C1%2C%22X%22%2C1%2C1%2C1%2C1%2C1%2C1%2C1%2C%22%E2%80%A2%22%2C%22%E2%80%A2%22%5D%2C%5B%22Z%22%2C1%2C1%2C1%2C1%2C1%2C1%2C1%2C1%2C%22%E2%80%A2%22%2C1%2C%22%E2%97%A6%22%5D%2C%5B1%2C%22Z%22%2C1%2C1%2C1%2C1%2C1%2C1%2C1%2C%22%E2%80%A2%22%2C%22%E2%97%A6%22%5D%2C%5B%22Z%22%2C1%2C1%2C1%2C1%2C1%2C1%2C1%2C%22%E2%80%A2%22%2C1%2C1%2C%22%E2%80%A2%22%5D%2C%5B1%2C%22Z%22%2C1%2C1%2C1%2C1%2C1%2C1%2C%22%E2%80%A2%22%2C1%2C%22%E2%97%A6%22%5D%2C%5B%22Z%22%2C1%2C1%2C1%2C1%2C1%2C1%2C%22%E2%80%A2%22%2C1%2C1%2C1%2C%22%E2%97%A6%22%5D%2C%5B1%2C%22Z%22%2C1%2C1%2C1%2C1%2C1%2C%22%E2%80%A2%22%2C1%2C1%2C%22%E2%80%A2%22%5D%2C%5B%22Z%22%2C1%2C1%2C1%2C1%2C1%2C%22%E2%80%A2%22%2C1%2C1%2C1%2C1%2C%22%E2%80%A2%22%5D%2C%5B1%2C%22Z%22%2C1%2C1%2C1%2C1%2C%22%E2%80%A2%22%2C1%2C1%2C1%2C%22%E2%80%A2%22%5D%2C%5B%22Z%22%2C1%2C1%2C1%2C1%2C%22%E2%80%A2%22%2C1%2C1%2C1%2C1%2C1%2C%22%E2%80%A2%22%5D%2C%5B1%2C%22Z%22%2C1%2C1%2C1%2C%22%E2%80%A2%22%2C1%2C1%2C1%2C1%2C%22%E2%80%A2%22%5D%2C%5B%22Z%22%2C1%2C1%2C%22%E2%80%A2%22%5D%2C%5B1%2C%22Z%22%2C1%2C1%2C%22%E2%80%A2%22%5D%2C%5B1%2C1%2C%22Z%22%2C1%2C1%2C%22%E2%80%A2%22%5D%2C%5B%22%E2%80%A6%22%5D%2C%5B%22%E2%80%A2%22%2C%22%E2%80%A2%22%2C%22X%22%5D%2C%5B%22Y%5E-t%22%2C%22X%5E-t%22%5D%5D%2C%22init%22%3A%5B0%2C0%2C%22%2B%22%2C%22%2B%22%2C%22%2B%22%2C0%2C%22%2B%22%2C%22%2B%22%2C%22%2B%22%2C%22%2B%22%5D%7D}{Quirk}) & $8$ & $6$ & $1$ & homogeneous \\
        Z/X-type \gateautoand, this work (\cref{lem:fanout_autoand}, \href{https://algassert.com/quirk#circuit=%7B%22cols%22%3A%5B%5B%22Y%5Et%22%2C%22X%5Et%22%5D%2C%5B1%2C1%2C%22X%22%2C%22%E2%80%A2%22%2C%22%E2%80%A2%22%5D%2C%5B1%2C1%2C1%2C%22zpar%22%2C1%2C%22zpar%22%2C1%2C1%2C1%2C1%2C1%2C1%2C1%2C%22X%22%5D%2C%5B1%2C1%2C1%2C1%2C%22zpar%22%2C1%2C%22zpar%22%2C1%2C1%2C1%2C1%2C1%2C1%2C1%2C%22X%22%5D%2C%5B1%2C1%2C%22%E2%80%A6%22%5D%2C%5B1%2C1%2C%22xpar%22%2C1%2C1%2C%22xpar%22%2C1%2C%22xpar%22%2C1%2C1%2C1%2C%22X%22%5D%2C%5B1%2C1%2C%22xpar%22%2C1%2C1%2C1%2C%22xpar%22%2C1%2C%22xpar%22%2C1%2C1%2C1%2C%22X%22%5D%2C%5B1%2C1%2C%22%E2%80%A6%22%5D%2C%5B%22zpar%22%2C1%2C1%2C%22zpar%22%2C1%2C1%2C1%2C1%2C1%2C%22X%22%5D%2C%5B1%2C%22zpar%22%2C1%2C1%2C%22zpar%22%2C1%2C1%2C1%2C1%2C1%2C%22X%22%5D%2C%5B1%2C1%2C1%2C%22H%22%2C%22H%22%5D%2C%5B1%2C1%2C1%2C%22Measure%22%2C%22Measure%22%2C1%2C1%2C1%2C1%2C%22Measure%22%2C%22Measure%22%2C%22Measure%22%2C%22Measure%22%2C%22Measure%22%2C%22Measure%22%5D%2C%5B1%2C1%2C1%2C1%2C1%2C1%2C%22H%22%2C1%2C%22H%22%2C%22%E2%80%A2%22%5D%2C%5B1%2C1%2C1%2C1%2C1%2C%22H%22%2C1%2C%22H%22%2C1%2C1%2C%22%E2%80%A2%22%5D%2C%5B1%2C1%2C1%2C1%2C1%2C%22H%22%2C%22H%22%5D%2C%5B1%2C1%2C1%2C1%2C1%2C%22Measure%22%2C%22Measure%22%2C%22Measure%22%2C%22Measure%22%5D%2C%5B%22Z%22%2C1%2C1%2C%22%E2%80%A2%22%5D%2C%5B1%2C%22Z%22%2C1%2C1%2C%22%E2%80%A2%22%5D%2C%5B%22Z%22%2C1%2C1%2C1%2C1%2C1%2C1%2C1%2C1%2C1%2C%22%E2%80%A2%22%2C%22%E2%80%A2%22%5D%2C%5B1%2C%22Z%22%2C1%2C1%2C1%2C1%2C1%2C1%2C1%2C%22%E2%80%A2%22%2C1%2C1%2C%22%E2%80%A2%22%5D%2C%5B1%2C1%2C%22X%22%2C1%2C1%2C1%2C1%2C1%2C1%2C1%2C%22%E2%80%A2%22%2C1%2C1%2C%22%E2%80%A2%22%5D%2C%5B1%2C1%2C%22X%22%2C1%2C1%2C1%2C1%2C1%2C1%2C%22%E2%80%A2%22%2C1%2C1%2C1%2C1%2C%22%E2%80%A2%22%5D%2C%5B%22Z%22%2C1%2C1%2C1%2C1%2C1%2C1%2C1%2C1%2C1%2C%22%E2%97%A6%22%5D%2C%5B1%2C1%2C%22X%22%2C1%2C1%2C%22%E2%80%A2%22%2C1%2C1%2C1%2C1%2C%22%E2%80%A2%22%5D%2C%5B%22Z%22%2C1%2C1%2C1%2C1%2C%22%E2%97%A6%22%2C1%2C1%2C1%2C1%2C%22%E2%97%A6%22%5D%2C%5B1%2C1%2C%22X%22%2C1%2C1%2C1%2C%22%E2%80%A2%22%2C1%2C1%2C%22%E2%80%A2%22%5D%2C%5B1%2C%22Z%22%2C1%2C1%2C1%2C1%2C%22%E2%80%A2%22%2C1%2C1%2C%22%E2%97%A6%22%5D%2C%5B1%2C1%2C%22X%22%2C1%2C1%2C1%2C1%2C%22%E2%80%A2%22%2C1%2C1%2C%22%E2%97%A6%22%5D%2C%5B%22Z%22%2C1%2C1%2C1%2C1%2C1%2C1%2C%22%E2%80%A2%22%2C1%2C1%2C%22%E2%80%A2%22%5D%2C%5B%22%E2%80%A6%22%2C1%2C%22X%22%2C1%2C1%2C1%2C1%2C1%2C%22%E2%80%A2%22%2C%22%E2%97%A6%22%5D%2C%5B1%2C%22Z%22%2C1%2C1%2C1%2C1%2C1%2C1%2C%22%E2%80%A2%22%2C%22%E2%80%A2%22%5D%2C%5B1%2C1%2C%22X%22%2C1%2C1%2C1%2C1%2C1%2C1%2C%22%E2%80%A2%22%2C%22%E2%80%A2%22%5D%2C%5B%22%E2%80%A2%22%2C%22%E2%80%A2%22%2C%22X%22%5D%2C%5B%22Y%5E-t%22%2C%22X%5E-t%22%5D%5D%2C%22init%22%3A%5B0%2C0%2C0%2C%22%2B%22%2C%22%2B%22%2C%22%2B%22%2C%22%2B%22%5D%7D}{Quirk}) & $7$ & $4$ & $1$ & homogeneous \\
    \hline
    \end{tabular}
    \caption{Comparison of new and prior implementations for computing the auto-corrected \gateand gate. Space is the number of qubits of the state, not including the two input qubits. Depth is the quantum depth of the circuit, not including input states. cDepth is the contributive quantum depth, that is the quantum depth that is contributed by the circuit to a much larger algorithm (see~\cref{def:contributive_depth}). For both depth metrics, Hadamards performed immediately prior to single-qubit measurements are subsumed into the measurement, and Paulis are tracked in frame. \textcite[Figure 4]{gidney2019flexible} construct an AutoCCZ gate, which is more powerful than an \gateand, via a series of $CZ$ gates performed on the ancilla state to act as delayed-choice corrections. Note also that this construction employs unitary gates rather than just PPMs and thus may require additional qubits to implement via PPMs. \textcite[Figure 8]{gidney2025how} instead teleports in a $CZ$ gate up to $S$ phase errors via the use of $\ket{+i}$ resource states; thus, it is not fully auto-corrected and requires additional resources. In \textcite[Figures 14b, 15a]{litinski2022active}, delayed-choice $CZ$ gates are performed on the pass-through qubits of the \gateand, rather than on the ancilla system, reducing depth while increasing contributive depth. All implementations have the same reaction depth, and require a $\ket{CCZ}$ or $\toffstate$ state.}
    \label{tab:comparison}
\end{table}

As special cases of Thm~\ref{thm:const_reaction_depth} we describe new implementations of auto-corrected \gateand (\gateautoand) gates. These implementations use only a single Toffoli state and homogeneous PPMs of $Z$- or $X$-type, and make use of further novel implementations of in-place and out-of-place \gatexor gates. The implementations are compared against prior \gateand gates in Table~\ref{tab:comparison}.

In Section~\ref{sec:compilation_strategy} we give examples of compiled Boolean subroutines with low space and depth, including ripple-carry adders~\cite{cuccaro2004new} and schoolbook multipliers~\cite{litinski2024quantum}. We compare our ripple-carry adder to previous implementations in~\cref{tab:adder_comparison}.

\begin{table}[]
    \centering
    \begin{tabular}{|l|c|c|c|c|}
    \hline
        & \textbf{Space} & \textbf{Depth} & \textbf{cDepth} & \textbf{PPMs}\\
    \hline
        Gidney adder \cite[Fig.~8]{gidney2025how} & $9\ell$ & $\sim 3\ell$ & $5$ & non-homogeneous \\
        Litinski adder \cite[Fig.~15, 16]{litinski2022active} & $\geq 9\ell$ & $\sim 6\ell$ & $5$ & non-homogeneous \\
    \hline
        This work (\cref{sec:ripple_carry}) & $10\ell$ & $\sim 3\ell$ & $4$ & almost homogeneous \\
    \hline
    \end{tabular}
    \caption{Comparison of new and prior implementations of the $\ell$-bit ripple-carry adder. Space is the number of qubits, beyond the original two $\ell$-qubit inputs, required. Depth is the quantum depth of the circuit, not including the original state inputs. cDepth is the contributive quantum depth, that is the quantum depth that is contributed by the circuit to a much larger algorithm (see~\cref{def:contributive_depth}). For both depth metrics, Hadamards performed immediately prior to single-qubit measurements are subsumed into the measurement, and Paulis are tracked in frame. All implementations have reaction depth $2\ell$, and require $\ell$ $\ket{CCZ}$ or $\toffstate$ states. Refs.~\cite{gidney2025how,litinski2022active} both use many non-homogeneous PPMs, while our implementation uses entirely homogeneous PPMs, except for at most $\ell$ non-homogeneous PPMs (and an average of $\ell/2$ non-homogeneous PPMs) required for measurement-based uncomputation.}
    \label{tab:adder_comparison}
\end{table}

\subsection{Related work}

Reaction depth is closely related to magic depth. \textcite{selinger2013quantum} studied the $T$-depth of circuits, and found that a broad class of circuits can be implemented in $T$-depth 1. Selinger's result does not imply our~\cref{thm:const_reaction_depth} (or vice versa), but it is similar in spirit.

The closest works to our own are Refs.~\cite{gidney2025how,low2026denser}, which prove a very similar result to our~\cref{thm:const_reaction_depth}; they prove that any Boolean circuit admits an implementation with constant reaction depth, leaving some qubits dirty. We prove that a restrictive gateset of homogeneous PPMs is sufficient for such an implementation, and in that case the quantum depth is also constant. 


\section{Preliminaries}

\subsection{Graph theory}
Throughout this work we rely on elementary graph theory. For convenience we give a brief summary here.

\begin{define}[DAG]
    A Directed Acyclic Graph (DAG) is a directed graph $G = (V,E)$, where $V$ is the set of vertices and $E$ is the set of directed edges, such that $G$ contains no directed cycles. Let $\mathrm{Size}_v(G) = |V|$, the number of vertices in $G$, and $\mathrm{Size}_e(G) = |E|$, the number of edges.
\end{define}

\begin{define}[Path]
    A path in a DAG $G$ is a subgraph $p(V', E')$ of $G$ such that $p$ is a path graph. That is, $V'$ is a set of vertices which can be listed in order
    $(v_1, v_2, \cdots, v_k)$
    with edges exactly $(v_i, v_{i+1}) \in E'$, $\forall i = 1,2,\cdots,k-1$.
    The vertex-length $\mathrm{Length}_v(p) = \mathrm{Size}_v(p)$. The edge-length $\mathrm{Length}_e(p) = \mathrm{Size}_e(p)$.
\end{define}

\begin{define}[Depth]
    The vertex-depth $\mathrm{Depth}_v(G)$ of a DAG $G$ is the vertex-length of the longest path, and similar for edge-depth,
    \begin{align*}
    \mathrm{Depth}_v(G) = \max_{p \in G} \mathrm{Length}_v(p); \qquad
    \mathrm{Depth}_e(G) = \max_{p \in G} \mathrm{Length}_e(p)
    \end{align*}
\end{define}

Note that $\mathrm{Length}_e(p) = \mathrm{Length}_v(p) - 1$, and $\mathrm{Depth}_e(G) = \mathrm{Depth}_v(G) - 1$.

\begin{define}[Labels]
    A labeled DAG is a DAG $G = (V,E, \ell_V, \ell_E)$ where $\ell_V$ and $\ell_E$ are sets of labels assigned to vertices $V$ and edges $E$ respectively.
\end{define}

\begin{define}
    Let $x$ be a label associated only to vertices. The $x$-vertex-size of a DAG is the number of vertices labeled $x$,
    \[\mathrm{Size}_v^x(G) = |\{v \in V\ |\ \ell_V(v) = x \}| .\] 
    Let $y$ be a label associated only to edges. The $y$-edge-size of a DAG is similarly,
    \[\mathrm{Size}_e^y(G) = |\{e \in E\ |\ \ell_E(e) = y\}|.\]
    
    The $x$-vertex-length $\mathrm{Length}_v^x(p)$ of a path $p$ is the number of vertices labeled $x$ in $p$, $\mathrm{Length}_v^x(p) = \mathrm{Size}_v^x(p)$.
    Similarly, given a label $y$ associated only to edges, the $y$-edge-length $\mathrm{Length}_e^y(p)$ of a path $p$ is the number of edges labeled $y$ in $p$, $\mathrm{Length}_e^y(p) = \mathrm{Size}_e^y(p)$.
\end{define}

The $x$-vertex-depth $\mathrm{Depth}_v^x(G)$ and $y$-edge-depth $\mathrm{Depth}_e^y(G)$ are defined accordingly as
\[\mathrm{Depth}_v^x(G) = \max_{p \in G} \mathrm{Length}_v^x(p); \qquad \mathrm{Depth}_e^y(G) = \max_{p \in G} \mathrm{Length}_e^y(p).\]

\subsection{Quantum circuits}\label{sec:quantum_circuits}

\subsubsection{Depth and count of circuits}

Let $\mathcal{C}$ be a quantum circuit, consisting of quantum logic gates and measurements. Assume for now that $\mathcal{C}$ has no classically controlled quantum gates.

\begin{notation}
For simplicity, from now on we shall include measurements, initializations and resets in the term `quantum gates', as well as unitary operations; we do not include idling as a gate.
\end{notation}

We can always view $\mathcal{C}$ as a DAG, where each quantum gate is a vertex and the qubit between two adjacent gates is a directed edge.

\begin{define}[Quantum circuit DAG]
    Let $\mathcal{C}$ be a quantum circuit. Label each quantum gate in $\mathcal{C}$ with a unique index $v$. Let the set of all such indices be $V$. For each pair of vertices $u, v$ corresponding to sequential quantum gates with common support on qubits, let $(u, v)$ be a directed edge from gate $u$ to $v$. Let the set of all edges be $E$. Then $G(\mathcal{C}) = (V, E)$ is the DAG associated to $\mathcal{C}$. 
\end{define}

Note that the quantum circuit DAG omits classical logic and dependencies entirely, and does not incorporate the action of the gates in $V$ on the qubits; it includes only the fact that they exist.

\begin{define}\label{def:quantum_depth}
    The depth $d(\mathcal{C})$ of a quantum circuit is $\mathrm{Depth}_v(G(\mathcal{C}))$. The gate count of a quantum circuit is $\mathrm{Size}_v(G(\mathcal{C}))$.
\end{define}

\begin{remark}
    The above definition of depth presumes that gates which have overlapping support cannot be applied simultaneously. There are some models of fault-tolerant quantum computation in which some gates can be applied simultaneously even when they overlap on logical qubits~\cite{zhang2024time,cowtan2025parallel,litinski2019game,litinski2022active}.
\end{remark}

One can specialize the gate count or depth to specific gate types. For example, the  $T$-count and $T$-depth are $\mathrm{Size}_v^T(G(\mathcal{C}))$ and $\mathrm{Depth}_v^T(G(\mathcal{C}))$ respectively. As high-fidelity magic states are typically more expensive than Clifford operations~\cite{gidney2024magic,sahay2026fold,litinski2019magic,beverland2022assessing}, optimization of magic count has been studied extensively, including in unitary synthesis~\cite{bocharov2015efficient,selinger2014efficient,kliuchnikov2023shorter} and Boolean logic~\cite{low2024trading}.
Magic depth has also been extensively studied~\cite{selinger2013quantum,mukhopadhyay2025quantum}. As we shall see in~\cref{sec:reaction_limited_defs}, in the common setting where we are simulating a unitary by:
\begin{itemize}
    \item Clifford computation,
    \item with Paulis tracked in frame,
    \item augmented by magic states,
\end{itemize}
then the magic depth of that unitary is an upper bound for the reaction depth, as the only corrections which have to be implemented are those incurred by magic state injection \footnote{Strictly speaking, corrections can also be incurred by measurement-based uncomputation (see Sec.~\ref{sec:compilation_strategy}) but this is used as a replacement for magic gates.}. The magic depth of the unitary is, however, \textit{not} a lower bound on the reaction depth, see Appendix~\ref{app:reaction_depth}.

\subsubsection{Pauli-based computation on qLDPC codes}\label{sec:pbc}

Rather than performing unitary gates, a common approach to fault-tolerant computation is to perform logical measurements to evolve the quantum state~\cite{litinski2019game,litinski2022active, he2025extractors, gidney2025how}. In Pauli-based computation (PBC), states are prepared in stabilizer or magic states, and then logic is accomplished by performing Pauli product measurements (PPMs)~\cite{bravyi2016trading}, which take the following form:
\[\ket{\psi} \mapsto \frac{1}{2} (I + (-1)^m P) \ket{\psi},\]
where $P$ is a tensor product of single-qubit Paulis and $m \in \{0, 1\}$ is some potentially nondeterministic measurement outcome. 
The PBC model can differ greatly from the traditional circuit model in which computation is performed by e.g. unitary operations and single-qubit Pauli measurements.

\begin{example}[CNOT via PPMs]\label{ex:CNOT_PPMs}
    \begin{align*}
        \begin{quantikz}[wire types = {q,q,n,n}, row sep={0.9cm,between origins}]
        \lstick{$q_1$}      & \ctrl{1}  &  \\
        \lstick{$q_2$}      & \targ{}   &  \\
        & & \\
        & &
        \end{quantikz} 
    \,&=\,\,\,
    \begin{quantikz}[wire types = {q,q,q,n}, row sep={0.9cm,between origins}]
        \lstick{$q_1$}      & \gate{Z}\wire[d]{b}                   &  &  &                                   & \gate{Z} & & &\\
        \lstick{$q_2$}      & \wire[d]{b}                   & \gate{X} & & \gate{X}\wire[d]{b}                                      & \wire[u]{c} & & \gate{X} &\\
        \lstick{$\ket{+}$}  & \gate{Z}\wire[d]{b}                   & \wire[u]{c}  & & \gate{X}\wire[d]{b}                                      & \wire[u]{c} & \meter{} & \phase{}\wire[u]{c}\setwiretype{c} & \setwiretype{n}\\
                            & \meter{}                              & \phase{}\wire[u]{c}\setwiretype{c}     & \setwiretype{n} & \meter{} & \phase{}\wire[u]{c}\setwiretype{c} & \setwiretype{n}& &
    \end{quantikz}
    \end{align*}
\end{example}

\begin{example}[$Z$ PPM via CNOTs]
    \begin{align*}
    \begin{quantikz}[wire types = {q,q,n,q,n}, row sep={0.9cm,between origins}]
        \lstick{$q_1$}      & \gate{Z}\wire[d]{b}                   &                                       & \\
        \lstick{$q_2$}      & \gate{Z}\wire[d]{b}                   &                                       & \\
                            & \raisebox{0.5em}{$\vdots$}\wire[d]{b} &                                       & \\
        \lstick{$q_n$}      & \gate{Z}\wire[d]{b}                   &                                       & \\
                            & \meter{}                              &  &
    \end{quantikz}
    \,\, &= \,\,\,
    \begin{quantikz}[wire types = {q,q,n,q,q}, row sep={0.9cm,between origins}]
        \lstick{$q_1$}      & \ctrl{4}  &           &                               &           & \\
        \lstick{$q_2$}      &           & \ctrl{3}  &                               &           & \\
                            &           &           & \raisebox{0.5em}{$\ddots$}    &           & \\
        \lstick{$q_n$}      &           &           &                               & \ctrl{1}  & \\
        \lstick{$\ket{0}$}  & \targ{}   & \targ{}   &                               & \targ{}   & \meter{}
    \end{quantikz} 
    \end{align*}
\end{example}

In particular, a single PPM with support on $n$ qubits can, in certain circumstances, perform an operation broadly equivalent to $n$ CNOTs. Note that we describe PPMs using boxes and bold vertical lines connected to a measurement; a glossary for the nonstandard circuit notation in this paper is presented in Appendix~\ref{app:circuit_notation}.

With lattice surgery, surface codes can perform arbitrary PPMs, although, when compared to the abstract circuit, there may be additional overhead due to routing on a 2D grid. Optimally compiling arbitrary quantum programs to lattice surgery in 2D while minimizing spacetime volume is known to be hard~\cite{tan2024sat}.\footnote{Indeed, most optimization problems on quantum circuits are NP-hard, including finding the optimal $T$-count and depth~\cite{vandewetering2024optimising}.}

More general LDPC codes have recently been studied for the purpose of reducing the overhead of fault-tolerance~\cite{breuckmann2021ldpc}. To this end, the notion of surgery has been generalized to arbitrary LDPC codes~\cite{cohen2022low,cowtan2024css,ide2024fault,williamson2024low}, along with other methods of logical measurement using transversal gates~\cite{huang2023homomorphic}. In particular, extractor architectures permit any PPM on any set of logical qubits~\cite{he2025extractors, blue2026extractors}, although extractor-style architectures based on `partial extractors', such as the Tour de gross architecture~\cite{yoder2025tour}, generally require additional compilation into measurements and automorphisms to perform arbitrary PPMs.

Extractor-style architectures are predicted to substantially reduce the space overhead required for fault-tolerant quantum algorithms, such as Shor's algorithm for cryptanalysis~\cite{cain2026shor,webster2026pinnacle}. However, a core problem with extractor-style architectures is that they can have substantially longer runtimes than their surface code equivalents, as the logical measurements may need to be serialized even when acting on disjoint logical qubits in the same codeblock.

A variety of logical measurement approaches for LDPC codes enjoy substantially lower runtimes, including `fast' or `constant-time' surgery~\cite{cowtan2025fast,baspin2025fast,hillmann2024single}, and homomorphic measurements~\cite{huang2023homomorphic,xu2025fast}, along with other approaches that reduce the amortized time cost~\cite{cowtan2025parallel,zheng2025high,gu2026qgpu,zhang2024time}. Such approaches commonly come with a significant caveat: they tend to be only capable of Pauli product measurements in the all-$X$ or all-$Z$ basis.\footnote{For reasons of brevity we defer a discussion of why this caveat is pervasive in Appendix~\ref{app:fast_ldpc_logic}. Note that this caveat does not apply to some schemes, such as Ref.~\cite{cowtan2025parallel}; Refs.~\cite{zheng2025high,cowtan2026nearoptimal} can also be extended to heterogeneous PPMs.} 
For universal computation, an architectural approach based on mixing `fast' and `slow' logical measurement methods may be optimal~\cite{zheng2026logical,bhardwaj2026high}. Nevertheless, this restriction motivates the notion of homogeneous measurements.

\begin{define}
    A ($Z$ or $X$-type) homogeneous PPM is a PPM of the form:
    \[\ket{\psi} \mapsto \frac{1}{2} (I + (-1)^m P) \ket{\psi}\]
    where $P = \bigotimes_i Z_i$ or $P = \bigotimes_i X_i$. 
\end{define}

We shall refer to a homogeneous $Z$-type PPM as belonging to the set $\mathtt{PPM}_Z$, and a homogeneous $X$-type PPM to the set $\mathtt{PPM}_X$. Note that we assume that the PPMs are still \textit{addressable}, meaning that the PPMs can have nontrivial support on any arbitrary set of qubits.

It transpires that a large class of quantum subroutines, namely those consisting of Boolean arithmetic, are amenable to efficient compilation to homogeneous PPMs, assuming a supply of $\toffstate = CCX\ket{+}\ket{+}\ket{0}$ states. 

\begin{remark}
One subtlety is that while the \textnormal{computation} can be done using homogeneous PPMs and $\toffstate$ states, the \textnormal{uncomputation} of variables afterwards (for example in a ripple-carry adder circuit~\cite{cuccaro2004new,gidney2018halving}) can require non-homogeneous PPMs or $\ket{+i}$ states to avoid using further magic. This is why, in Thm~\ref{thm:const_reaction_depth}, we leave some qubits dirty.
\end{remark}
 
\subsection{Boolean logic in quantum computing}

Coherent Boolean logic is a key subroutine for many quantum algorithmic primitives. Quantum adders are used in modular arithmetic for factoring~\cite{gidney2025how}, ladder operators for Hamiltonian simulation~\cite{motlagh2025quantum}, and to implement $Z$ rotations via addition into a resource state~\cite{kitaev2002classical,gidney2018halving}. Primitives like QROM and QRAM, which enable the loading of classical data to be used in a quantum program~\cite{babbush2018encoding}, are Boolean functions. QROM in particular is a key primitive to implement block encodings~\cite{dalzell2025quantum}, which are critical for Hamiltonian simulation~\cite{babbush2018encoding,vonburg2021quantum,lee2021even,low2025fast,low2026denser,mukhopadhyay2024quantum}, solving linear systems~\cite{dalzell2024shortcut}, and other matrix arithmetic~\cite{dalzell2025quantum,}. 

Boolean logic is concerned with Boolean functions operating over binary variables: 
\begin{define}[Boolean and vectorial Boolean functions]
    A Boolean function $f: \{ 0, 1 \}^n \mapsto \{0, 1 \}$ takes in some Boolean inputs $x_1, ... x_n$ and produces a single Boolean output. A vectorial Boolean function $f : \{ 0,1 \}^n \mapsto \{0,1 \}^m$ instead outputs $m$ Boolean values.
\end{define}

A vectorial Boolean function can always be thought of as $m$ Boolean functions.
When describing a Boolean function, it suffices to describe the value of $f$ on all $2^n$ inputs in a truth table. Boolean functions can also be written in the Algebraic Normal Form (ANF), which expresses $f$ as the \gatexor of a series of monomials, e.g., $f(x_1, x_2, x_3) = x_1 x_2 \oplus x_3$. The ANF of \gatexor is $x_1 \oplus x_2$ and of \gateand is $x_1 x_2$. 

We can express various algorithmic primitives via this framework. For example, a single carry-bit binary adder takes in three bits $x, y, c$ and outputs their sum. Then,
\begin{align}
    \texttt{ADD}(x, y, c) = ( \texttt{MAJ}(x, y,c), \gatexor(x, y, c)),
\end{align}
where $\texttt{MAJ}(x, y,c) = xy \oplus xc \oplus yc$ is the majority function on $x,y,c$~\cite{blais2025boolean} and $\gatexor(x, y, c) = x \oplus y \oplus c$.  An $n$-bit adder can be generalized as such.

Similarly, QROM can be thought of as the following unitary transformation:
\begin{align}
    \textnormal{QROM} \ket{k} \ket{y} = \ket{k} \ket{f(k) \oplus y},
\end{align}
for some Boolean function $f: \{0, 1\}^n \mapsto \{0, 1\}^b$ where $n$ is the number of indexing bits and $b$ is the size of the output register. If the output register is initialized to $0$ as $\ket{y} = \ket{0}$, then this transformation loads our classical data.

We can also represent Boolean functions as a circuit comprised of primitive gates (Boolean functions) and wires (inputs/outputs), see e.g. Ref.~\cite{blais2025boolean} for an introduction. We define a circuit as follows:
\begin{define}[Boolean circuits]
A Boolean circuit $\mathtt{C}$ is $n$ input bits $x_1, ... x_n$ and a sequence of gates $g_1, ... g_m$ represented as a directed acyclic graph (DAG). The input of every gate $g_k $ is either an input bit or the output of a prior gate $g_1, ... g_{k -1}$; in DAG notation, every vertex is a gate and every edge is a wire with either an input bit or output of a gate.
\end{define}

In our setting, we only allow $g_i$ to be either \gatexor or \gateand. We permit \textit{unbounded fan-in} for the \gatexor gate, meaning that it can take any number of inputs, and \textit{unbounded fan-out} for both gates, meaning that they can copy the outputs arbitrarily many times. 

\begin{fact}[Universality of $\{\gatexor, \gateand\}$]\label{fact: universality xor and}
    \gatexor and \gateand, along with the constant 1, are sufficient for universal Boolean computation, assuming unbounded fan-out. That is, any Boolean function $f$ can be written as a circuit comprised only of \gatexor and \gateand gates. 
\end{fact}
This fact is immediate upon expressing an arbitrary Boolean function $f$ in ANF. The constant 1 is required to express the \gatenot gate,
\[\texttt{NOT}(x) = \gatexor(x, 1).\]

\gatexor/\gateand circuits have previously been studied as an intermediate representation for Boolean function optimization~\cite{soeken2020determining}.
There exist a few relevant metrics of interest for a Boolean circuit $\mathtt{C}$. 


\begin{define}[Size of circuit]
The number of gates in a Boolean circuit $\mathtt{C}$ is the vertex-size of the DAG,
$\size_v(\mathtt{C}) = |V|$.
\end{define}

\begin{define}[$\andsize$ of circuit (multiplicative complexity)]
The multiplicative complexity of a Boolean circuit $\mathtt{C}$ is the number of \gateand gates it has, 
$\andsize(\mathtt{C}).$
\end{define}

\begin{define}[$\anddepth$ of circuit (multiplicative depth)]
The multiplicative depth of a Boolean circuit $\mathtt{C}$ is the maximum number of $\gateand$ gates that must be evaluated serially to evaluate the circuit, $\anddepth(\mathtt{C})$.
\end{define}

Boolean circuit optimization has been extensively studied~\cite{turan2024optimizing, soeken2018lut,soeken2020determining}. Note that minimizing $\andsize$ or $\anddepth$ are both NP-hard problems~\cite{levin2003tale,find2014complexity}, similar to minimizing $T$-count and $T$-depth in quantum circuits~\cite{vandewetering2024optimising}. Both $\andsize$ and 
$\anddepth$ are crucial metrics for the cost of a coherent Boolean circuit performed on a fault-tolerant quantum computer, as \gateand
 gates require magic states, while \gatexor gates require only Clifford gates and \gatenot gates are Pauli $X$s. Thus when a Boolean circuit $\mathtt{B}$ is implemented on a quantum computer by a quantum circuit $\mathcal{C}$ which contains only input states, Clifford gates and Pauli measurements, \gatenot gates are free, as they require only Pauli frame updates \footnote{See Fact~\ref{fact:pauli_frame}.}.
 
In quantum computing, using a generic compiler to convert Boolean circuits to PPMs can increase overheads unnecessarily. Boolean circuits are often expressed first as quantum circuits composed of unitaries and single-qubit measurements, which incurs overheads when then compiling to PPMs~\cite{gidney2025classical}. For example, CNOT gates, which are essentially \gatexor gates, are described as requiring $\pi/4$ rotations, immediately requiring the use of $Y$-basis measurements or $\ket{i}$ states~\cite[Fig.~11(b)]{litinski2019game}, or additional measurements, see Example~\ref{ex:CNOT_PPMs}. Additionally, compilation of Boolean logic has largely been specialized to the surface code rather than general QEC or qLDPC codes. Surface code compilation focuses primarily on minimizing routing overheads by planning lattice surgery interactions manually or via numerical approaches ~\cite{de2020zx,caesura2025faster,gidney2025how,litinski2022active,gidney2019flexible,khattar2025verifiable}. In contrast, qLDPC codes can be engineered to perform all-to-all PPMs within a code block~\cite{cowtan2025parallel, gu2026qgpu}, and routing constraints on qLDPC code-based architectures are vastly different~\cite{he2025extractors}, where they are present at all. Non-Pauli unitary Clifford gates in qLDPC codes can also be much more expensive or less addressable than their surface code equivalents~\cite{chakraborty2026nogo}.
In this work we compile Boolean circuits directly to PPMs.

\section{Reaction-limited computation}\label{sec:reaction_limited_defs}

\subsection{Decoding and reaction time}

In fault-tolerant quantum computing, there is always some latency between a given syndrome extraction cycle being performed and the classical coprocessor successfully decoding that cycle and being able to apply corrections on the quantum computer~\cite{Terhal2015}. This latency comes partly from the \textit{physical} latency of communication between the quantum processor and the classical coprocessor, but primarily from the difficulty of the decoding problem itself~\cite{battistel2023realtime, chamberland2023techniques}. Should the time between a given syndrome extraction circuit being performed and the decoder outputting a correction be too long, the logical program may need to stall until the correction is determined. This is most relevant for magic state injection, where the correction is not merely a Pauli frame update.

Indeed, let $r_{\rm gen}$ be the rate at which syndrome bits are generated from the quantum circuit and let $r_{\rm proc}$ be the rate at which syndrome bits are processed by the classical coprocessor. Then if $r_{\rm gen}/r_{\rm proc} > 1$ then the quantum computer will undergo a slowdown exponential in the depth of magic state injections~\cite[Sec.~G.3]{Terhal2015}. This is because, during a magic state injection, the decoder must ascertain the result of a logical measurement which induces a non-Pauli Clifford correction. One can delay the Clifford correction while the coprocessor finishes decoding, but this Clifford must generally be applied before the next magic state injection on those qubits. While the logical quantum circuit is idling to wait for the Clifford correction, it generates further syndrome data, further slowing the decoding required for the \textit{next} correction, and so on.

While this result could naively be interpreted as implying that the classical coprocessor must be fast enough to decode each layer of syndrome extraction before the next layer is performed, that is a sufficient but not necessary condition, and is too strong for most realistic devices.\footnote{Particularly, practical decoders decode a spacetime code consisting of multiple syndrome rounds simultaneously, rather than decoding each syndrome round 1-by-1~\cite{hillmann2025localized, hillmann2024single, higgott2025sparseblossom, skoric2023parallel}.} Instead, the relevant metric is the \textit{throughput} of the coprocessor, which can be high even when the latency is high.

Let $\tau$ be the time that a bit spends being processed by the system, before it is ejected. Then
    \[r_{\rm proc} = L/\tau ,\]
where $L$ is the number of bits processed by the system simultaneously.

If we consider $\tau$ to be the time for the data from a given syndrome extraction round to be sent to the classical coprocessor, decoded, and then corrective instructions returned to the quantum computer, then $r_{\rm proc}$ is the processing rate of the coprocessor, which must be sufficiently large to avert the backlog problem. Thus if $\tau$ is large then $L$ must also be large to compensate.

An important consequence of the above is that there may still be some delay after a magic state injection is performed, in order for the decoder to deduce a Clifford correction before the next magic state injection, as there is a non-negligible latency $\tau$ between the final syndrome round of a logical measurement and its correction being ascertained and applied. But, crucially, this delay is constant (on average), and there is no backlog problem when $r_{\rm proc}$ is sufficiently high. In other words, while the logical circuit may need to idle for time $\tau$ between magic state injection rounds, those idling rounds are processed at a high enough rate to avoid an exponential build-up of syndrome data in the queue.\footnote{In general the logical circuit will typically be performing logical Cliffords between rounds of magic state injection, rather than merely idling; idling is necessary only if there are no more Cliffords to be performed before the next injection, and time $\tau$ has not elapsed yet.}

The regime in which (a) the latency of the coprocessor is non-negligible but (b) the throughput is still high enough to avoid the backlog problem is exactly where the notion of reaction-limited computation becomes relevant.
In short, given a quantum circuit $\mathcal{C}$ with fast enough logical quantum operations, and a sufficient density of magic state injections, the time to perform $\mathcal{C}$ can be dominated by the reaction time of the coprocessor:

\begin{define}[Reaction-limited]\label{def:reaction_limited_computation}
    Suppose a quantum architecture $\mathcal{A}$ has classical reaction time $\tau$ and a quantum circuit $\mathcal{C}$ has reaction depth $r(\mathcal{C})$ and runtime $t$ on $\mathcal{A}$. Then, the execution of $\mathcal{C}$ on $\mathcal{A}$ is \textit{reaction-limited} if and only if
    \begin{align}
        t \sim r(\mathcal{C}) \cdot \tau.
    \end{align}
\end{define}

Of course, this definition only makes sense once the reaction depth $r(\mathcal{C})$ is well-defined, which is the purpose of the following section. \footnote{Def.~\ref{def:reaction_limited_computation} is essentially a rewording of a statement in the introduction of Ref.~\cite{gidney2019flexible}.}

Fast hardware modalities, such as superconducting and photonic qubits~\cite{acharya2024quantumerrorcorrectionsurface,aghaee2025scaling}, are more likely to be in the reaction-limited regime, as the cycle times are on the scale of $\mu s$. Thus unless the decoding is extremely straightforward, $\tau$ is likely to be non-negligible. Neutral atoms and ion traps, conversely, are slower, with cycle times on the scale of $ms$ or higher at the time of writing~\cite{ransford2026helios,bluvstein2026quera}. However, accurate decoding at logical error rates suitable for the teraquop regime may still lead to high $\tau$, particularly if expensive post-processing on belief propagation is used to extend the waterfall region when decoding qLDPC codes~\cite{roffe2020decoding,panteleev2021degenerate,bhardwaj2026high}, and as cycle times improve the reaction time may become a concern.

\subsection{Reaction depth}


We will now generalize our definitions from Section~\ref{sec:quantum_circuits} to include classical logic.

\begin{define}[Classical-quantum circuit DAG]\label{def:cq_circuit_dag}
     Let $\mathcal{C}$ be a quantum circuit, potentially including classically conditioned quantum gates. Associate to each quantum gate in $\mathcal{C}$ a unique index $v$. Let the set of all such indices be $V$. For each pair of vertices $u, v$ corresponding to sequential quantum gates with common support on qubits, let $(u, v)$ be a directed edge from gate $u$ to $v$. Label each edge with the string $\mathtt{q}$ (for quantum).

     For each measurement $m$, let $\{v_1, v_2, \cdots, v_k\}$ be the set of quantum gates conditioned on the outcome of $m$. Then let $\{h_1, h_2,\cdots h_k\} = \{(m, v_1), (m, v_2), \cdots, (m, v_k)\}$ be a set of directed edges from $m$ to each classically conditioned gate. Label each such edge $h_i$ with the string $\mathtt{c}$ (for classical).

     Let $E$ be the set of all edges, let $E_Q$ be the subset of $E$ whose labels are $\mathtt{q}$ and let $E_C$ be the subset of $E$ whose labels are $\mathtt{c}$, such that $E = E_Q \sqcup E_C$.

     Then $H(\mathcal{C}) = (V, E_Q \sqcup E_C)$ is the classical-quantum DAG associated to $\mathcal{C}$.
\end{define}

Note that just as the quantum circuit DAG ignores the action of the quantum gates on qubits, the classical-quantum circuit DAG also ignores the nature of the \textit{classical} coprocessing: the only thing it accounts for is which quantum gates are classically dependent on the outcomes of which prior measurements.

Consider a path $p$ in a classical-quantum circuit DAG. $\mathrm{Length}_e^\mathtt{c}(p)$ counts the number of sequential times the quantum device may need to consult the classical coprocessor before applying a further quantum gate. So, for example, a measurement followed by a gate conditioned on the outcome adds 1 to the length.

\begin{define}[Reaction depth of a circuit]\label{def:reaction_depth}
    The reaction depth $r(\mathcal{C})$ of a quantum circuit $\mathcal{C}$ is,

    \[r(\mathcal{C}) := \mathrm{Depth}_e^\mathtt{c}(H(\mathcal{C})).\]
\end{define}

\begin{example}[Qubit teleportation]
In the standard qubit teleportation circuit, the reaction depth is 1. There are two measurements with corrections, and the corrections are performed sequentially, but the two measurements may be performed simultaneously.

    \begin{align*}
        \mathcal{C} = 
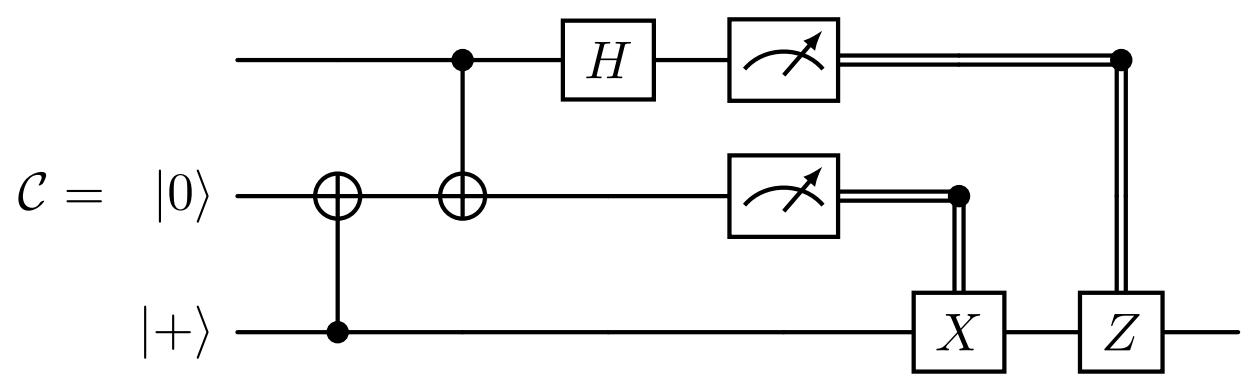
        \begin{quantikz}[wire types = {q,q,q}, row sep={0.9cm,between origins}]
                                & &  \ctrl{1} & \gate{H} & \meter{} & \setwiretype{c} & \phase{}\wire[d]{c} & \setwiretype{n} \\
        \lstick{$\ket{0}$}      & \targ{} & \targ{} & & \meter{} & \phase{}\wire[d]{c}\setwiretype{c} & \wire[d]{c}\setwiretype{n} & \\
        \lstick{$\ket{+}$}       & \ctrl{-1} & & & & \gate{X} & \gate{Z}& 
        \end{quantikz}
    \end{align*}
The DAG $H(\mathcal{C})$ is shown below. Quantum dependencies are solid edges, and classical dependencies are dashed edges. There is no path through the DAG which includes both classical wires, and so the reaction depth is 1.
    \begin{align*}
        H(\mathcal{C})\ =\ 
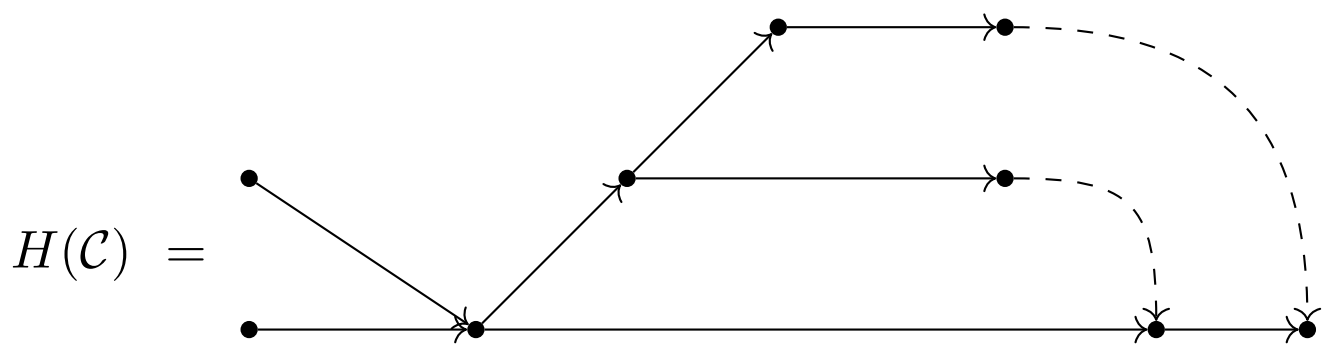
\begin{tikzpicture}
	\begin{pgfonlayer}{nodelayer}
		\node [style=boundary vertex] (0) at (-4, -1) {};
		\node [style=boundary vertex] (1) at (-2, 1) {};
		\node [style=boundary vertex] (2) at (0, 3) {};
		\node [style=boundary vertex] (3) at (3, 3) {};
		\node [style=boundary vertex] (4) at (3, 1) {};
		\node [style=boundary vertex] (5) at (5, -1) {};
		\node [style=boundary vertex] (6) at (7, -1) {};
		\node [style=boundary vertex] (7) at (-7, 1) {};
		\node [style=boundary vertex] (8) at (-7, -1) {};
	\end{pgfonlayer}
	\begin{pgfonlayer}{edgelayer}
		\draw [style=arrow] (0) to (1);
		\draw [style=arrow] (1) to (4);
		\draw [style=arrow] (1) to (2);
		\draw [style=arrow] (2) to (3);
		\draw [style=arrow] (0) to (5);
		\draw [style=arrow] (5) to (6);
		\draw [style=dashed arrow, bend left=45, looseness=1.50] (4) to (5);
		\draw [style=dashed arrow, bend left=45, looseness=1.25] (3) to (6);
		\draw [style=arrow] (8) to (0);
		\draw [style=arrow] (7) to (0);
	\end{pgfonlayer}
\end{tikzpicture}
    \end{align*}
\end{example}

\begin{remark}
    The input and output wires of $\mathcal{C}$ are not included in the DAG $H(\mathcal{C})$. This is because they do not affect the quantum or reaction depths; the only reason to include them would be to compose DAGs directly, as opposed to composing circuits and then mapping to the DAG. It is straightforward to extend the formalism to include input and output wires, but the definitions for composition require further machinery~\cite{fong2015decoratedcospans}, which we do not need in this paper.
\end{remark}

~\cref{def:quantum_depth} extends immediately to classical-quantum DAGs, by letting $d(\mathcal{C}) = \mathrm{Depth}_v(H(\mathcal{C}))$; now, however, the longest path can traverse classical wires as well as quantum ones. 
The reaction depth lower bounds the quantum circuit depth, that is $r(\mathcal{C}) \leq d(\mathcal{C})$, as the longest path along classical wires is a path through the DAG $H(\mathcal{C})$, with edges between quantum gates.

\begin{fact}
    Both quantum circuit depth and reaction depth are subadditive in general, and additive in the worst case. That is, if we compose two quantum circuits $\mathcal{C}_1$ and $\mathcal{C}_2$,

    \[d(\mathcal{C}_2 \circ \mathcal{C}_1) \leq d(\mathcal{C}_1) + d(\mathcal{C}_2),\]
    \[r(\mathcal{C}_2 \circ \mathcal{C}_1) \leq r(\mathcal{C}_1) + r(\mathcal{C}_2).\]
\end{fact}
These facts are immediate from DAG composition. We illustrate the subadditivity of reaction depth in Appendix~\ref{app:reaction_depth} using an example of a small Toffoli circuit from Ref.~\cite{gidney2026eat}. There, we also discuss several other subtleties of reaction depth.

Not all shots of the same circuit will have to consult the classical coprocessor the same number of times; nor will all shots have to perform the same number of quantum operations. For example, a measurement result could determine whether a large subcircuit, itself containing many measurements with classically controlled gates, is performed or not. When the measurement result is 0, that entire circuit does not need to be performed, and so the classical coprocessor needs to be consulted far fewer times, and the quantum program employs far fewer quantum operations. To accommodate for this possibility, we also define the DAG associated to a circuit \textit{run}.

Let $\mathfrak{M}$ be the $\F_2$-vector space of all possible measurement outcomes with classically conditioned quantum gates in $\mathcal{C}$; observe that $\dim \mathfrak{M} \leq \dim \F_2^{E_C}$. The reaction depth of any particular shot of $\mathcal{C}$ is determined by a vector $b \in \mathfrak{M}$, which we view as a bitstring.
    
\begin{define}[Circuit run DAG]\label{def:circuit_run_dag}
    Let $b \in \mathfrak{M}$ be the string of measurement outcomes in $\mathcal{C}$ which have other quantum gates dependent on them. For each classical computation $f$ which takes as input a substring of $b$ and whose output controls at least one quantum gate in $\mathcal{C}$, compute the output of $f$.

    Then $H_b(\mathcal{C})$, the DAG for a circuit run with outcomes $b$, is constructed in the following manner:
    \begin{enumerate}
        \item For each quantum gate which is applied given the output of each $f$, associate a vertex $u$ with label $\mathtt{a}$ for `applied'.
        \item For each quantum gate which is not applied given the output of each $f$, associate a vertex $v$ labeled $\mathtt{n}$ for `not applied'.
        \item Add directed edges labeled $\mathtt{q}$ and $\mathtt{c}$ as in Def.~\ref{def:cq_circuit_dag}, but with the modifications:
        \begin{enumerate}
            \item Any $\mathtt{c}$-labeled edges whose source is a measurement that is not applied, and hence labeled $\mathtt{n}$, are deleted.
            \item Any target of a deleted $\mathtt{c}$-labeled edge is labeled $\mathtt{n}$.
        \end{enumerate}
    \end{enumerate}
\end{define}

The definition of $H_b(\mathcal{C})$ demands knowledge of the actual actions of the classical operations to determine the outputs of each $f$, rather than just the combinatorial structure of $H(\mathcal{C})$.

\begin{remark}
There are generally other measurement outcomes, which do not have further quantum gates conditioned upon them, in $\mathcal{C}$. These are ignored in our definitions as they do not affect reaction depth.

We further ignore the possibility that there are detectors~\cite{derks2025designing} at the level of the logical circuit, which preclude some output bitstrings (without a logical fault having occurred). It is straightforward to modify the definition to account for this possibility, by searching only over bitstrings $b$ which do not contradict the detectors.
\end{remark}

\begin{define}[Quantum depth of a run]
    The quantum depth of the run of a quantum circuit $\mathcal{C}$ with measurement outcomes $b$ is
    \[d_b(\mathcal{C}) =  \mathrm{Depth}_v^\mathtt{a}(H_b(\mathcal{C})).\]
\end{define}

\begin{define}[Reaction depth of a run]\label{def:reaction_depth_run}
    The reaction depth of the run of a quantum circuit $\mathcal{C}$ with measurement outcomes $b$ is 
    \[r_b(\mathcal{C}) = \mathrm{Depth}_e^\mathtt{c}(H_b(\mathcal{C})).\]
\end{define}

Practical quantum algorithms tend to rely on simple classical computations, such as low-depth \gatexor and \gateand circuits, to determine the quantum corrections~\cite{low2025fast, gidney2021factoring, gidney2025how}. In general, however, the reaction depth of any given run with outcomes $b$ could be uncomputable; for example, the halting problem could be embedded into one of the classical computations $f$. 

\begin{example}[Repeat-until-success reset]\label{ex:rus_reset}
A straightforward, albeit contrived, example of a circuit whose runs have differing reaction depths is as follows. The aim is to output a qubit in the $\ket{0}$ state, using only $\ket{+}$ states and measurements in the $Z$ basis. The circuit measures a qubit in the $Z$ basis; if it outputs 1, a qubit is initialized in the $\ket{+}$ state and measured in the $Z$ basis, etc.
    \begin{align*}
    \mathcal{C} = 
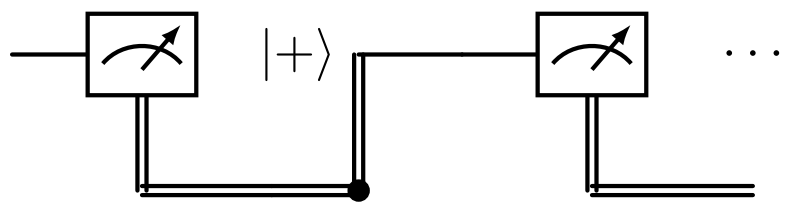
    \begin{quantikz}[wire types = {q,n}, row sep={0.9cm,between origins}]
              &      \meter{}             & \setwiretype{n} & \lstick{$\ket{+}$} & \setwiretype{q}                     & \meter{} &  \cdots \setwiretype{n}\\
        & \wire[u]{c} & \setwiretype{c} & \phase{}\wire[u]{c}& \setwiretype{n} & \wire[u]{c} & \setwiretype{c} 
    \end{quantikz}
    \end{align*}
The DAG for the reset circuit is shown below,
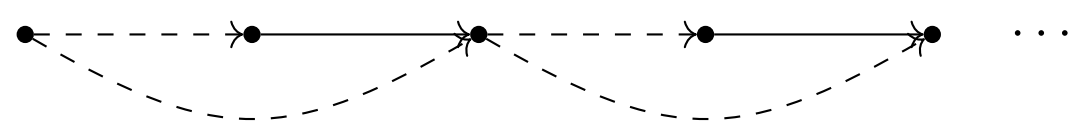
\begin{align*}H(\mathcal{C})\ =\ \begin{tikzpicture}
	\begin{pgfonlayer}{nodelayer}
		\node [style=boundary vertex] (0) at (-6, 0) {};
		\node [style=boundary vertex] (1) at (-3, 0) {};
		\node [style=boundary vertex] (2) at (0, 0) {};
		\node [style=boundary vertex] (3) at (3, 0) {};
		\node [style=boundary vertex] (4) at (6, 0) {};
		\node [style=none] (5) at (7.5, 0) {$\cdots$};
	\end{pgfonlayer}
	\begin{pgfonlayer}{edgelayer}
		\draw [style=arrow] (1) to (2);
		\draw [style=arrow] (3) to (4);
		\draw [style=dashed arrow] (0) to (1);
		\draw [style=dashed arrow] (2) to (3);
		\draw [style=dashed arrow, bend right, looseness=1.25] (0) to (2);
		\draw [style=dashed arrow, bend right, looseness=1.25] (2) to (4);
	\end{pgfonlayer}
\end{tikzpicture}\end{align*}
Conditioned on the output of the first measurement, the following initialization and measurement occur; conditioned on the output of the second measurement, the following initialization and measurement occur, and so on.

If any of the measurements return 0 in a given run, the DAG $H_b(\mathcal{C})$ is truncated there and all vertices and edges past that point are labeled $\mathtt{n}$.
If, for example, the reset circuit returns 0 at the first measurement, then the quantum depth $d_b(\mathcal{C}) = 1$, as there is only one measurement. 

The reaction depth $r_b(\mathcal{C}) = 1$ also. The initialization in $\ket{+}$ is labeled $\mathtt{n}$, as are all subsequent gates, and the following classical edges are deleted entirely. Explicitly, the DAG $H_b(\mathcal{C})$, where $b_0 = 0$, is shown below:
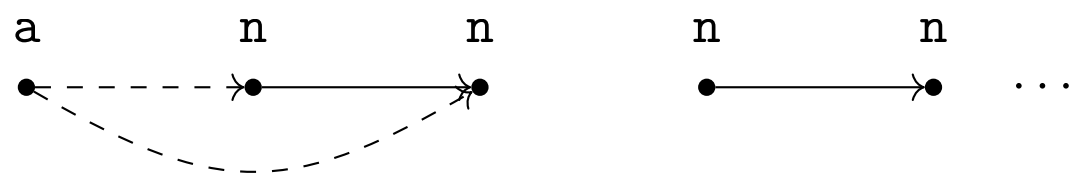
\begin{align*}H_b(\mathcal{C})\ =\ \begin{tikzpicture}
	\begin{pgfonlayer}{nodelayer}
		\node [style=boundary vertex] (0) at (-6, 0) {};
		\node [style=boundary vertex] (1) at (-3, 0) {};
		\node [style=boundary vertex] (2) at (0, 0) {};
		\node [style=boundary vertex] (3) at (3, 0) {};
		\node [style=boundary vertex] (4) at (6, 0) {};
		\node [style=none] (5) at (7.5, 0) {$\cdots$};
		\node [style=none] (6) at (-6, 0.75) {$\mathtt{a}$};
		\node [style=none] (7) at (-3, 0.75) {$\mathtt{n}$};
		\node [style=none] (8) at (0, 0.75) {$\mathtt{n}$};
		\node [style=none] (9) at (3, 0.75) {$\mathtt{n}$};
		\node [style=none] (10) at (6, 0.75) {$\mathtt{n}$};
	\end{pgfonlayer}
	\begin{pgfonlayer}{edgelayer}
		\draw [style=arrow] (1) to (2);
		\draw [style=arrow] (3) to (4);
		\draw [style=dashed arrow] (0) to (1);
		\draw [style=dashed arrow, bend right, looseness=1.25] (0) to (2);
	\end{pgfonlayer}
\end{tikzpicture}\end{align*}
\end{example}

The above example illustrates the particularities of Def.~\ref{def:reaction_depth_run}: even if an applied measurement's output results in no quantum gates being applied, the classical edges are still present, and still contribute to the reaction depth of a run. That is, the classical co-processor still needs to be consulted in order to know whether the quantum gates need to be applied or not. But the following measurements, which are not applied and labeled $\mathtt{n}$, do not have any classical edges. 

\begin{example}[$R_Z$ injection]~\label{ex:rz_injection}
For a more practical example, magic state injection of a $R_Z(\pi/{2^\ell})\ket{+}$ state, for $\ell \in \mathbb{N}$, leads to circuit runs with different reaction depths when $\ell \geq 3$. Depending on the outcome of the first injection measurement, a corrective $R_Z(\pi/{2^{\ell-1}})\ket{+}$ injection may be required, and so on, descending down the Clifford hierarchy. The reaction depth of the circuit is $\ell-1$, ignoring Pauli corrections, but the reaction depth of a given circuit run can be as low as 1.
\end{example}

See~\cref{ex:T_consumption} below for the case where $\ell = 2$.
Note that $r(\mathcal{C}) \geq \max_b\{r_b(\mathcal{C})\}$, and so the reaction depth for the circuit can be considered an approximation of the maximal number of times the classical co-processor must be consulted during execution of $\mathcal{C}$. 

\begin{remark}
    The reason why $r(\mathcal{C})$ is defined purely at the level of $\mathcal{C}$, rather than as $\max_b\{r_b(\mathcal{C})\}$, is because the latter would generally take time exponential in $\dim \mathfrak{M} \leq |E_C|$ to compute, even if the classical functions which determine quantum corrections take $\CO(1)$ time to compute (and they may be uncomputable). For quantum algorithms with potentially billions or trillions of magic state injections~\cite{babbush2018encoding, low2026denser, gidney2025how}, this is a prohibitive metric to compute. By contrast, $r(\mathcal{C})$ is computable in $\CO(|V|+|E|)$ time by topsorting $H(\mathcal{C})$. Thus in a sense our definition favors computability at the potential expense of accuracy, as the bound $r(\mathcal{C}) \geq \max_b\{r_b(\mathcal{C})\}$ is not always tight.
\end{remark}

As can be seen by Example~\ref{ex:rus_reset}, the reaction depth $r(\mathcal{C})$, or even the maximal reaction depth of a circuit run $\max_b\{r_b(\mathcal{C})\}$, may not be the most relevant metrics in all circumstances. The repeat-until-success state preparation circuit has
\[r(\mathcal{C}) = \max_b\{r_b(\mathcal{C})\} = \infty,\]
but the \textit{average} reaction depth of a circuit run is only 2, assuming that the first measurement has a 50\% chance of outputting 0, as the expected number of times the $Z$ measurement must be performed in order to acquire a $\ket{0}$ state is 2.

Similarly, in~\cref{ex:rz_injection} the injection of a $R_Z(\pi/{2^\ell})\ket{+}$ state, when $\ell \geq 2$, has $r(\mathcal{C}) = \max_b\{r_b(\mathcal{C})\} = \ell-1$, ignoring Pauli corrections. But the average reaction depth of a given run is $2- 2^{2-\ell}$. In particular, when $\ell = 2$ the circuit injects $\ket{T}$ and the average reaction depth of a circuit run is 1 the same as its maximal reaction depth. When $\ell=3$ the circuit injects $\ket{\sqrt{T}}$ and the average reaction depth of a  circuit run is 3/2, but its maximal reaction depth is 2.

\begin{fact}[Pauli frame tracking]\cite{knill2005quantum}\label{fact:pauli_frame}
    Let $\mathcal{C}$ be a circuit composed of arbitrary input states, Pauli measurements (including products) and Clifford gates, such that all qubits are measured out at the end. Then all classically-controlled Pauli gates in $\mathcal{C}$ can be commuted to the end of the circuit to become purely classical corrections.
\end{fact}

The set of classical corrections which must be applied is known as the Pauli frame, which is updated as the quantum circuit runs~\cite{knill2005quantum}.\footnote{A similar principle underpins the Pauli frame in stabilizer quantum error correction.} When commuting a Pauli through the circuit, if it encounters any anticommuting Pauli (product) measurement, any further correction dependent on that measurement will pick up a dependency on the initial Pauli.

Note that, according to our definitions, commuting the Paulis to the end and updating the Pauli frame correspondingly is an \textit{optimization} on a circuit which reduces the reaction depth, rather than in-built into the definition of reaction depth. 
As commuting Paulis to the end of a Clifford circuit is a trivial and canonical optimization, we give a modified definition of reaction depth, which is more useful in practice.

\begin{define}[$\mathtt{p}$-reaction depth]\label{def:p_reaction_depth}
    Let $\mathcal{C}$ be a circuit composed of arbitrary input states, Pauli measurements (including products) and Clifford gates.
    Define the {\rm Pauli-deferred circuit} $\mathcal{C}^\mathtt{p}$ as the modification of $\mathcal{C}$ by commuting all Pauli gates to the end of the circuit, past any final measurements.

    Then the $\mathtt{p}$-reaction depth $r^\mathtt{p}(\mathcal{C})$ is,
    \[r^\mathtt{p}(\mathcal{C}) := r(\mathcal{C}^\mathtt{p}),\]
    and the $\mathtt{p}$-reaction depth of a circuit run with outcomes $b$ is,
    \[r_b^\mathtt{p}(\mathcal{C}) := r_b(\mathcal{C}^\mathtt{p}).\]
\end{define}

Note that when a measurement $M_1$ has a Pauli gate $P$ dependent on its outcome which is commuted past a future measurement $M_2$ of Pauli $Q$ such that $[P, Q] \neq 0$, and $M_2$ also has a gate $U$ dependent on its outcome, $M_1$ picks up a classical wire with target $U$.

\begin{notation}
    From now on we shall abuse notation and always use the term {\rm reaction depth} to refer to $\mathtt{p}$-{\rm reaction depth}, as the circuits will all be Clifford, with Pauli measurements and arbitrary input states. The same holds for quantum depth. There will still be some Pauli gates present in a circuit when there are output wires, but it is assumed that when this circuit is run as part of a quantum program the Paulis can be commuted past any future measurements, and so never have to be applied explicitly.
\end{notation}

A common approach to reducing the quantum depth of a quantum circuit which includes magic state injection is to use \textit{auto-correction} circuits. These circuits typically trade off quantum circuit depth for extra qubits, and can result in reaction-limited circuits. Critically, these circuits do not reduce the reaction depth; rather, they shift quantum depth to auxiliary qubits, thereby reducing the overall depth of a quantum circuit (see Appendix~\ref{app:reaction_depth} for a worked out example). Rather than performing injection and then applying a Clifford correction to the output qubits, they instead apply the Clifford correction to ancilla qubits so that the only correction necessary on the output qubits is Pauli. 

\begin{example}[$\ket{T}$ state consumption circuit]\label{ex:T_consumption}
A standard $\ket{T}$ state injection circuit, using a $Z\otimes Z$ measurement, requires an $S$ gate correction. 
The auto-corrected $\ket{T}$ state injection circuit~\cite{litinski2022active} requires only a $Z$ correction on the output qubit, using an adaptive measurement (shown below using a $V = HSH$ gate to change the measurement basis) on the ancilla qubit hosting the $\ket{T}$ state instead.
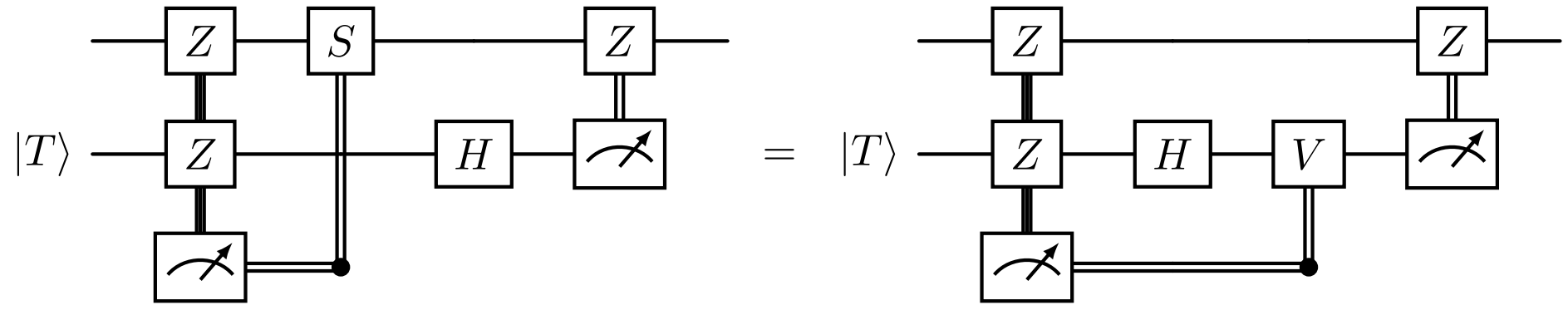
\begin{align*}
        \begin{quantikz}[wire types = {q,q,n}, row sep={0.9cm,between origins}]
                                & \gate{Z}\wire[d]{b} &  \gate{S} & & \gate{Z} & \\
        \lstick{$\ket{T}$}      & \gate{Z}\wire[d]{b} & \wire[u]{c}  & \gate{H} & \meter{}\wire[u]{c} & \setwiretype{n}\\
                                & \meter{} & \phase{}\wire[u]{c}\setwiretype{c} & \setwiretype{n} & &
        \end{quantikz}
        \, &= \,\,
        \begin{quantikz}[wire types = {q,q,n}, row sep={0.9cm,between origins}]
                                 & \gate{Z}\wire[d]{b} &  & & \gate{Z} & \\
        \lstick{$\ket{T}$}      & \gate{Z}\wire[d]{b} & \gate{H}  & \gate{V} & \meter{}\wire[u]{c} & \setwiretype{n}\\
                                & \meter{} & \setwiretype{c} & \phase{}\wire[u]{c}&\setwiretype{n}  &
        \end{quantikz}
\end{align*}
The overall quantum depth of the r.h.s. circuit is not lower than on the l.h.s., and neither is the reaction depth, but crucially the quantum depth has been moved into the ancilla system rather than the output qubits; the longest path terminates on the destructive $Z$-measurement, as the corrective $Z$ gate will be absorbed into future measurements and hence applied classically. Therefore, composing the auto-corrected circuit with further quantum operations will result in a circuit with lower quantum depth overall.
\end{example}

To formalize the notion of ``offloading'' depth from the main pathway of the computation onto ancillae, we define \textit{contributive depth}.

\begin{define}[Contributive depth]\label{def:contributive_depth}
    Let $\mathcal{C}$ be a quantum circuit. We say that the input qubits to $\mathcal{C}$ are the wires which enter the circuit from the left; output qubits are wires which exit the circuit on the right. Let the classical-quantum circuit DAG $H(\mathcal{C})$ be initially defined in Pauli-deferred form, as in~\cref{def:p_reaction_depth}.
    \begin{itemize}
        \item Consider the set of input qubits to $\mathcal{C}$. For each input qubit, label the vertex corresponding to the first gate in $H(\mathcal{C})$ that qubit encounters $\mathtt{i}$.
        \item Consider the set of output qubits from $\mathcal{C}$. For each output qubit, label the vertex corresponding to the last gate in $H(\mathcal{C})$ that qubit comes from $\mathtt{o}$.
    \end{itemize}

    Let $S_{\mathtt{io}}$ be the set of all paths from vertices labeled $\mathtt{i}$ to vertices labeled $\mathtt{o}$ in $H(\mathcal{C})$. Then the contributive quantum circuit depth is,
    \[\mathrm{cd}(\mathcal{C}) = \max_{p \in S_{\mathtt{io}}} \mathrm{Length}_v(p).\]

    Vertices may be labeled both $\mathtt{i}$ and $\mathtt{o}$, such that $\mathrm{cd}(\mathcal{C})$ may be 1; if there are no vertices labeled $\mathtt{i}$ and $\mathtt{o}$ then $\mathrm{cd}(\mathcal{C}) = 0$.
\end{define}
Note that state initializations are not inputs, and measured-out qubits are not outputs, as they are internal to the circuit diagram.

Continuing with Example~\ref{ex:T_consumption}, let the l.h.s. circuit be $\mathcal{C}$ and the r.h.s. circuit be $\mathcal{D}$. Counting the $\ket{T}$ state initialization but discounting the $Z$ gate correction, the l.h.s. circuit has $d(\mathcal{C}) = 4$, while the r.h.s. circuit has $d(\mathcal{D}) = 5$. However, $\mathrm{cd}(\mathcal{C}) = 2$ and $\mathrm{cd}(\mathcal{D}) = 1$; the depth has been offloaded onto the ancilla system to minimize the contributive depth. Because the $Z$ correction on the pass-through qubit in $\mathcal{D}$ is Pauli, it is ignored by the definition.
 
The intuition behind contributive quantum circuit depth is as follows. Consider a circuit $\mathcal{C}$, which is a subcircuit of a circuit $\mathcal{A}$ such that $\mathcal{A} = \mathcal{D}\circ\mathcal{C}\circ \mathcal{B}$ for some circuits $\mathcal{B}, \mathcal{D}$ which both have much higher depth than $\mathcal{C}$, and which have no measurements with classically conditioned gates. Then the depth of $\mathcal{A}$ satisfies,
\[d(\mathcal{A}) \leq d(\mathcal{D}) + \mathrm{cd}(\mathcal{C}) + d(\mathcal{B}).\]
This is because the longest path through $H(\mathcal{A})$ must pass through $H(\mathcal{C})$ from its input to output qubits. Because $\mathcal{B}$ and $\mathcal{D}$ have no measurements with classically conditioned gates, there are no additional long paths which come from the classical dependencies; the Pauli corrections on the outputs of $\mathcal{C}$ can be commuted to the end of the circuit without hitting other measurements whose outcomes must be determined before the circuit can continue.

In practice, we can consider $\mathcal{A}$ to be a large, fault-tolerant algorithm with billions of gates. $\mathcal{B}$ and $\mathcal{D}$ are still going to have classically conditioned gates in them, but if there is a sufficiently large number of gates between the end (beginning) of $\mathcal{C}$ and the next (previous) measurement whose classically controlled gates are non-Pauli, the program's speed through $\mathcal{C}$ is determined by the contributive quantum depth, rather than the reaction depth or `raw' quantum depth. Furthermore, even if there are neighboring measurements whose classically controlled gates are non-Pauli, if the corrections are applied to ancillae (as with auto-corrected state injection), the pass-through qubits can continue through the algorithm. 

In other words, unless the program has to stall to wait for the Pauli frame to be updated correctly, the contribution of $\mathcal{C}$ to the overall circuit depth is $\mathrm{cd}(\mathcal{C})$.

One can define a similar notion of contributive reaction depth, but this is not as useful. Typically, there are measurements of ancillae whose dependent gates will affect the Pauli frame on the main pathway of the computation, and so classical paths from the inputs will continue past the outputs to the rest of the computation. 

Minimizing contributive depth can be a useful heuristic for compilation. Circuits with high contributive depth must take time proportional to that contributive depth to execute, whereas if a circuit has low contributive depth then, given the right circumstances, the execution time can be much lower than its full quantum depth would indicate.

There are several other subtleties of reaction depth. For brevity, we defer them to Appendix~\ref{app:reaction_depth}.

\section{Fast Boolean logic}\label{sec:boolean_results}

In this section we describe techniques for implementing coherent Boolean logic with low quantum- and reaction-depth, while using only homogeneous PPMs and $\ket{CCX}$ states. In~\Cref{ss: effects of homogenous ppms} we show how to construct multi-qubit \gatexor and \gatefanout gates using only these basic operations. In~\Cref{ss:delayed_choice_xor_fanout} we show how to modify the results of the previous section to construct \textit{delayed-choice} versions of these gates. These delayed choice gates allow for a choice of measurement basis on ancilla qubits to control the operations effected on the remaining circuit. Our delayed-choice constructions are distinct from prior art~\cite{fowler2013timeoptimal,gidney2019flexible,gidney2025how,litinski2022active} and can be advantageous for qLDPC architectures. In~\Cref{ss: and_and_auto_and} we discuss techniques for implementing \gateand gates using our basic operations. In particular, this section constructs two new \gateautoand states: resource states which allow for the implementation of \gateand gates with low contributive depth. These \gateautoand states are based on the delayed choice gates constructed in the previous section.\footnote{See \Cref{tab:comparison} for a more detailed comparison.} Finally, in~\Cref{ss:boolean_logic_constant_depth} we prove~\Cref{thm:const_reaction_depth}, showing it is possible to implement arbitrary Boolean logic with constant quantum depth and reaction depth, at the cost of exponentially many ancilla which are left dirty after the computation.

\subsection{Implementing \gatexor and \gatefanout with homogeneous PPMs}\label{ss: effects of homogenous ppms}

In this subsection we analyze simple circuits built from a few basic ingredients: input qubits in some arbitrary quantum state, ancilla qubits initialized in the $\ket{0}$ or $\ket{+}$ state, homogeneous PPMs, and adaptive single-qubit Pauli operations. Though we show the Pauli gates explicitly, note that they need not be implemented in hardware and instead may be tracked in frame. We show circuits of this form with only a single multi-qubit Z-type PPM can implement a PARITY gate whose target is an ancilla qubit, while circuits with only a single multi-qubit X-type PPM can implement a \gatefanout gate with controls and targets both lying on data qubits.

There are a variety of ways to analyze these circuits, including via Pauli webs over the ZX Calculus~\cite{bombin2024unifying}. For accessibility, we prove the following relations algebraically, but the ZX proofs are straightforward and are included in Appendix~\ref{app:proofs}. 




\begin{lemma}[Out-of-place \gatexor]\label{lem:ztype_ppm}
Let $\sum_{\vec x} c_{\vec x}\ket{\vec{x}}$ be an input state in a superposition of computational basis vectors $\ket{\vec{x}} = \ket{x_1 ... x_n}$, along with a qubit state $\ket{+}$ which shall host the output. Then, by measurement of the Pauli operator ${Z_1...Z_{n+1}}$ we effect an out-of-place \gatexor, up to an $X$ Pauli correction:

\begin{align}
    \sum_{\vec x} c_{\vec x} \ket{\vec x}\ket{+} \xmapsto{(I + (-1)^mZ_1...Z_{n+1})} \sum_{\vec x} c_{\vec x} \ket{\vec x} \otimes X^m \ket{\bigoplus_{i} x_i}, 
\end{align}
where $\bigoplus_{i} x_i$ is the \gatexor of the $x_i$ states, and so the output qubit now stores the coherent parity. This operation has width $n + 1$ qubits.
\end{lemma}
\proof
\begin{align*}
    (I + (-1)^m Z...Z)  \ket{\vec x} \ket{+} &= \ket{\vec x} \ket{+} + (-1)^{m + \oplus_i x_i} \ket{\vec x} \ket{-} \\
    &= \ket{\vec x} {\otimes X^m} \ket{\bigoplus_i x_i},
\end{align*}
using the facts that $Z\ket{x_i} = (-1)^{x_i}\ket{x_i}$ for a computational basis state $\ket{x_i}$ and $\ket{+} + (-1)^a \ket{-} = \ket{a}$, up to normalization.
\endproof

An immediate consequence of~\cref{lem:ztype_ppm} is the circuit equivalence in Figure~\ref{fig:parity_Z_type}. Note that we use some non-standard circuit notation, which is explained in Appendix~\ref{app:circuit_notation}.

\begin{figure}[h]
    \centering
    \begin{align*}
    \begin{quantikz}[wire types = {q,q,n,q,q,n}, row sep={0.9cm,between origins}, align equals at=3]
        \lstick{$q_1$}      & \ctrl{4}  &           &                               &           & \\
        \lstick{$q_2$}      &           & \ctrl{3}  &                               &           & \\
                            &           &           & \raisebox{0.5em}{$\ddots$}    &           & \\
        \lstick{$q_n$}      &           &           &                               & \ctrl{1}  & \\
        \lstick{$\ket{0}$}  & \targ{}   & \targ{}   &                               & \targ{}   & \\
                            &           &           &                               &           &
    \end{quantikz} 
    \,\,\, &= \,\,\,
    \begin{quantikz}[wire types = {q,q,n,q,q,n}, row sep={0.9cm,between origins}, align equals at=3]
        \lstick{$q_1$}      & \pctrl{0}\wire[d]{q}  & \\
        \lstick{$q_2$}      & \pctrl{0}\wire[d]{q}  & \\
              & \raisebox{0.5em}{$\vdots$}\wire[d]{q}   & \\
        \lstick{$q_n$}      &    \pctrl{0}\wire[d]{q}  &\\
        \lstick{$\ket{0}$}  & \gate{X}   & \\
                            &           & 
    \end{quantikz}\hspace{-17mm}
    \,\,\, &= \,\,\,
    \begin{quantikz}[wire types = {q,q,n,q,q,n}, row sep={0.9cm,between origins}, align equals at=3]
        \lstick{$q_1$}      & \gate{Z}\wire[d]{b}                   &                                       & \\
        \lstick{$q_2$}      & \gate{Z}\wire[d]{b}                   &                                       & \\
                            & \raisebox{0.5em}{$\vdots$}\wire[d]{b} &                                       & \\
        \lstick{$q_n$}      & \gate{Z}\wire[d]{b}                   &                                       & \\
        \lstick{$\ket{+}$}  & \gate{Z}\wire[d]{b}                   & \gate{X}                              & \\
                            & \meter{}                              & \phase{}\wire[u]{c}\setwiretype{c}    & \setwiretype{n}
    \end{quantikz}
    \end{align*}
    \caption{A \gateparity gate implemented using a single $Z$-type PPM. The black circles with $\oplus$ denote an XOR of the qubits, while the $Z$-boxes connected with thick, filled wires represent a multi-qubit PPM. The unfilled wire connecting to the $X$ box on the rightmost diagram represents a classically controlled $X$ gate (which may be tracked in the Pauli frame).}
    \label{fig:parity_Z_type}
\end{figure}
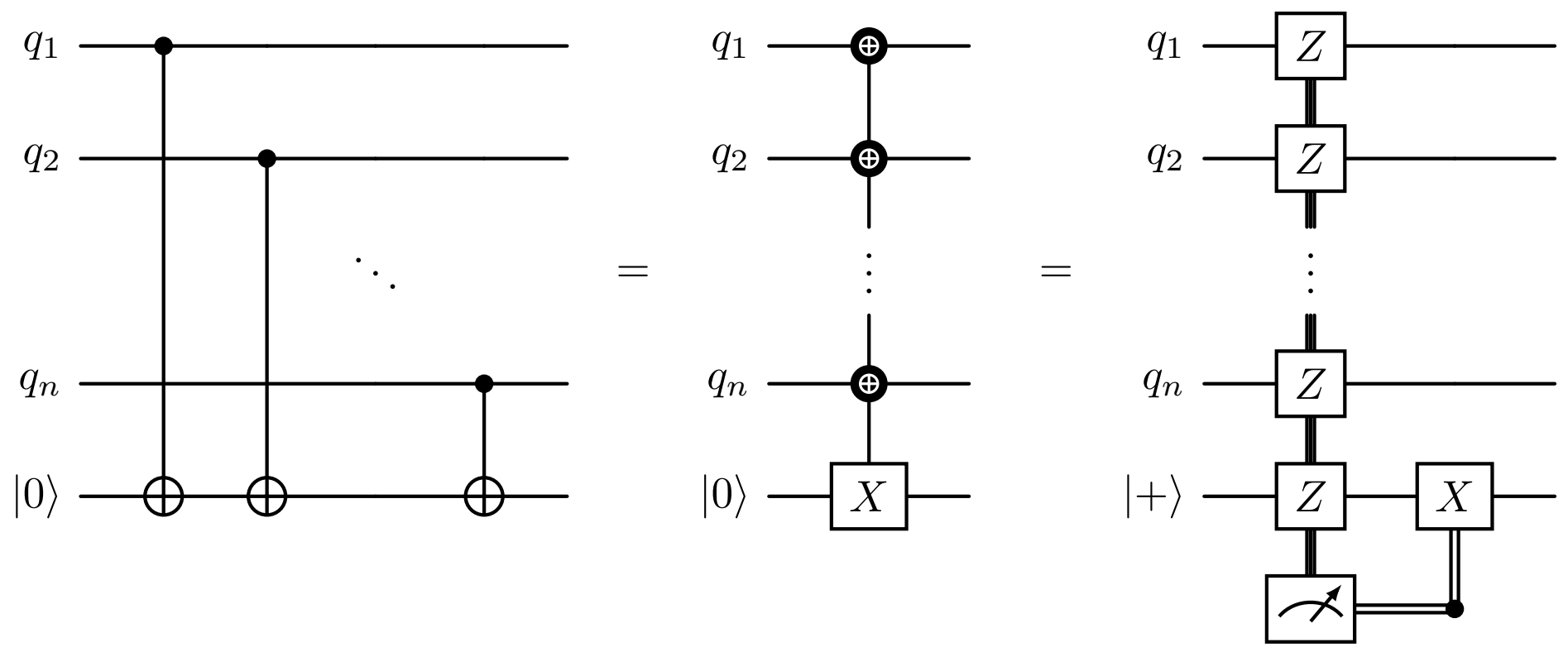

Similarly, we analyze the effect of an $X$-type PPM, finding it implements a set of parallel \textit{in-place} \gatexor gates, i.e. a \gatefanout, where an ancilla qubit initialized in the $\ket{0}$ state is required for Pauli corrections. 

\begin{lemma}[PPM-based \gatefanout]\label{lem:xtype_ppm}

Consider a $n+1$-qubit state $\ket{\psi}=\sum_{\vec x, y} c_{\vec x, y}\ket{\vec{x}}\ket{y}\ket{0}$ where each $\ket{\vec{x}} = \ket{x_1 ... x_{n-1}}$ and $\ket{y}$ are computational basis states. Then applying the homogeneous X-type PPM ${X_1...X_{n+1}}$ to the whole system and observing outcome $m$ followed by measuring out the $n$-th qubit in the $Z$ basis and observing outcome $\ell$ produces the $n$ qubit state:
\begin{align}
     \ket{\psi'} \propto (Z_{n})^m(X^{\otimes n})^\ell\sum_{\vec x, y} c_{\vec x,  y}\ket{\vec{x}\oplus  y{\vec{1}}}\ket{y}.
\end{align}
on the unmeasured $n$ qubits. We note this is equivalent to a \gatefanout operation controlled on the $n$-th qubit of the state, up to Pauli corrections. 
\end{lemma}

\begin{proof}
We first analyze the effect of the $(n+1)$ qubit PPM:
\begin{align*}
     \sum_{\vec x, y} c_{\vec x, y}\ket{\vec x}\ket{y}\ket{0} &\xmapsto{(I + (-1)^m X_1, ..., X_{n+1})} \ket{\vec x}\ket{y}\ket{0} + (-1)^m \ket{\vec x+1}\ket{y+1}\ket{1} \\
     &\hspace{66pt}= \sum_{\vec x, y} c_{\vec x, y}\left(\sum_{j=0}^1 (-1)^{mj}\ket{\vec{x}\oplus j {\vec{1}}}\ket{y+j}\ket{j}\right) \\
     &\hspace{66pt}= \sum_{\vec x, y} c_{\vec x, y}\left(\sum_{j'=0}^1 (-1)^{m(j' \oplus y)}\ket{\vec{x}\oplus j'{\vec{1}}\oplus y {\vec{1}}}\ket{j'}\ket{j'\oplus y}\right)
\end{align*}
where the last line comes from reindexing $j' = j \oplus y$. 

Then making a computational basis measurement on the $n$-th qubit and observing outcome $\ell$ produces the state:
\begin{align*}
    \sum_{\vec x, y} c_{\vec x, y}\left(\sum_{j'=0}^1 (-1)^{m(j' \oplus y)}\ket{\vec{x}\oplus j'{\vec{1}}\oplus y{\vec{1}}}\ket{j'\oplus y} \braket{\ell|j'}\right) 
    &=
    \sum_{\vec x, y} c_{\vec x, y}\left( (-1)^{m(\ell \oplus y)}\ket{\vec{x}\oplus \ell{\vec{1}} \oplus y{\vec{1}}}\ket{\ell \oplus y}\right) \\
    &= (Z_{n})^m(X^{\otimes n})^\ell\sum_{\vec x, y} c_{\vec x, y}\ket{\vec{x}\oplus y{\vec{1}}}\ket{y},
\end{align*}
as claimed.
\end{proof}

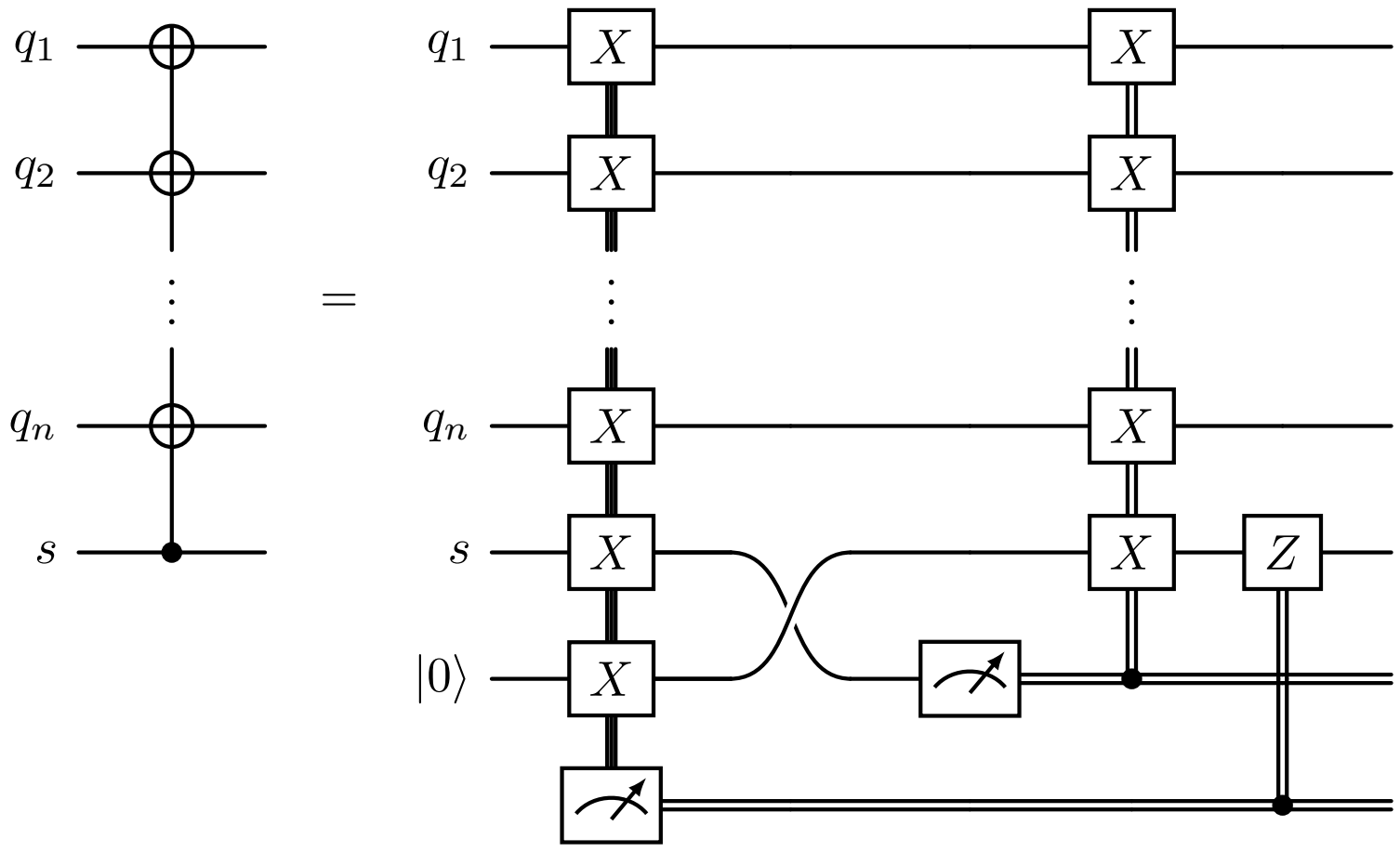
\begin{figure}[h]
    \centering
    \begin{align*}
    \begin{quantikz}[wire types = {q,q,n,q,q,n}, row sep={0.9cm,between origins}, align equals at=3]
        \lstick{$q_1$}      & \targ{}                       & \\
        \lstick{$q_2$}      & \targ{}\wire[u]{q}            & \\
                            & \raisebox{0.5em}{$\vdots$}\wire[u]{q}    & \\
        \lstick{$q_n$}      & \targ{}\wire[u]{q}            & \\
        \lstick{$s$}        & \ctrl{-1}                     & \\
                            &                               &
    \end{quantikz} 
    \,\,\, &= \,\,\,
    \begin{quantikz}[wire types = {q,q,n,q,q,q,n}, row sep={0.9cm,between origins}, align equals at=3]
        \lstick{$q_1$}      & \gate{X}\wire[d]{b}                   &                   &                   & \gate{X}                              & & \\
        \lstick{$q_2$}      & \gate{X}\wire[d]{b}                   &                   &                   & \gate{X}\wire[u]{c}                   & & \\
                            & \raisebox{0.5em}{$\vdots$}\wire[d]{b} &                   &                   & \raisebox{0.5em}{$\vdots$}\wire[u]{c} & & \\
        \lstick{$q_n$}      & \gate{X}\wire[d]{b}                   &                   &                   & \gate{X}\wire[u]{c}                   & & \\
        \lstick{$s$}        & \gate{X}\wire[d]{b}                   & \permute{2,1}     &                   & \gate{X}\wire[u]{c}                   & \gate{Z} & \\
        \lstick{$\ket{0}$}  & \gate{X}\wire[d]{b}                   &                   & \meter{}          & \phase{}\wire[u]{c}\setwiretype{c}    & \wire[u]{c} & \\
                            & \meter{}                              & \setwiretype{c}   &                   &                                       & \phase{}\wire[u]{c} & 
    \end{quantikz}
    \end{align*}
    \caption{A \gatefanout gate implemented using a single $X$-type PPM.}
    \label{fig:fanout_X_type}
\end{figure}

We can summarize the previous result with the circuit identity in Figure~\ref{fig:fanout_X_type}. Note that the PPM implementation of the \gatefanout gate also teleports the information initially stored in the $s$ qubit to the ancilla qubit. 

\subsection{Delayed choice \gatexor and \gatefanout}
\label{ss:delayed_choice_xor_fanout}

We now show how the gates presented in the previous subsection can be made \textit{delayed-choice}, meaning that they allow for classical control which can occur \textit{after} any multi-qubit PPMs. This is enabled by the addition of an ancilla qubit measured in either the $X/Z$ basis. Note that delayed-choice measurements are distinct from gate teleportation~\cite{litinski2022active}, as they do not require Bell pairs to implement. Instead, delayed-choice PPMs either implement a target PPM or implement no measurement, depending on the measurement chosen for an ancilla qubit.

The next definition and theorem illustrate the basic idea used to construct delayed choice operations. 


\begin{define}[Delayed-choice PPMs]
    Suppose $P$ is a PPM acting on an $n$-qubit state $\ket{\phi}$ as:
    \begin{align*}
        \ket{\phi} \mapsto \frac{1}{2}(I + (-1)^m P) \ket{\phi}
    \end{align*}  Then the $Z$-delayed-choice $P$ is a PPM acting on $\ket{\psi}$ and an ancilla qubit in the $\ket{+}$ state as: 
    \[\ket{\phi}\otimes \ket{+} \mapsto \frac{1}{2}(I + (-1)^m (P \otimes Z))(\ket{\phi}\otimes \ket{+})\]
    and the $X$-delayed-choice  $P$ is a PPM acting on $\ket{\psi}$ and an ancilla qubit in the $\ket{0}$ state as: 
    \[\ket{\phi}\otimes \ket{0} \mapsto \frac{1}{2}(I + (-1)^m (P \otimes X))(\ket{\phi}\otimes \ket{0}).\]
\end{define}

Delayed-choice PPMs have previously been used in the context of magic state injection~\cite{litinski2019magic, litinski2019game}, typically for auto-corrected injection~\cite{litinski2022active,gidney2019flexible}.

\begin{lemma}
\label{lem: delayed-choice}
    For any PPM A, one may choose whether to apply $A$ by measuring the ancilla qubit involved in a $Z$-delayed-choice A in the $Z$ or $X$ basis. If this qubit is measured in the $Z$ basis, then $A$ is applied to the remaining qubits, up to Pauli correction controlled on the measurement outcome. If this qubit is measured in the $X$ basis, $I$ is applied instead, again up to classically-controlled Pauli corrections. 

    In equations, let $A$ be a PPM of the form 
    \begin{align*}
        \ket{\phi} \mapsto \frac{1}{2}(I + (-1)^m P) \ket{\phi}.
    \end{align*} 
    Then for any one-qubit computational basis state $\ket{\ell}$ we have:
    \[(I \otimes \bra{\ell})\frac{1}{2}(I + (-1)^m (P \otimes Z))(\ket{\phi}\otimes \ket{+}) \propto (I + (-1)^{\ell+m} P)\ket{\phi}\]
    and
    \[(I \otimes \bra{\ell}H)\frac{1}{2}(I + (-1)^m (P \otimes Z))(\ket{\phi}\otimes \ket{+}) \propto P^\ell\ket{\phi}.\]
\end{lemma}

A similar result holds for $X$-delayed-choice A measurements, with the computational and $\ket{\pm}$ measurements switched.

\begin{proof}
Consider the case that we measure the last qubit in the $Z$ basis. Then we find: 
\begin{align*}
    (I \otimes \bra{\ell})\frac{1}{2}(I + (-1)^m (P\otimes Z)) (\ket{\phi}\otimes \ket{+}) 
    &= (I \otimes \bra{\ell})\frac{1}{2}(\ket{\phi}\otimes \ket{+} + (-1)^{m}(P\ket{\phi}\otimes \ket{-})) \\
    &\propto\left(\ket{\phi} + (-1)^{m+\ell}P\ket{\phi}\right) \\
    &= (I+(-1)^{m+\ell}P)\ket{\phi}.
\end{align*}
To address the case where the last qubit is measured in the $X$ basis, we compute:
\begin{align*}
    (I \otimes \bra{\ell}H)\frac{1}{2}(I + (-1)^m (P \otimes Z))(\ket{\phi}\otimes \ket{+})
    &= (I \otimes \bra{\ell}H)\frac{1}{2}(\ket{\phi}\otimes \ket{+} + (-1)^{m}(P\ket{\phi}\otimes \ket{-})) \\
    &= (I \otimes \bra{\ell})\frac{1}{2}(\ket{\phi}\otimes \ket{0} + (-1)^{m}(P\ket{\phi}\otimes \ket{1})) \\
    &\propto (-1)^{\ell m} P^\ell \ket{\phi} \\
    &\propto P^\ell \ket{\phi}.
\end{align*}
Note that the factor of $(-1)^{\ell m}$ contributes only a global scalar irrespective of the input state, and so can be ignored.
\end{proof}

With some care, these delayed choice measurements can be converted to \emph{delayed choice unitaries}: classically controlled unitaries where the classical control can be effected by changing the basis in which some ancilla qubits are measured. As in the case of delayed choice measurements, these delayed choice unitaries allow for ``backward-in-time'' control: where the decision about \emph{what} unitary to implement (up to Pauli corrections) can be made after the interaction with the system on which the unitary acts.

\begin{lemma}[Delayed-Dirty-\gateparity] 
\label{lem:delayed-dirty-parity}
Given a state $\ket{\psi} =\sum_{\vec{x}} \alpha_{x}\ket{\vec{x}}$ and two ancilla qubits in the $\ket{+}$ state, a homogeneous Z-type PPM applied to the full system followed by a single-qubit measurement in the $X$ basis applied to the final ancilla qubit implements the (identity) operation
\begin{align*}
    \ket{\psi}\ket{+} &\mapsto \ket{\psi}\ket{+}
\end{align*}
up to Pauli corrections on the system consisting of the initial state and unmeasured ancilla qubit. The same setup ending with a single-qubit measurement in the $Z$ basis applied to the final ancilla qubit implements the operation
\begin{align*}
\qquad
    \ket{\psi}\ket{+} &\mapsto \sum_{\vec{x}} \alpha_{\vec{x}}\ket{\vec{x}}\ket{\mathrm{Parity}(\vec{x})}
\end{align*}
up to Pauli corrections on the initial state and unmeasured ancilla qubit.
\end{lemma}

\begin{proof}
This follows from \Cref{lem:ztype_ppm,lem: delayed-choice}.
\end{proof}

\begin{figure}[h]
    \centering
    \begin{align*}
    \begin{quantikz}[wire types = {q,q,n,q,q,c}, row sep={0.9cm,between origins}, align equals at=3]
        \lstick{$q_1$}      &                       & \pctrl{0}\wire[d]{q}                   & \\
        \lstick{$q_2$}      &                       & \pctrl{0}\wire[d]{q}                   & \\
                            &                       & \raisebox{0.5em}{$\vdots$}\wire[d]{q} & \\
        \lstick{$q_n$}      &                       & \pctrl{0}\wire[d]{q}                   & \\
        \lstick{$\ket{+}$}  & \gate{H}              & \gate{X}                              & \\
        \lstick{$c$}        & \phase{}\wire[u]{c}   & \phase{}\wire[u]{c}                   & 
    \end{quantikz} 
    \,\,\, &= \,\,\,
    \begin{quantikz}[wire types = {q,q,n,q,q,q,n,c}, row sep={0.9cm,between origins}, align equals at=3]
        \lstick{$q_1$}      & \gate{Z}\wire[d]{b}                   &                               &           &                              
                            & \gate{Z}                              & \\
        \lstick{$q_2$}      & \gate{Z}\wire[d]{b}                   &                               &           &                                                                  & \gate{Z}\wire[u]{c}                   & \\
                            & \raisebox{0.5em}{$\vdots$}\wire[d]{b} &                               &           &                               
                            & \raisebox{0.5em}{$\vdots$}\wire[u]{c} & \\
        \lstick{$q_n$}      & \gate{Z}\wire[d]{b}                   &                               &           &                                                                  & \gate{Z}\wire[u]{c}                   & \\
        \lstick{$\ket{+}$}  & \gate{Z}\wire[d]{b}                   &                               &           & \gate{X}                                                         & \gate{Z}\wire[u]{c}                   & \\
        \lstick{$\ket{+}$}  & \gate{Z}\wire[d]{b}                   & \gate{H}                      & \meter{}  & \pctrl{0}\wire[u]{c}\setwiretype{c}  
                            & \phase{}\wire[u]{c}               & \\    
                            & \meter{}                              & \setwiretype{c}\wire[u]{c}    &           & \pctrl{0}\wire[u]{c}           
                            & \wire[u]{c}       & \\
        \lstick{$c$}        &                                       & \phase{}\wire[u]{c}           &           & \phase{}\wire[u]{c}           
                            & \octrl{0}\wire[u]{c}      &   
    \end{quantikz}
    \end{align*}
    \caption{A delayed-dirty-\gateparity gate implemented using a single $Z$-type PPM. The bottommost qubit, when measured in the $Z$ basis, activates the \gateparity operation; otherwise when measured in the $\pm$ basis implements the identity operation. \href{https://algassert.com/quirk\#circuit=\%7B\%22cols\%22\%3A\%5B\%5B\%22Y\%5Et\%22\%2C\%22X\%5Et\%22\%5D\%2C\%5B\%22zpar\%22\%2C\%22zpar\%22\%2C\%22zpar\%22\%2C\%22zpar\%22\%2C\%22X\%22\%5D\%2C\%5B1\%2C1\%2C1\%2C\%22H\%22\%2C1\%2C\%22\%E2\%80\%A2\%22\%5D\%2C\%5B1\%2C1\%2C1\%2C1\%2C\%22Measure\%22\%5D\%2C\%5B1\%2C1\%2C1\%2C\%22Measure\%22\%5D\%2C\%5B1\%2C1\%2C1\%2C\%22\%E2\%80\%A6\%22\%5D\%2C\%5B1\%2C1\%2C\%22X\%22\%2C\%22zpar\%22\%2C\%22zpar\%22\%2C\%22\%E2\%97\%A6\%22\%5D\%2C\%5B\%22zpar\%22\%2C\%22zpar\%22\%2C\%22X\%22\%2C1\%2C1\%2C\%22\%E2\%97\%A6\%22\%5D\%2C\%5B1\%2C1\%2C1\%2C\%22\%E2\%80\%A6\%22\%5D\%2C\%5B\%22Z\%22\%2C\%22Z\%22\%2C1\%2C\%22\%E2\%80\%A2\%22\%2C1\%2C\%22\%E2\%80\%A2\%22\%5D\%2C\%5B\%22Y\%5E-t\%22\%2C\%22X\%5E-t\%22\%5D\%5D\%2C\%22init\%22\%3A\%5B0\%2C0\%2C\%22\%2B\%22\%2C\%22\%2B\%22\%5D\%7D}{Quirk.}}
    

    \label{fig:delayed_parity_Z_type}
\end{figure}
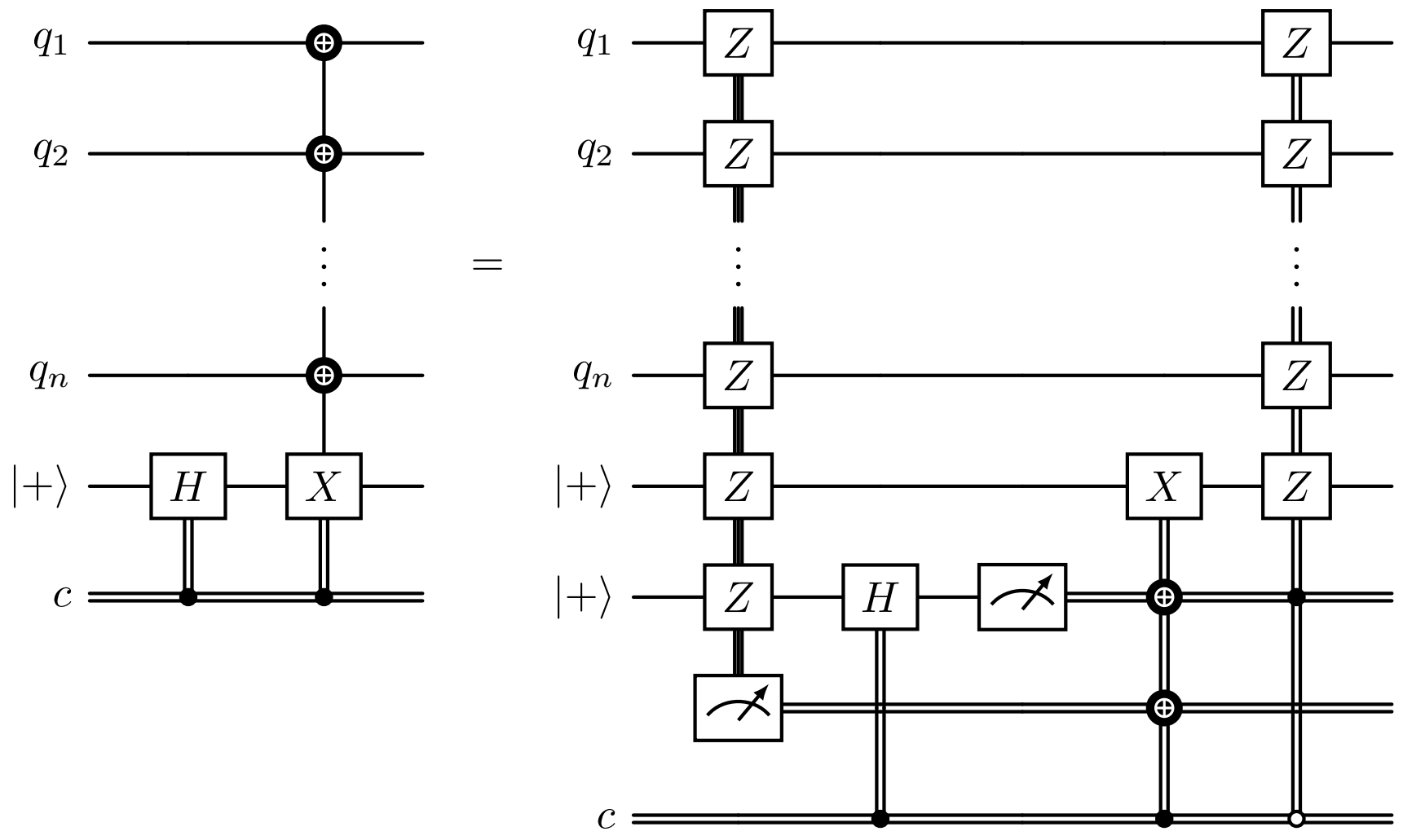

We call this gate a ``Delayed-Dirty-\gateparity'' gate. We say delayed because, as in the case of delayed PPMs discussed above, the choice of whether to implement a gate or the identity operation can be made after applying the $Z$-type PPM, by choosing an appropriate measurement on the ancilla system.  We call this gate ``dirty'' because, in the case where the parity gate is not applied, the ancilla qubit is left in the $\ket{+}$ state rather than a clean $\ket{0}$ ancilla. The circuit in~\Cref{fig:delayed_parity_Z_type} summarizes the content of~\Cref{lem:delayed-dirty-parity} and makes explicit any Pauli-corrections.  

It is possible to remove the ``dirty'' qualifier from the delayed-dirty-\gateparity gate while preserving the delayed structure: for example by introducing an extra ancilla qubit in the $\ket{0}$ state and writing the parity of that qubit into the target ancilla using a second delayed-dirty-\gateparity gate which activates only when the original gate does not. In the circuit constructions occurring later in this paper we typically work directly with the delayed-dirty-\gateparity gate to minimize overall ancilla use.

Similarly, the \gatefanout gate described in~\Cref{lem:xtype_ppm}  can be made delayed-choice by introducing an ancilla qubit and employing a delayed-choice measurement. However, there is one significant flaw: in a delayed-choice setting where the $X$-type PPM did not occur, the information in the $s$ qubit would be measured in the computational basis (see the circuit presented on the right hand side of the equality in~\Cref{fig:fanout_X_type}). This measurement would thus collapse the qubit's superposition and destroy the computation. 

Consequently, any delayed-choice \gatefanout circuit based on the construction in~\Cref{fig:fanout_X_type} must be modified to avoid this collapse. Perhaps the simplest modification is to introduce a SWAP gate controlled on the same classical bit that controls the delayed-choice $X$-type PPM, so that the subsequent measurement is applied to the ancilla qubit whenever the PPM is not applied. Unfortunately, this classically controlled SWAP negates any advantage obtained by making the $X$-type PPM delayed choice.\footnote{Unless the SWAP is also made delayed choice!} An alternate construction, which uses a total of two ancilla qubits to ``hide" the information in the $s$ qubit when the PPM is not applied, is presented later in this section. However, we first discuss an approach which uses only a single ancilla qubit and produces a delayed-choice fanout operation \emph{in the special case} where the qubit controlling the fanout is measured in the $\ket{\pm}$ basis immediately after the fanout. We call this operation a delayed-consumptive-\gatefanout gate. 


\begin{lemma}[Delayed-Consumptive-\gatefanout]
\label{lem:delayed-consumptive-fanout}
    Consider an $n$-qubit state $\ket{\psi} = \sum_{\vec{x}, y} \alpha_{\vec{x},y} \ket{\vec{x}}\ket{y}$, along with a single ancilla qubit in the $\ket{0}$ state. We will refer to the final qubit of the $n$-qubit state as the ``control" qubit. Then, applying a homogeneous $X$-type PPM to the state and ancilla system and measuring: 
    \begin{enumerate}
        \item \label{itm:delayed_consumptive_fanout_identity}
        The control qubit in the $\ket{\pm}$ basis and the ancilla qubit in the computational basis is equivalent (up to Pauli corrections) to applying the identity operation and measuring the control qubit in the $\ket{\pm}$ basis:
        \begin{align}
            \sum_{\vec{x}, y} \alpha_{\vec{x},y} \ket{\vec{x}}\ket{y} \rightarrow \sum_{\vec{x}, y} \alpha_{\vec{x},y} \ket{\vec{x}}\bra{m}H\ket{y}.
        \end{align}
        The measurement outcome $m$ is given by the control qubit measurement. 
        \item \label{itm:delayed_consumptive_fanout_fires}
        The control qubit in the computational basis and the ancilla qubit in the $\ket{\pm}$ basis is equivalent (up to Pauli corrections) to the consumptive fanout operation:
        \begin{align}
            \sum_{\vec{x}, y} \alpha_{\vec{x},y} \ket{\vec{x}}\ket{y} \rightarrow \sum_{\vec{x}, y} \alpha_{\vec{x},y} \ket{\vec{x} \oplus y\vec{1}}\bra{m}H\ket{y}.
        \end{align}
        The measurement outcome $m$ is given by the XOR of the X-type PPM and ancilla qubit measurements. 
    \end{enumerate}
\end{lemma}

\begin{proof}
    The proof follows from \Cref{lem: delayed-choice,lem:xtype_ppm}. We analyze the two cases. In case~\labelcref{itm:delayed_consumptive_fanout_identity}, \Cref{lem: delayed-choice} gives that after measuring the ancilla qubit in the computational basis and observing the outcome $m_2$, the state on the remaining system is 
    \begin{align*}
        \left(X^{\otimes n}\right)^{m_2}\sum_{\vec{x}, y} \alpha_{\vec{x},y} \ket{\vec{x}}\ket{y}.
    \end{align*}
    Then we see measuring the control qubit in the $\ket{\pm}$ basis and observing outcome $m_1$ produces the state
    \begin{align*}
        \left(X^{\otimes (n-1)}\right)^{m_2}\sum_{\vec{x}, y} \alpha_{\vec{x},y} \ket{\vec{x}}\bra{m_1}HX^{m_2}\ket{y} = (-1)^{m_1m_2}\left(X^{\otimes (n-1)}\right)^{m_2}\sum_{\vec{x}, y} \alpha_{\vec{x},y} \ket{\vec{x}}\bra{m_1}H\ket{y}.
    \end{align*}
    as claimed. 

    In case~\labelcref{itm:delayed_consumptive_fanout_fires}, \Cref{lem:xtype_ppm} gives that measuring the control qubit in the computational basis and observing outcome $m_1$ produces the state:
    \begin{align*}
    \sum_{\vec x, y} \alpha_{\vec x, y}(X^{\otimes n -1})^{m_1} \ket{\vec{x} \oplus y\vec{1}}\otimes Z^{m_3}X^{m_1}\ket{y}
    \end{align*}
    where $m_3$ gives the outcome of the $X$-type PPM. Then measuring the ancilla qubit in the $\ket{\pm}$ basis and observing outcome $m_2$ produces the state
    \begin{align*}
    &\sum_{\vec x, y} \alpha_{\vec x, y}(X^{\otimes n -1})^{m_1} \ket{\vec{x} \oplus y\vec{1}}\otimes \bra{m_2}HZ^{m_3}X^{m_1}\ket{y} \\
    &\hspace{80pt}= (-1)^{m_1(m_2 \oplus m_3)}\sum_{\vec x, y} \alpha_{\vec x, y}(X^{\otimes n -1})^{m_1} \ket{\vec{x} \oplus y\vec{1}}\otimes \bra{m_2\oplus m_3}H\ket{y}.
    \end{align*}   
    Noting the factor of $(-1)^{m_1(m_2 \oplus m_3)}$ gives a non-physical global phase we see this is of the desired form. 
\end{proof}

\begin{figure}[h]
    \centering
    \begin{align*}
    \begin{quantikz}[wire types = {q,q,n,q,q,c}, row sep={0.9cm,between origins}, align equals at=3]
        \lstick{$q_1$}      & \targ{}                                   &           &           &                            \\
        \lstick{$q_2$}      & \targ{}\wire[u]{q}                        &           &           &                            \\
                            & \raisebox{0.5em}{$\vdots$}\wire[u]{q}     &           &           &                            \\
        \lstick{$q_n$}      & \targ{}\wire[u]{q}                        &           &           &                            \\
        \lstick{$s$}        & \ctrl{-1}                                 & \gate{H}  & \meter{}  & \setwiretype{c} \rstick{$m$} \\
        \lstick{$c$}        & \phase{}\wire[u]{c}                       &           &           &                            \\
    \end{quantikz} 
    \,\, &= \,\,\,
    \begin{quantikz}[wire types = {q,q,n,q,q,q,c,n}, row sep={0.9cm,between origins}, align equals at=3]
        \lstick{$q_1$}      & \gate{X}\wire[d]{b}                   &                       &           &           & \gate{X}                              & \gate{X}                              &                       &                       &\\
        \lstick{$q_2$}      & \gate{X}\wire[d]{b}                   &                       &           &           & \gate{X}\wire[u]{c}                   & \gate{X}\wire[u]{c}                   &                       &                       &\\
                            & \raisebox{0.5em}{$\vdots$}\wire[d]{b} &                       &           &           & \raisebox{0.5em}{$\vdots$}\wire[u]{c} 
        & \raisebox{0.5em}{$\vdots$}\wire[u]{c} &                       &                       &\\
        \lstick{$q_n$}      & \gate{X}\wire[d]{b}                   &                       &           &           & \gate{X}\wire[u]{c}                   & \gate{X}\wire[u]{c}                   &                       &                       &\\
        \lstick{$s$}        & \gate{X}\wire[d]{b}                   & \gate{H}              & \gate{H}  & \meter{} \arrow[r,"m_1"{yshift=2pt},draw=none] &  \wire[u]{c}\setwiretype{c}            & \phase{}\wire[u]{c}                   &                       & \swap{0}              & \rstick{$m$} \\
        \lstick{$\ket{0}$}  & \gate{X}\wire[d]{b}                   & \gate{H}\wire[u]{c}   &           & \meter{} \arrow[r,"m_2"{yshift=2pt},draw=none] & \phase{}\wire[u]{c}\setwiretype{c}    & \wire[u]{c}                           & \targ{}               & \swap{0}\wire[u]{c}   &\\
        \lstick{$c$}        & \wire[d]{b}                           & \phase{}\wire[u]{c}   &           &           & \octrl{0}\wire[u]{c}               
        & \phase{}\wire[u]{c}                   & \wire[u]{c}           & \phase{}\wire[u]{c}   &\\
                            & \meter{} \arrow[r,"m_3"{yshift=2pt},draw=none]                             & \setwiretype{c}       &           &           &        
        &                                       & \phase{}\wire[u]{c}   &                   &
    \end{quantikz}
    \end{align*}
    \caption{A delayed-consumptive-\gatefanout gate implemented using a single $X$-type PPM and adaptive single-qubit measurements. When $c=0$, the ancilla qubit is measured in the computational basis, and the overall effect up to Pauli corrections is a $\pm$ basis measurement of the control qubit and identity elsewhere. When $c = 1$ the ancilla qubit is measured in the $\pm$ basis and the effect is a consumptive fanout.
    Note all the operations to the right of the adaptive measurements are either classically controlled single-qubit Pauli gates or purely classical post-processing.}
    \label{fig:delayed_consumptive_fanout}
\end{figure}
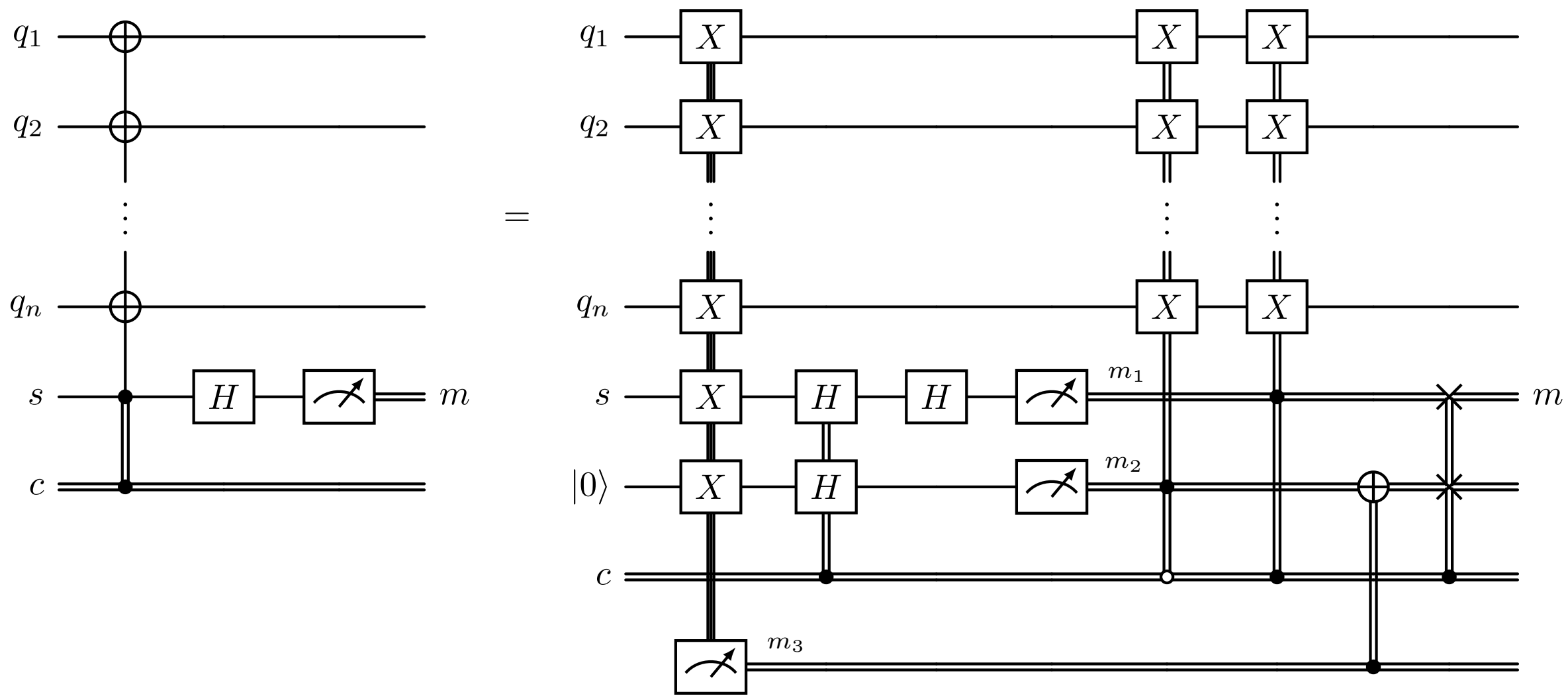

We can summarize the result of the above lemma with~\Cref{fig:delayed_consumptive_fanout}, which should be compared with the identity presented in~\Cref{fig:fanout_X_type}.

It is also possible to build on the delayed-consumptive-\gatefanout operation to construct a delayed- (non-consumptive-) \gatefanout gate. The construction presented below uses one extra ancilla qubit to store a temporary copy of the control qubit. 

\begin{lemma}[Delayed-\gatefanout]
\label{lem:delayed-fanout}
Consider an $n$-qubit state $\ket{\psi} = \sum_{\vec{x}, y} \alpha_{\vec{x},y} \ket{\vec{x}}\ket{y}$. We refer to the final qubit of the n-qubit state as the ``control'' qubit. Then there is a depth-two sequence of homogeneous PPMs acting on $\ket{\psi}$ along with an ancilla qubit in the $\ket{0}$ state and an ancilla qubit in the $\ket{+}$ state  which implements:
\begin{enumerate}
    \item The identity operation 
    \begin{align*}
        \sum_{\vec{x}, y} \alpha_{\vec{x},y} \ket{\vec{x}}\ket{y}\ket{+}\ket{0} \rightarrow \sum_{\vec{x}, y} \alpha_{\vec{x},y} \ket{\vec{x}}\ket{y}
    \end{align*}
    up to Pauli corrections when the control qubit is measured in the $\ket{\pm}$ basis and the $\ket{0}$ ancilla qubit is measured in the computational basis.
    \item The fanout operation
    \begin{align*}
        \sum_{\vec{x}, y} \alpha_{\vec{x},y} \ket{\vec{x}}\ket{y}\ket{+}\ket{0} \rightarrow \sum_{\vec{x}, y} \alpha_{\vec{x},y} \ket{\vec{x} \oplus y\vec{1}}\ket{y}
    \end{align*}
    up to Pauli corrections when the control qubit is measured in the computational basis and the $\ket{0}$ ancilla qubit is measured in the $\ket{\pm}$ basis.
\end{enumerate}
\end{lemma}

\begin{proof}
    The idea is to first use a Z-type PPM to ``copy'' the computational basis state of the control qubit onto the ancilla qubit in the $\ket{+}$ state, then perform the delayed-consumptive-\gatefanout operation. Applying a $ZZ$ PPM to the control qubit and ancilla qubit in the $\ket{+}$ state and observing outcome $m_1$ effects the transformation: 
    \begin{align*}
    \sum_{\vec{x}, y} \alpha_{\vec{x},y} \ket{\vec{x}}\ket{y}\ket{+} \rightarrow \sum_{\vec{x}, y} \alpha_{\vec{x},y} \ket{\vec{x}}\ket{y}\ket{y \oplus m_1}
    \end{align*}
    by~\Cref{lem:ztype_ppm}. Next, we consider the effect of a delayed-consumptive-\gatefanout operation controlled from the qubit storing the $\ket{y}$ state and targeting the qubits storing the $\vec{x}$ state. Let $m_2$ be the outcome observed when measuring the control qubit. Then, by~\Cref{lem:delayed-consumptive-fanout}, when measurement outcomes are chosen so that the fanout does not occur this effects the transformation
    \begin{align*}
        \sum_{\vec{x}, y} \alpha_{\vec{x},y} \ket{\vec{x}}\ket{y}\ket{y \oplus m_1} \rightarrow &\sum_{\vec{x}, y} \alpha_{\vec{x},y} \ket{\vec{x}}\ket{y \oplus m_1}\bra{m_2} H\ket{y} \\
        &\propto \sum_{\vec{x}, y} \alpha_{\vec{x},y} \ket{\vec{x}}\ket{y \oplus m_1}(-1)^{m_2 y} \\
        &=\sum_{\vec{x}, y} \alpha_{\vec{x},y} \ket{\vec{x}}X^{m_1}Z^{m_2}\ket{y}
    \end{align*}
    up to Pauli corrections. Similarly, when the fanout does occur, the transformation is 
    \begin{align*}
        \sum_{\vec{x}, y} \alpha_{\vec{x},y} \ket{\vec{x}}\ket{y}\ket{y \oplus m_1} \rightarrow &\sum_{\vec{x}, y} \alpha_{\vec{x},y} \ket{\vec{x} \oplus y\vec{1}}\ket{y \oplus m_1}\bra{m_2} H\ket{y} \\
        &\propto\sum_{\vec{x}, y} \alpha_{\vec{x},y} \ket{\vec{x} \oplus y\vec{1}}\ket{y \oplus m_1}(-1)^{m_2 y} \\
        &=\sum_{\vec{x}, y} \alpha_{\vec{x},y} \ket{\vec{x} \oplus y\vec{1}}X^{m_1}Z^{m_2}\ket{y}
    \end{align*}
    up to Pauli corrections. These two cases describe a delayed-\gatefanout up to Pauli corrections, and we are done. 
\end{proof}

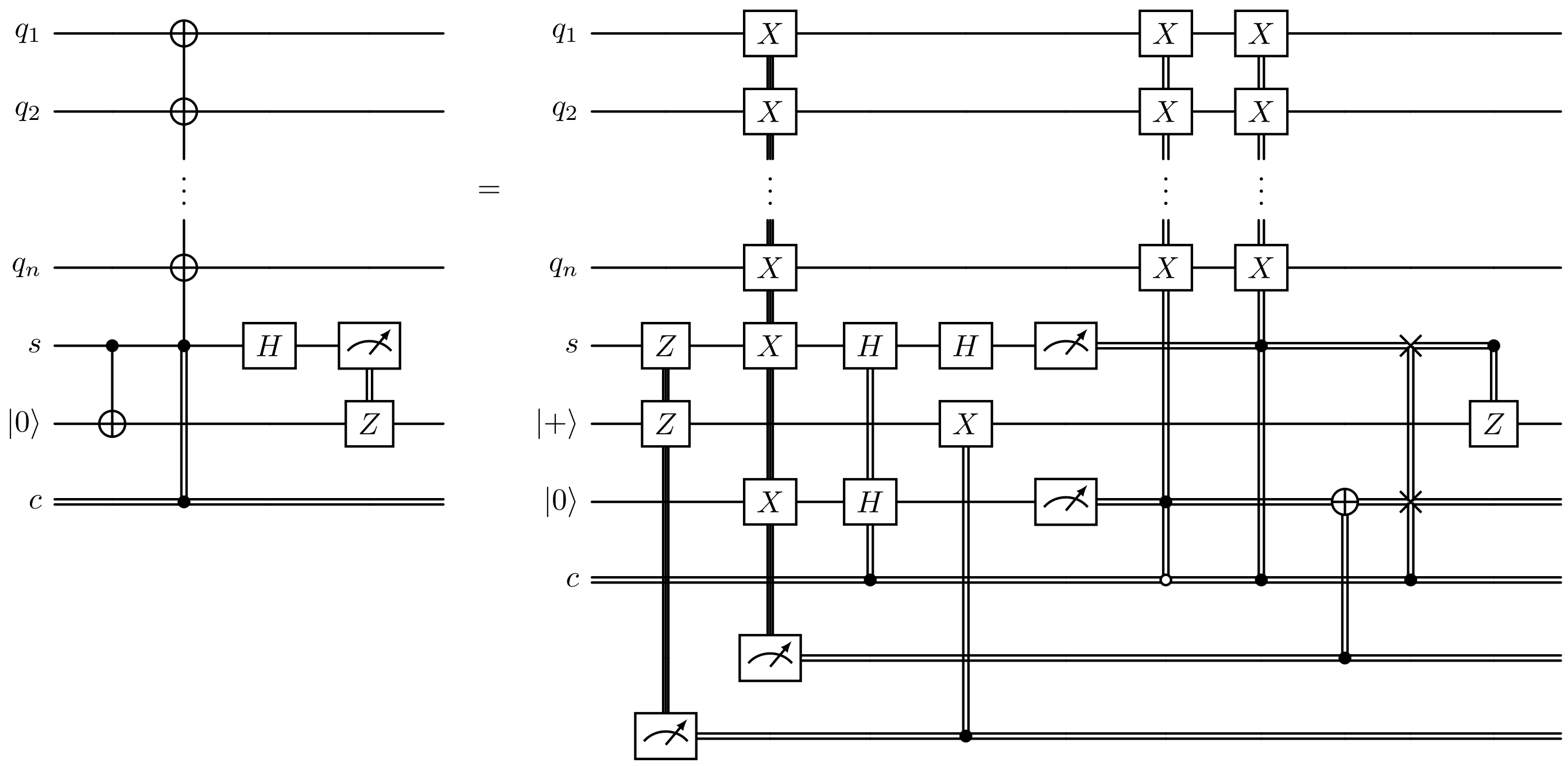
\begin{figure}[h]
    \centering
    \begin{align*}
    \hspace{-8mm}\begin{quantikz}[wire types = {q,q,n,q,q,q,c}, row sep={0.9cm,between origins}, align equals at=3]
        \lstick{$q_1$}    &  & \targ{}                       &           &           &                            \\
        \lstick{$q_2$}    &  & \targ{}\wire[u]{q}            &           &           &                            \\
                          &  & \raisebox{0.5em}{$\vdots$}\wire[u]{q}    &           &           &                            \\
        \lstick{$q_n$}    &  & \targ{}\wire[u]{q}            &           &           &                            \\
        \lstick{$s$}      & \ctrl{1} & \ctrl{-1}                     & \gate{H}  & \meter{}\wire[d]{c}  & \setwiretype{n} \\
        \lstick{$\ket{0}$} & \targ{} & \wire[u]{c} & & \gate{Z} & \\
        \lstick{$c$}      &  & \phase{}\wire[u]{c}           &           &           &                           
    \end{quantikz} 
    \,\,\, &= \,\,\,
    \begin{quantikz}[wire types = {q,q,n,q,q,q,q,c,n,n}, row sep={0.9cm,between origins}, align equals at=3]
        \lstick{$q_1$}   &   & \gate{X}\wire[d]{b}                   &                       &           &           & \gate{X}                              & \gate{X}                              &                       &                       & &\\
        \lstick{$q_2$}   &   & \gate{X}\wire[d]{b}                   &                       &           &           & \gate{X}\wire[u]{c}                   & \gate{X}\wire[u]{c}                   &                       &                       & & \\
                        &    & \raisebox{0.5em}{$\vdots$}\wire[d]{b} &                       &           &           & \raisebox{0.5em}{$\vdots$}\wire[u]{c} 
        & \raisebox{0.5em}{$\vdots$}\wire[u]{c} &                       &                       & & \\
        \lstick{$q_n$}   &   & \gate{X}\wire[d]{b}                   &                       &           &           & \gate{X}\wire[u]{c}                   & \gate{X}\wire[u]{c}                   &                       &                       & &\\
        \lstick{$s$}     & \gate{Z}\wire[d]{b}  & \gate{X}\wire[d]{b}                   & \gate{H}              & \gate{H}  & \meter{}  & \wire[u]{c}\setwiretype{c}            & \phase{}\wire[u]{c}                   &                       & \swap{0}              & \phase{}\wire[d]{c} & \setwiretype{n}\\
        \lstick{$\ket{+}$} & \gate{Z}\wire[d]{b} & \wire[d]{b} & \wire[u]{c} & \gate{X} & & \wire[u]{c} & \wire[u]{c} & & \wire[u]{c} & \gate{Z} &\\
        \lstick{$\ket{0}$}  & \wire[d]{b} & \gate{X}\wire[d]{b}                   & \gate{H}\wire[u]{c}   &   \wire[u]{c}    & \meter{}  & \phase{}\wire[u]{c}\setwiretype{c}    & \wire[u]{c}                           & \targ{}               & \swap{0}\wire[u]{c}   & &\\
        \lstick{$c$}     & \wire[d]{b}   & \wire[d]{b}                           & \phase{}\wire[u]{c}   &   \wire[u]{c}     &           & \octrl{0}\wire[u]{c}                  
        & \phase{}\wire[u]{c}                   & \wire[u]{c}           & \phase{}\wire[u]{c}   & &\\
                        &  \wire[d]{b}  & \meter{}                              & \setwiretype{c}       &  \wire[u]{c}         &           &        
        &                                       & \phase{}\wire[u]{c}   &                   & & \\
        & \meter{} & \setwiretype{c} & & \phase{}\wire[u]{c} & & & & & & &
    \end{quantikz}
    \end{align*}
    \caption{A delayed-(non-consumptive-)\gatefanout gate implemented using a $Z$-type PPM, an $X$-type PPM and adaptive single-qubit measurements. All the operations to the right of the adaptive measurements are either classically controlled single-qubit Pauli gates or purely classical post-processing.}
    \label{fig:delayed_fanout}
\end{figure}

The circuit for the above lemma is shown in~\Cref{fig:delayed_fanout}.

It is also useful to consider variants of the delayed-dirty-\gateparity and delayed-consumptive-\gatefanout constructions where multiple delayed-choice PPMs are applied to the input qubits. If measurement bases are chosen so that exactly one of the delayed choice PPMs affects the input bits the resulting circuit implements a \gateparity gate with delayed-choice control over inputs, or a \gatefanout gate with delayed-choice control over targets. We present the delayed-\gateparity first. 



\begin{lemma}[Multi-delayed-\gateparity]\label{lem:delayed-gateparity}
Let $\ket{\psi} = \sum_{\vec{x}} \alpha_{\vec{x}}\ket{\vec{x}}$ be an $n$-qubit quantum state, and $S_1, S_2, ... S_\ell \subseteq [n]$. Then, for any $i \in [\ell]$, there is a circuit consisting of $\ell$ multi-qubit homogeneous Z-type PPMs, $\ell - 1$ $X$-basis measurements, and precisely one $Z$-basis measurement which acts on $\ket{\psi}$ along with $\ell +1$ ancilla in the $\ket{+}$ state and effects the transformation
\begin{align}
    \ket{\psi} \ket{+}^{\otimes (\ell+1)} \mapsto \sum_{\vec{x}} \alpha_{\vec{x}} \ket{\vec{x}}\ket{  \mathrm{Parity}_{S_i}(\vec{x })}
\end{align}
up to classically-controlled Pauli corrections. Furthermore, these $\ell$ circuits can be chosen so that they differ only in the location of the single-qubit $X$ and $Z$ basis measurements on the ancilla: that is, the choice of which subset of qubits on which to control the parity operation is affected can be made delayed-choice.
\end{lemma}

\begin{proof}
We first consider a sequence of $\ell$ delayed-dirty-\gateparity gates, each of which shares a common target but use fresh ancilla in the $\ket{+}$ state as a delayed-choice control, and with the $j$-th such gate implementing a parity operation controlled on the $S_j$ subset of qubits. It is clear that this sequence of gates, with appropriate delayed-choice measurements on the ancilla and Pauli corrections, implements the operation described in~\Cref{lem:delayed-gateparity}. Commuting all the Pauli corrections to the end of the circuit (so that they are applied after the sequence of multi-qubit PPMs) completes the proof. 



\end{proof}

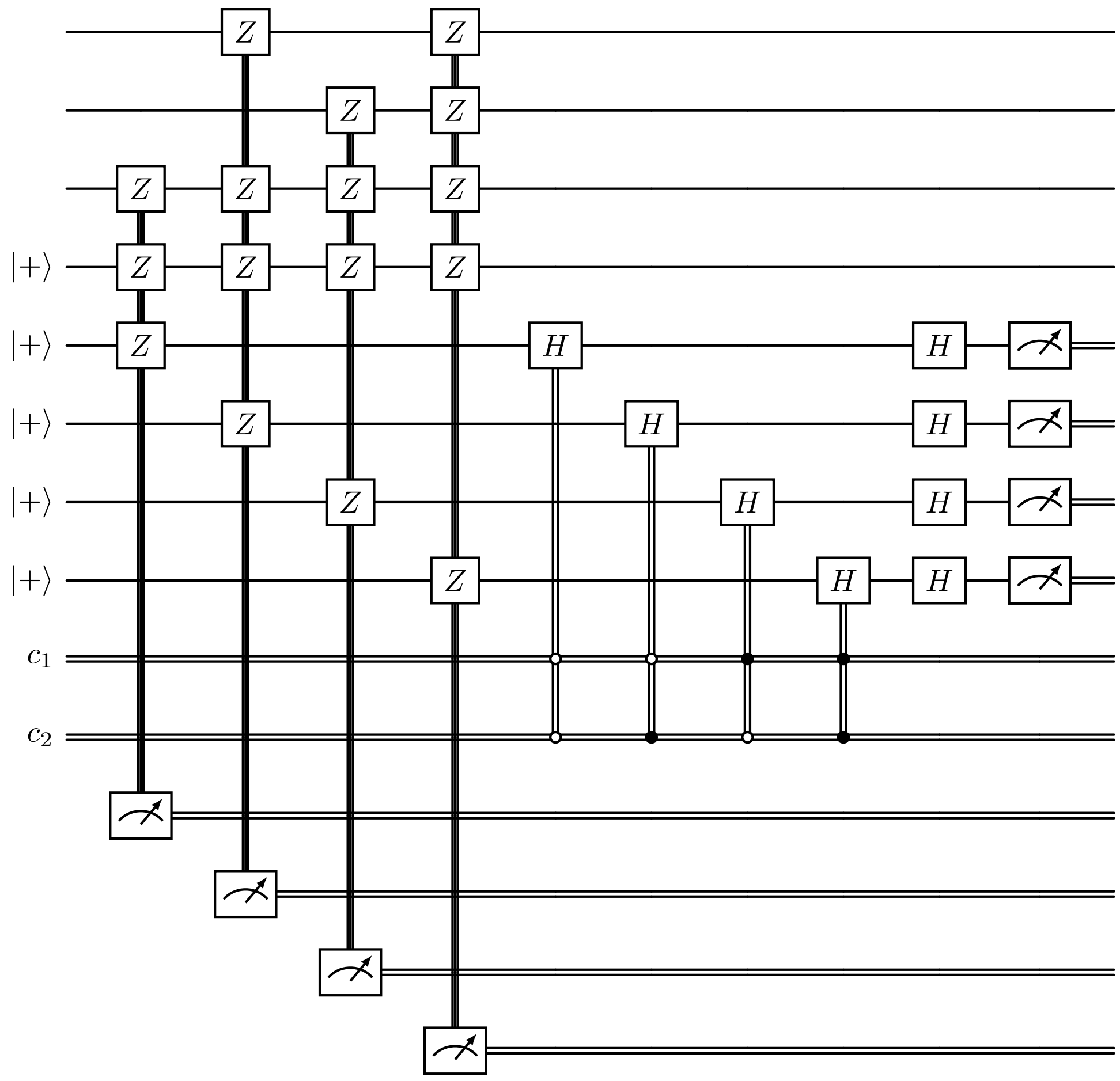
\begin{figure}[h]
    \centering
\begin{align*}
    \begin{quantikz}[wire types = {q,q,q,q,q,q,q,q,c,c,n,n,n,n}, row sep={0.9cm,between origins}]
                        & & \gate{Z}\wire[d]{b} & & \gate{Z}\wire[d]{b} & & & & & & & \\
                        & & \wire[d]{b} & \gate{Z}\wire[d]{b} & \gate{Z}\wire[d]{b} & & & & & & & \\
                        & \gate{Z}\wire[d]{b} & \gate{Z}\wire[d]{b} & \gate{Z}\wire[d]{b} & \gate{Z}\wire[d]{b} & & & & & & & \\
        \lstick{$\ket{+}$} & \gate{Z}\wire[d]{b} & \gate{Z}\wire[d]{b} & \gate{Z}\wire[d]{b} & \gate{Z}\wire[d]{b} & & & & & & & \\  
        \lstick{$\ket{+}$} & \gate{Z}\wire[d]{b} & \wire[d]{b} & \wire[d]{b} & \wire[d]{b} & \gate{H}\wire[d]{c} & & & & \gate{H} & \meter{} & \setwiretype{c}\\
        \lstick{$\ket{+}$} & \wire[d]{b} & \gate{Z}\wire[d]{b} & \wire[d]{b} & \wire[d]{b} & \wire[d]{c} & \gate{H}\wire[d]{c} & & & \gate{H} & \meter{} & \setwiretype{c}\\
        \lstick{$\ket{+}$} & \wire[d]{b} & \wire[d]{b} & \gate{Z}\wire[d]{b} & \wire[d]{b} & \wire[d]{c} & \wire[d]{c} & \gate{H}\wire[d]{c} & & \gate{H} & \meter{} & \setwiretype{c}\\
        \lstick{$\ket{+}$} & \wire[d]{b} & \wire[d]{b} & \wire[d]{b} & \gate{Z}\wire[d]{b} & \wire[d]{c} & \wire[d]{c} & \wire[d]{c} & \gate{H}\wire[d]{c} & \gate{H} & \meter{} & \setwiretype{c}\\
        \lstick{$c_1$}                & \wire[d]{b} & \wire[d]{b} & \wire[d]{b} & \wire[d]{b} & \octrl{0}\wire[d]{c} & \octrl{0}\wire[d]{c} & \ctrl{0}\wire[d]{c} & \ctrl{0}\wire[d]{c} & & & \\
        \lstick{$c_2$}                & \wire[d]{b} & \wire[d]{b} & \wire[d]{b} & \wire[d]{b} & \octrl{0} & \ctrl{0} & \octrl{0} & \ctrl{0} & & & \\
                        & \meter{} & \wire[d]{b}\setwiretype{c} & \wire[d]{b} & \wire[d]{b} & & & & & & & \\
                        & & \meter{} & \wire[d]{b}\setwiretype{c} & \wire[d]{b} & & & & & & & \\
                        & & & \meter{} & \wire[d]{b}\setwiretype{c} & & & & & & & \\
                        & & & & \meter{} & \setwiretype{c} & & & & & &
        \end{quantikz}
\end{align*}
        \caption{A multi-delayed-\gateparity gate implemented using $Z$-type PPMs and adaptive single-qubit measurements. Pauli corrections are omitted for brevity, but can be computed from~\Cref{fig:delayed_parity_Z_type}.}\label{fig:multi_delayed_parity}
\end{figure}

\Cref{fig:multi_delayed_parity} shows an example of a multi-delayed-\gateparity gate, where $n=3$ and $\ell = 4$. In this case, 
\[S_1 = \{3\};\quad S_2 = \{1,3\};\quad S_3 = \{2,3\};\quad S_4 = \{1,2,3\},\]
and $S_i$ for $i=2c_1+c_2+1$ is selected by the controls $c_1,c_2$.

A comparable idea can be applied for the Delayed-consumptive-\gatefanout gate, producing a consumptive \gatefanout gate with a delayed-choice selection of targets:

\begin{lemma}[Multi-delayed-consumptive-\gatefanout]
\label{lem:multi-delayed-consumptive-fanout}
Let $\ket{\psi} = \sum_{\vec{x}, y} \alpha_{\vec{x}, y} \ket{\vec{x}} \ket{y}$ be an $(n + 1)$-qubit quantum state and $\vec{s}^{(1)}, \vec{s}^{(2)}, ... \vec{s}^{(\ell)} \in \{0,1\}^n$ be arbitrary $n$-bit strings. Then, for any $i \in [\ell]$, there is a circuit consisting of $\ell$ multi-qubit homogeneous X-type PPMs, $\ell-1$ $Z$-basis measurements, and precisely one $X$-basis measurement which acts on $\ket{\psi}$ along with $\ell$ ancilla in the $\ket{0}$ state and effects the transformation
\begin{align}
    \ket{\psi} \ket{0}^{\otimes (\ell)} \mapsto \sum_{\vec{x},y} \alpha_{\vec{x},y} \ket{\vec{x} \oplus (y\cdot \vec{s}^{(i)})} \bra{m} H\ket{y}.
\end{align}
up to classically-controlled Pauli corrections, and with $m$ recoverable from the classical measurement outcomes.
Furthermore, these $\ell$ circuits can be chosen so that they differ only in the location of the single-qubit $X$- and $Z$- basis measurements on the ancilla: that is, the choice of which subset of qubits on which to implement the fanout operation can be made delayed-choice. 
\end{lemma}


\begin{proof}
This proof follows similar logic as the proof of~\Cref{lem:delayed-gateparity}, but due to some subtleties with the consumptive measurement we present it somewhat differently. First, consider a sequence of $\ell$ different $X$-type PPMs, with the $j$-th such PPM acting on the qubits indicated by $\vec{s}^{(j)}$ along with the coherent control qubit (i.e. the $n+1$-th input qubit) and a fresh ancilla in the $\ket{0}$ state which will act as a delayed-choice control. Also imagine that all but the $i$-th PPM are followed by a computational basis measurement on the corresponding delayed-choice control qubit and subsequent classically-controlled Pauli corrections such that they have no effect on the input qubits, as per~\Cref{lem: delayed-choice}. After this sequence of PPMs and corrections, we measure the one remaining ($i$-th) ancilla qubit in the $\ket{\pm}$ basis, the coherent control qubit in the computational basis, and apply Pauli corrections and classical post-processing as in the delayed-consumptive-fanout operation (\Cref{fig:delayed_consumptive_fanout}) when the classical control bit $c=1$. It is clear from~\Cref{lem:delayed-consumptive-fanout} that this circuit as described effects the desired transformation; commuting all the classically controlled Pauli corrections to the end of the circuit then shows it has the desired form, and we are done.




\end{proof}

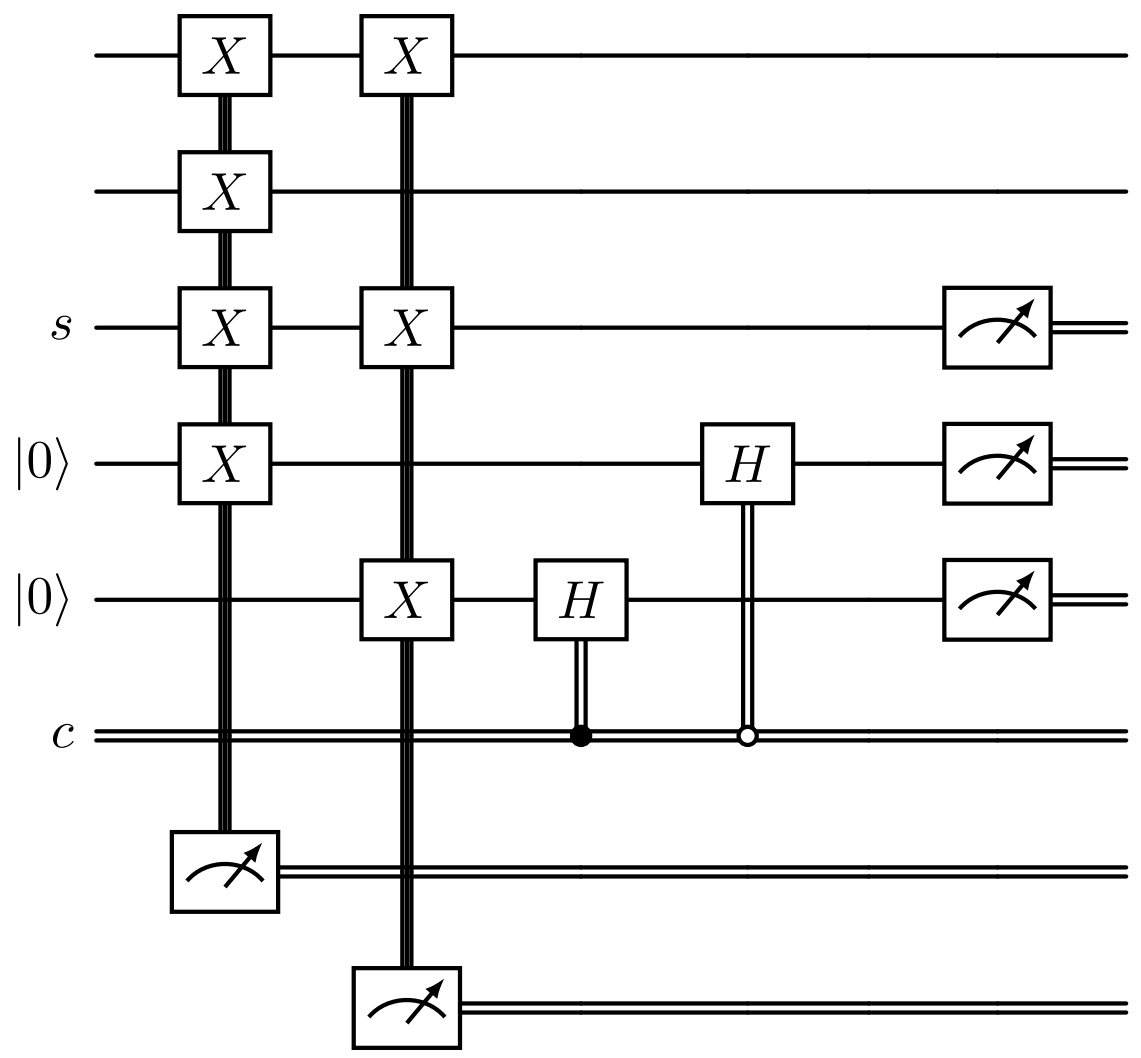
\begin{figure}[h]
    \centering
\begin{align*}
    \begin{quantikz}[wire types = {q,q,q,q,q,c,n,n}, row sep={0.9cm,between origins}]
                        & \gate{X}\wire[d]{b} & \gate{X}\wire[d]{b} & & & & & \\
                        & \gate{X}\wire[d]{b} & \wire[d]{b} & & & & & \\
        \lstick{$s$}               & \gate{X} \wire[d]{b} & \gate{X}\wire[d]{b} & & & & \meter{} & \setwiretype{c} \\
        \lstick{$\ket{0}$}                & \gate{X}\wire[d]{b} & \wire[d]{b} & & \gate{H}\wire[d]{c} & & \meter{} & \setwiretype{c} \\
        \lstick{$\ket{0}$}                & \wire[d]{b} & \gate{X}\wire[d]{b} & \gate{H}\wire[d]{c} & \wire[d]{c} & & \meter{} & \setwiretype{c} \\
        \lstick{$c$}                & \wire[d]{b} & \wire[d]{b} & \ctrl{0} & \octrl{0} & & & \\
                        & \meter{} & \wire[d]{b}\setwiretype{c} & & & & & \\
                        & & \meter{} & \setwiretype{c} & & & & 
        \end{quantikz}
\end{align*}
        \caption{A multi-delayed-consumptive-\gatefanout gate implemented using $X$-type PPMs and adaptive single-qubit measurements. Pauli corrections are omitted for brevity, but can be computed from~\Cref{fig:delayed_consumptive_fanout}.}\label{fig:multi_delayed_fanout}
\end{figure}

\Cref{fig:multi_delayed_fanout} shows an example of a multi-delayed-consumptive-\gatefanout gate, where $n=\ell = 2$. $\vec{s}^{(1)} = (1, 1)$ and $\vec{s}^{(2)} = (1, 0)$. If $c = 0$ then only the first consumptive-\gatefanout gate, with support $\vec{s}^{(1)}$, is applied; if $c=1$ then the second consumptive-\gatefanout gate, with support $\vec{s}^{(2)}$, is applied instead.

\subsection{Implementing \gateand using homogeneous PPMs}\label{ss: and_and_auto_and}
So far, we have only shown how to implement \gatexor gates via PPMs. To achieve Boolean universality, we also require the \gateand gate. Notably, the \gateand gate requires a Toffoli magic state to implement, while the Boolean operations described so far do not. We begin by reviewing a standard circuit which consumes a Toffoli magic state and implements an \gateand gate up to Clifford corrections (\textcite{gidney2025how} denotes these as ``wandering" \gateand gates). 



\begin{lemma}[(Wandering) \gateand gate~\cite{gidney2025how}]\label{lem:gate_and}
    Let $\ket{\psi} = \sum_{i,j} c_{i,j} \ket{i, j}$ be a two-qubit state and define the  CCX state 
    \begin{align*}
        \ket{CCX} = \textrm{CCX} \ket{++0}.
    \end{align*} Then there is a circuit consisting only of $Z$-type PPMs and $\ket{\pm}$ basis measurements which acts on $\ket{\psi}$ and a $\ket{CCX}$ resource state and computes the \gateand of qubits in $\ket{\psi}$ up to two CNOT gates and Pauli corrections. More explicitly, there is a circuit consisting only of $Z$-type PPMs and $\ket{\pm}$ basis measurements which effects the transformation
    \begin{align}
        \ket{i, j}\ket{CCX} \mapsto \ket{i, j} \ket{i j \oplus m_2 i \oplus m_1 j \oplus m_1 m_2 },
    \end{align}
    up to Pauli $Z$ corrections, where $m_1$ and $m_2$ denote the PPM measurement outcomes. 
\end{lemma}


\begin{proof}

The full circuit implementing $\gateand$, including CNOT and Pauli corrections, is given in~\Cref{fig:standard_AND}. 
Note first that up to normalization, $\ket{CCX} \propto \sum_{k, l} \ket{k, l, kl}$. So we have that 
\begin{equation}
    \ket{\psi}\ket{CCX} \propto \sum_{i,j,k,l} c_{i,j} \ket{i,j}\ket{k,l, kl}.
\end{equation}
Measuring the operators $Z_1 Z_3$ and $Z_2 Z_4$ and obtaining the outcomes $(-1)^{m_1}$ and $(-1)^{m_2}$ respectively (with $m_1, m_2 \in \{0, 1\}$), we get (by applying the relevant projectors) a state
\begin{equation}
\propto \sum_{i,j} c_{i,j} \ket{i,j} \ket{i + m_1, j + m_2, m_1m_2 +ij + m_1j + m_2i},
\end{equation}
which can be rewritten as
\begin{equation}
    = \sum_{i,j} c_{i,j} \ket{i,j}\left(\ket{+} + (-1)^{i+m_1} \ket{-} \right) \left(\ket{+} + (-1)^{j+m_2} \ket{-} \right) \ket{ij + m_1m_2 + m_1j + m_2i}.
\end{equation}
Measuring further $X_3$ and $X_4$, obtaining outcomes $(-1)^{x_3}$ and $(-1)^{x_4}$ respectively, we get
\begin{equation}
    = \sum_{i,j} c_{i,j} (-1)^{(i+m_1)x_3} (-1)^{(j+m_2)x_4} \ket{i, j, ij + m_1m_2 + m_1j + m_2i},
\end{equation}
which in turn can be rewritten as
\begin{equation}
    = (-1)^{m_1x_3 + m_2x_4} Z_1^{x_3} Z_2^{x_4} X_3^{m_1 m_2} (CNOT_{1,3})^{m_2} (CNOT_{2,3})^{m_1} \sum_{i,j} c_{i,j} \ket{i, j, ij},
\end{equation}
precisely as claimed.
\end{proof}

\begin{figure}[h]
    \centering
    \begin{align*}
        \begin{quantikz}[wire types = {q,q,q,q,q,n,n}, row sep={0.7cm,between origins}, align equals at=3]
        \lstick[2]{$\ket{\psi}$}& \gate{Z} \wire[d]{b}  &                               &                   &           &                       & 
        \ctrl{4}                & \gate{Z}              &                               &\\
                                & \wire[d]{b}           & \gate{Z} \wire[d]{b}          &                   &           & \ctrl{3}              &           
                                & \wire[u]{c}           & \gate{Z}                      &\\
        \lstick[3]{$\ket{CCX}$} & \gate{Z} \wire[d]{b}  & \wire[d]{b}                   & \gate{H}          & \meter{}  & \setwiretype{c}       &                                     & \phase{}\wire[u]{c}   & \wire[u]{c}                   &\\
                                & \wire[d]{b}           & \gate{Z} \wire[d]{b}          & \gate{H}          & \meter{}  & \setwiretype{c}       &
                                &                       & \phase{}\wire[u]{c}           &\\ 
                                & \wire[d]{b}           & \wire[d]{b}                   &                   &           & \targ{}               & 
        \targ{}                 & \targ{}               &                               &\\
                                & \meter{}              & \wire[d]{b} \setwiretype{c}   &                   &           & \phase{}\wire[u]{c}   & 
        \wire[u]{c}             & \phase{}\wire[u]{c}   &                               &\\
                                &                       & \meter{}                      & \setwiretype{c}   &           &                       &
        \phase{}\wire[u]{c}     & \phase{}\wire[u]{c}   &                               &
    \end{quantikz} 
    \end{align*}
    \caption{A circuit implementing an \gateand operation using PPMs along with classically-controlled Pauli and CNOT corrections. \href{https://algassert.com/quirk\#circuit=\%7B\%22cols\%22\%3A\%5B\%5B\%22Y\%5Et\%22\%2C\%22X\%5Et\%22\%5D\%2C\%5B1\%2C1\%2C\%22X\%22\%2C\%22\%E2\%80\%A2\%22\%2C\%22\%E2\%80\%A2\%22\%5D\%2C\%5B\%22zpar\%22\%2C1\%2C1\%2C\%22zpar\%22\%2C1\%2C\%22X\%22\%5D\%2C\%5B1\%2C\%22zpar\%22\%2C1\%2C1\%2C\%22zpar\%22\%2C1\%2C\%22X\%22\%5D\%2C\%5B\%22\%E2\%80\%A6\%22\%5D\%2C\%5B1\%2C1\%2C1\%2C\%22H\%22\%2C\%22H\%22\%5D\%2C\%5B1\%2C1\%2C1\%2C\%22Measure\%22\%2C\%22Measure\%22\%2C\%22Measure\%22\%2C\%22Measure\%22\%5D\%2C\%5B1\%2C\%22\%E2\%80\%A2\%22\%2C\%22X\%22\%2C1\%2C1\%2C\%22\%E2\%80\%A2\%22\%5D\%2C\%5B\%22\%E2\%80\%A2\%22\%2C1\%2C\%22X\%22\%2C1\%2C1\%2C1\%2C\%22\%E2\%80\%A2\%22\%5D\%2C\%5B1\%2C1\%2C\%22X\%22\%2C1\%2C1\%2C\%22\%E2\%80\%A2\%22\%2C\%22\%E2\%80\%A2\%22\%5D\%2C\%5B1\%2C\%22Z\%22\%2C1\%2C1\%2C\%22\%E2\%80\%A2\%22\%5D\%2C\%5B\%22Z\%22\%2C1\%2C1\%2C\%22\%E2\%80\%A2\%22\%5D\%2C\%5B\%22\%E2\%80\%A6\%22\%5D\%2C\%5B\%22\%E2\%80\%A2\%22\%2C\%22\%E2\%80\%A2\%22\%2C\%22X\%22\%5D\%2C\%5B\%22Y\%5E-t\%22\%2C\%22X\%5E-t\%22\%5D\%5D\%2C\%22init\%22\%3A\%5B0\%2C0\%2C0\%2C\%22\%2B\%22\%2C\%22\%2B\%22\%5D\%7D}{Quirk.}
    }
    \label{fig:standard_AND}
\end{figure}

The previous section shows the CNOT gates in~\Cref{fig:standard_AND} can be re-written in terms of $X$-type PPMs. This gives a circuit which implements an \gateand operation (up to Pauli corrections) in two stages: first $Z$-type PPMs are applied to the system, and then $X$-type PPMs are applied dynamically to the same system depending on the outcome of the $Z$-type PPMs. We next discuss two modifications to this circuit which can make the \gateand gate \emph{automatic}: meaning that the state produced on the output qubits is correct up to Pauli corrections after a sequence of (possibly dynamic) measurements are made on ancilla qubits.


The modifications rely on delayed-\gateparity and delayed-\gatefanout gates. By adding appropriate delayed-choice operations to the $\ket{CCX}$ state we produce \texttt{AutoAND} states: states such that any non-Pauli errors incurred during an \gateand-injection can be corrected via appropriate measurements on ancilla qubits. We give both delayed-\gateparity and delayed-\gatefanout based implementations. In the case of \gateparity, we implement a delayed-\gateparity operation controlled on subsets of qubits $S_1, ... S_4$ which encompass the four potential non-Pauli corrections needed. In the case of \gatefanout, we use delayed-consumptive-\gatefanout operations to implement the two potential corrections that may need to be written into the output bit. 

\begin{lemma}[\gateparity-based \texttt{AutoAND}]\label{lem:parity_autoand}
The following \texttt{AutoAND} state requires $8$ qubits, on top of the original 2 qubits, and implements
\begin{align}
    \ket{i} \ket{j} \mapsto \ket{i} \ket{j} \ket{ij}
\end{align}
up to some Pauli corrections.

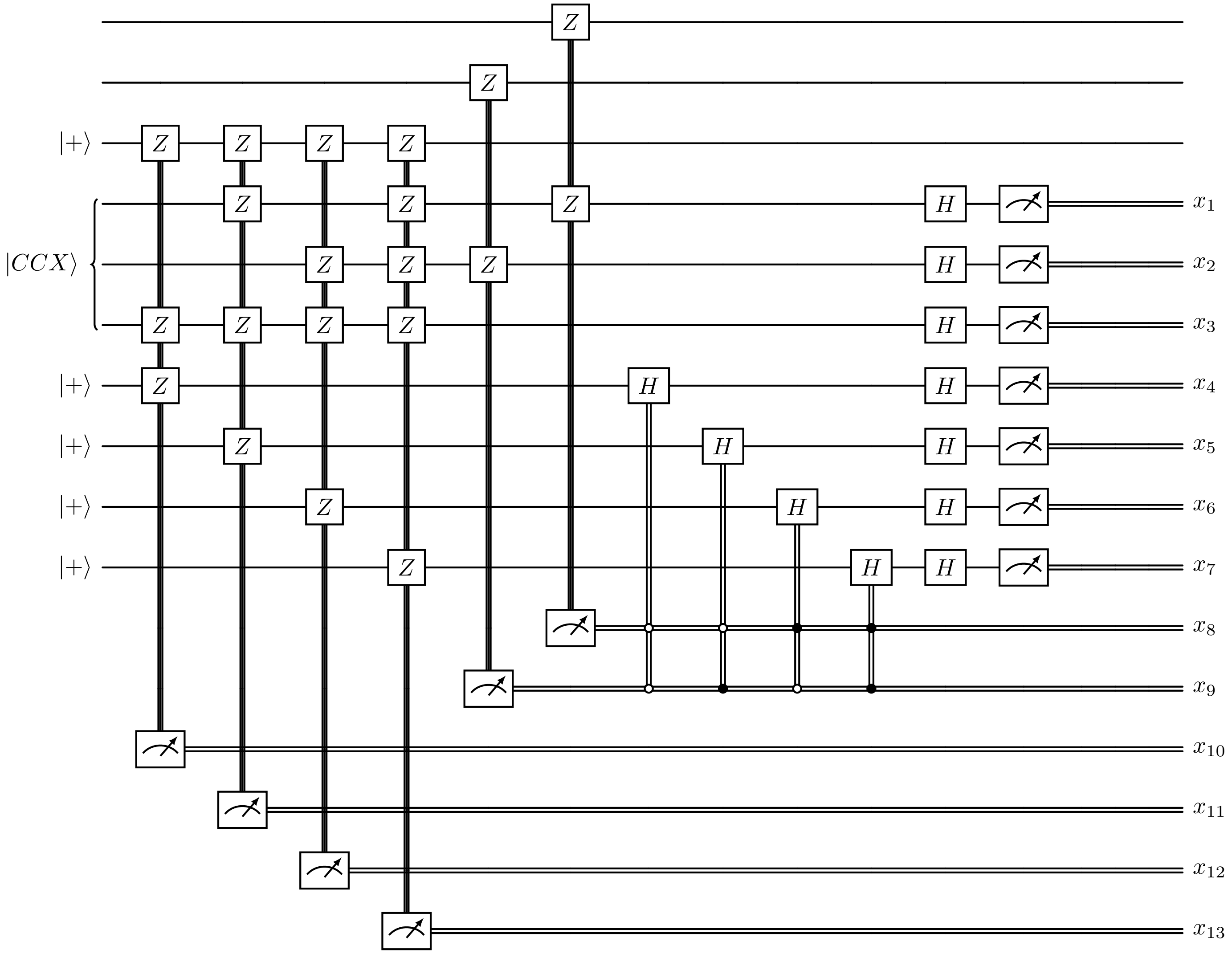
\begin{align*}\hspace{-10mm}
    \begin{quantikz}[wire types = {q,q,q,q,q,q,q,q,q,q,n,n,n,n,n,n}, row sep={0.9cm,between origins}]
                                & & & & &   & \gate{Z} \wire[d]{b} & & & &   & & & & & & \\
                                & & & & & \gate{Z} \wire[d]{b}  & \wire[d]{b} & & & &   & & & & & & \\
    \lstick{$\ket{+}$}          & \gate{Z} \wire[d]{b} & \gate{Z} \wire[d]{b}                              & \gate{Z} \wire[d]{b} & \gate{Z} \wire[d]{b} & \wire[d]{b}  & \wire[d]{b} & & & &   & & & & & & \\
    \lstick[3]{$\toffstate$}   & \wire[d]{b} & \gate{Z} \wire[d]{b} &                             \wire[d]{b} & \gate{Z} \wire[d]{b} & \wire[d]{b}  & \gate{Z} \wire[d]{b} & & & &   & \gate{H} & \meter{} & \setwiretype{c} & & & \rstick{$x_1$}\\
                                & \wire[d]{b} & \wire[d]{b} & \gate{Z} \wire[d]{b} & \gate{Z} \wire[d]{b} & \gate{Z} \wire[d]{b}  & \wire[d]{b} & & & &   & \gate{H} & \meter{} & \setwiretype{c} & & & \rstick{$x_2$}\\
                                & \gate{Z} \wire[d]{b} & \gate{Z} \wire[d]{b} & \gate{Z} \wire[d]{b} & \gate{Z} \wire[d]{b} & \wire[d]{b}  & \wire[d]{b} & & & &   & \gate{H} & \meter{} & \setwiretype{c} & & & \rstick{$x_3$}\\
    \lstick{$\ket{+}$}          & \gate{Z} \wire[d]{b} & \wire[d]{b} & \wire[d]{b} & \wire[d]{b} & \wire[d]{b}  & \wire[d]{b} & \gate{H} & & &   &                             \gate{H} & \meter{} & \setwiretype{c} & & & \rstick{$x_4$}\\
    \lstick{$\ket{+}$}          & \wire[d]{b} & \gate{Z} \wire[d]{b} & \wire[d]{b} & \wire[d]{b} & \wire[d]{b}  & \wire[d]{b} & \wire[u]{c} & \gate{H} & &   & \gate{H} & \meter{} & \setwiretype{c} & &                              & \rstick{$x_5$}\\
    \lstick{$\ket{+}$}          & \wire[d]{b} & \wire[d]{b} & \gate{Z} \wire[d]{b} & \wire[d]{b} & \wire[d]{b}  & \wire[d]{b} & \wire[u]{c} & \wire[u]{c} & \gate{H} &   & \gate{H} & \meter{} & \setwiretype{c} & &                              & \rstick{$x_6$}\\
    \lstick{$\ket{+}$}          & \wire[d]{b} & \wire[d]{b} & \wire[d]{b}& \gate{Z} \wire[d]{b} & \wire[d]{b}  & \wire[d]{b} & \wire[u]{c} & \wire[u]{c} & \wire[u]{c} & \gate{H}  & \gate{H} & \meter{} & \setwiretype{c} & &                              & \rstick{$x_7$}\\
                                & \wire[d]{b} & \wire[d]{b} & \wire[d]{b} & \wire[d]{b} & \wire[d]{b}  & \meter{} & \octrl{0}\wire[u]{c} \setwiretype{c} & \octrl{0}\wire[u]{c} & \ctrl{0}\wire[u]{c} & \ctrl{0}\wire[u]{c}  & & & & & & \rstick{$x_8$}\\
                                & \wire[d]{b} & \wire[d]{b} & \wire[d]{b} & \wire[d]{b} & \meter{}  & \setwiretype{c} & \octrl{0}\wire[u]{c} & \ctrl{0}\wire[u]{c} & \octrl{0}\wire[u]{c} & \ctrl{0}\wire[u]{c}  & & & & & & \rstick{$x_9$}\\
                                & \meter{} & \wire[d]{b}\setwiretype{c} & \wire[d]{b} & \wire[d]{b} &   & & & & &   & & & & & & \rstick{$x_{10}$}\\
                                & & \meter{} & \wire[d]{b}\setwiretype{c} & \wire[d]{b} &  & & & & & & & & & & & \rstick{$x_{11}$}\\
                                & & & \meter{} & \wire[d]{b}\setwiretype{c} &   & & & & & & & & & & & \rstick{$x_{12}$}\\
                                & & & & \meter{} & \setwiretype{c}  & & & & &   & & & & & & \rstick{$x_{13}$}\\
        \end{quantikz}
\end{align*}

This is shown in \href{https://algassert.com/quirk#circuit=%7B%22cols%22%3A%5B%5B%22Y%5Et%22%2C%22X%5Et%22%5D%2C%5B1%2C1%2C1%2C%22•%22%2C%22•%22%2C%22X%22%5D%2C%5B1%2C1%2C%22zpar%22%2C1%2C1%2C%22zpar%22%2C%22zpar%22%2C1%2C1%2C1%2C1%2C1%2C%22X%22%5D%2C%5B1%2C1%2C%22zpar%22%2C%22zpar%22%2C1%2C%22zpar%22%2C1%2C%22zpar%22%2C1%2C1%2C1%2C1%2C1%2C%22X%22%5D%2C%5B1%2C1%2C%22zpar%22%2C1%2C%22zpar%22%2C%22zpar%22%2C1%2C1%2C%22zpar%22%2C1%2C1%2C1%2C1%2C1%2C%22X%22%5D%2C%5B1%2C1%2C%22zpar%22%2C%22zpar%22%2C%22zpar%22%2C%22zpar%22%2C1%2C1%2C1%2C%22zpar%22%2C1%2C1%2C1%2C1%2C1%2C%22X%22%5D%2C%5B1%2C1%2C%22…%22%5D%2C%5B1%2C%22zpar%22%2C1%2C1%2C%22zpar%22%2C1%2C1%2C1%2C1%2C1%2C1%2C%22X%22%5D%2C%5B%22zpar%22%2C1%2C1%2C%22zpar%22%2C1%2C1%2C1%2C1%2C1%2C1%2C%22X%22%5D%2C%5B1%2C1%2C1%2C1%2C1%2C1%2C1%2C1%2C1%2C1%2C%22Measure%22%2C%22Measure%22%5D%2C%5B1%2C1%2C1%2C1%2C1%2C1%2C%22H%22%2C1%2C1%2C1%2C%22◦%22%2C%22◦%22%5D%2C%5B1%2C1%2C1%2C1%2C1%2C1%2C1%2C%22H%22%2C1%2C1%2C%22◦%22%2C%22•%22%5D%2C%5B1%2C1%2C1%2C1%2C1%2C1%2C1%2C1%2C%22H%22%2C1%2C%22•%22%2C%22◦%22%5D%2C%5B1%2C1%2C1%2C1%2C1%2C1%2C1%2C1%2C1%2C%22H%22%2C%22•%22%2C%22•%22%5D%2C%5B1%2C1%2C1%2C%22H%22%2C%22H%22%2C%22H%22%2C%22H%22%2C%22H%22%2C%22H%22%2C%22H%22%5D%2C%5B1%2C1%2C1%2C%22Measure%22%2C%22Measure%22%2C%22Measure%22%2C%22Measure%22%2C%22Measure%22%2C%22Measure%22%2C%22Measure%22%2C1%2C1%2C%22Measure%22%2C%22Measure%22%2C%22Measure%22%2C%22Measure%22%5D%2C%5B1%2C1%2C%22X%22%2C1%2C1%2C1%2C%22zpar%22%2C1%2C1%2C1%2C%22◦%22%2C%22◦%22%2C%22zpar%22%5D%2C%5B1%2C1%2C%22X%22%2C1%2C1%2C1%2C1%2C%22zpar%22%2C1%2C1%2C%22◦%22%2C%22•%22%2C1%2C%22zpar%22%5D%2C%5B1%2C1%2C%22X%22%2C1%2C1%2C1%2C1%2C1%2C%22zpar%22%2C1%2C%22•%22%2C%22◦%22%2C1%2C1%2C%22zpar%22%5D%2C%5B1%2C1%2C%22X%22%2C1%2C1%2C1%2C1%2C1%2C1%2C%22zpar%22%2C%22•%22%2C%22•%22%2C1%2C1%2C1%2C%22zpar%22%5D%2C%5B1%2C1%2C%22X%22%2C1%2C1%2C1%2C1%2C1%2C1%2C1%2C%22•%22%2C%22•%22%5D%2C%5B%22Z%22%2C1%2C1%2C1%2C1%2C1%2C1%2C1%2C1%2C%22•%22%2C1%2C%22◦%22%5D%2C%5B1%2C%22Z%22%2C1%2C1%2C1%2C1%2C1%2C1%2C1%2C%22•%22%2C%22◦%22%5D%2C%5B%22Z%22%2C1%2C1%2C1%2C1%2C1%2C1%2C1%2C%22•%22%2C1%2C1%2C%22•%22%5D%2C%5B1%2C%22Z%22%2C1%2C1%2C1%2C1%2C1%2C1%2C%22•%22%2C1%2C%22◦%22%5D%2C%5B%22Z%22%2C1%2C1%2C1%2C1%2C1%2C1%2C%22•%22%2C1%2C1%2C1%2C%22◦%22%5D%2C%5B1%2C%22Z%22%2C1%2C1%2C1%2C1%2C1%2C%22•%22%2C1%2C1%2C%22•%22%5D%2C%5B%22Z%22%2C1%2C1%2C1%2C1%2C1%2C%22•%22%2C1%2C1%2C1%2C1%2C%22•%22%5D%2C%5B1%2C%22Z%22%2C1%2C1%2C1%2C1%2C%22•%22%2C1%2C1%2C1%2C%22•%22%5D%2C%5B%22Z%22%2C1%2C1%2C1%2C1%2C%22•%22%2C1%2C1%2C1%2C1%2C1%2C%22•%22%5D%2C%5B1%2C%22Z%22%2C1%2C1%2C1%2C%22•%22%2C1%2C1%2C1%2C1%2C%22•%22%5D%2C%5B%22Z%22%2C1%2C1%2C%22•%22%5D%2C%5B1%2C%22Z%22%2C1%2C1%2C%22•%22%5D%2C%5B1%2C1%2C%22Z%22%2C1%2C1%2C%22•%22%5D%2C%5B%22…%22%5D%2C%5B%22•%22%2C%22•%22%2C%22X%22%5D%2C%5B%22Y%5E-t%22%2C%22X%5E-t%22%5D%5D%2C%22init%22%3A%5B0%2C0%2C%22%2B%22%2C%22%2B%22%2C%22%2B%22%2C0%2C%22%2B%22%2C%22%2B%22%2C%22%2B%22%2C%22%2B%22%5D%7D}{Quirk}.
\end{lemma}
\begin{proof}
Immediately after consuming a $\toffstate$ state, we are left with the following state:
\begin{align}
    \ket{i, j} \mapsto \ket{i, j}\ket{ij \oplus x_2 i \oplus x_1 j \oplus x_1x_2}
\end{align}
where $x_1, x_2$ are measurement outcomes and not known a priori. There are four potential cases: either we require no correction, we require $\oplus i$, $\oplus j$, or $\oplus i \oplus j \oplus 1$. This forms our four $Z$-type PPMs:  $S_1$ has support on just the Wandering \gateand (WAND) qubit, $S_2$ has support on both the WAND qubit and the $i$ qubit, $S_3$ has support on $j$ and WAND, and $S_4$ has support on the $i, j, $ and WAND qubit. Then, as $\ell = 4$ subsets, the overhead of performing delayed-\gateparity, is $\ell + 1 = 5$ ancilla qubits via~\cref{lem:delayed-gateparity}, yielding a total size of $8$ qubits.
\end{proof}

The Pauli correction on the first qubit is:
\begin{align*}
    Z^{x_7(x_9\oplus 1) \oplus x_6x_9\oplus x_5(x_9\oplus 1) \oplus x_4x_9 \oplus x_3x_9 \oplus x_1} = Z^{x_9(x_3\oplus x_4 \oplus x_5\oplus x_6 \oplus x_7) \oplus x_1\oplus x_5\oplus x_7}.
\end{align*}
The Pauli correction on the second qubit is:
\begin{align*}
    Z^{x_7(x_8\oplus 1) \oplus x_6(x_8\oplus 1) \oplus x_5x_8 \oplus x_4x_8 \oplus x_3x_8 \oplus x_2} = Z^{x_8(x_3\oplus x_4 \oplus x_5 \oplus x_6 \oplus x_7)\oplus x_2 \oplus x_6 \oplus x_7}.
\end{align*}
The Pauli correction on the third qubit is:
\begin{align*}
    X^{(x_8\oplus 1)(x_9\oplus 1)(x_4\oplus x_{10}) \oplus (x_8\oplus 1)x_9(x_5\oplus x_{11}) \oplus x_8(x_9\oplus 1)(x_6\oplus x_{12}) \oplus x_8x_9(x_7\oplus x_{13} \oplus 1)} Z^{x_3}.
\end{align*}
We do not prove the Pauli corrections here.

The contributive depth of the \gateparity-based \texttt{AutoAND} is 1 (see~\Cref{def:contributive_depth}); in a large quantum circuit the \texttt{AutoAND} contributes the same as applying a $CCX$ gate to the state $\ket{x}\ket{y}\ket{0}$. The reaction depth is also 1, ignoring Paulis.

\begin{remark}
    The parity-based \gateautoand is particularly efficient if implemented in a setting where $Z$-type PPMs can be applied simultaneously, even if overlapping on qubits. Such protocols have been designed, see e.g. Refs.~\cite{cowtan2025parallel,baspin20262fast2surgery}. In that case, the parity-based \gateautoand can use only two logical timesteps, not including the input states.
\end{remark}

\begin{lemma}[\gatefanout-based \texttt{AutoAND}]\label{lem:fanout_autoand}
The following \texttt{AutoAND} state requires $7$ qubits and implements
\begin{align}
    \ket{i} \ket{j} \mapsto \ket{i} \ket{j} \ket{ij}
\end{align}
up to some Pauli corrections.

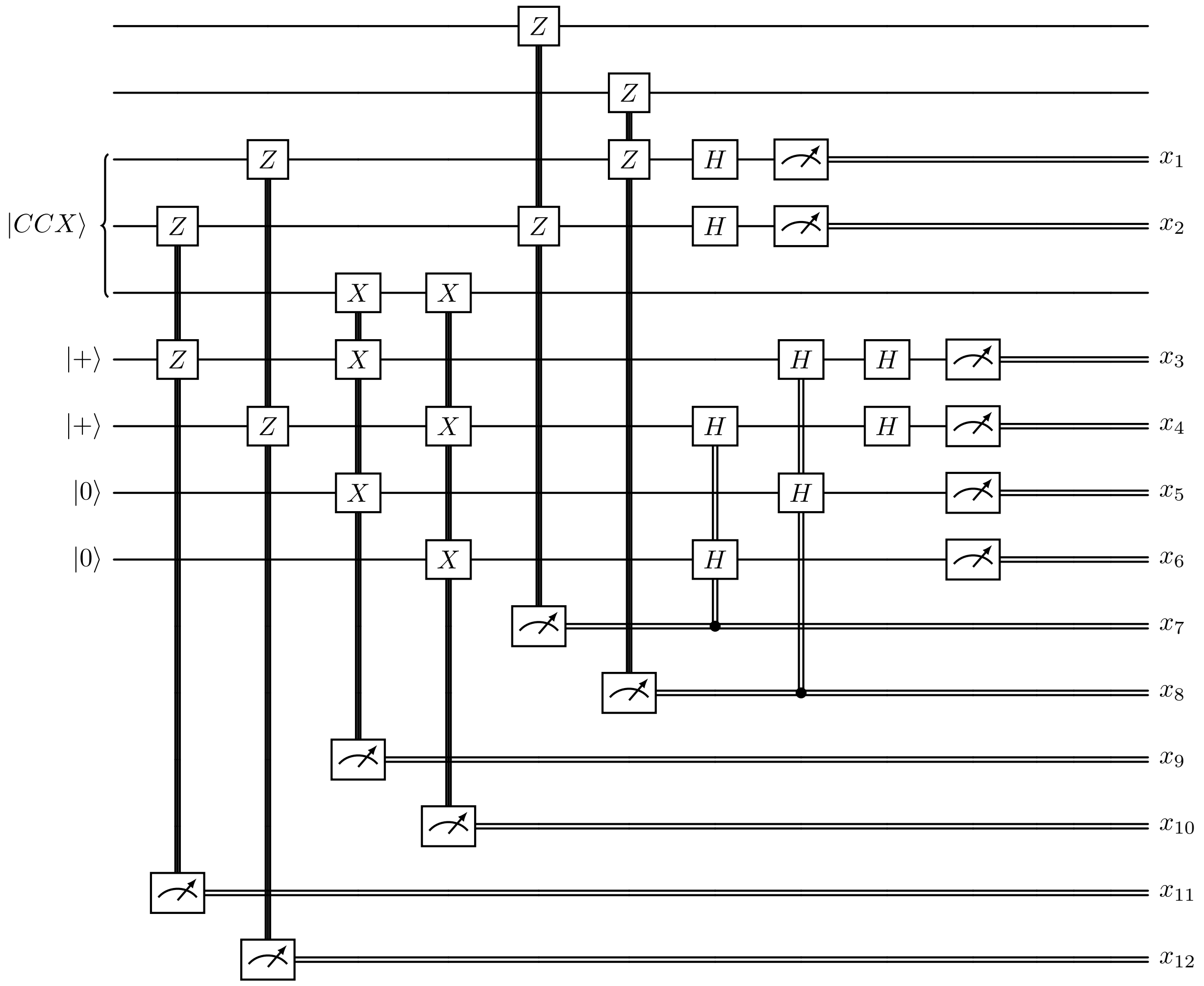
\begin{align*}
    \begin{quantikz}[wire types = {q,q,q,q,q,q,q,q,q,n,n,n,n,n,n}, row sep={0.9cm,between origins}]
        & & & & & \gate{Z}\wire[d]{b} & & & & & & & & & \\
        & & & & & \wire[d]{b} & \gate{Z}\wire[d]{b} & & & & & & & & \\
    \lstick[3]{$\toffstate$}    & & \gate{Z}\wire[d]{b} & & & \wire[d]{b} & \gate{Z}\wire[d]{b} & \gate{H} & \meter{} & \setwiretype{c} & & & & & \rstick{$x_1$}\\
                                        & \gate{Z}\wire[d]{b} & \wire[d]{b} & & & \gate{Z}\wire[d]{b} & \wire[d]{b} & \gate{H} & \meter{} & \setwiretype{c} & & & & & \rstick{$x_2$}\\
                                        & \wire[d]{b} & \wire[d]{b} & \gate{X}\wire[d]{b} & \gate{X}\wire[d]{b} & \wire[d]{b} & \wire[d]{b} & & & & & & & & \\
        \lstick{$\ket{+}$} & \gate{Z}\wire[d]{b} & \wire[d]{b} & \gate{X}\wire[d]{b} & \wire[d]{b} & \wire[d]{b} & \wire[d]{b} & & \gate{H} & \gate{H} & \meter{} & \setwiretype{c} & & & \rstick{$x_3$} \\
        \lstick{$\ket{+}$} & \wire[d]{b} & \gate{Z}\wire[d]{b} & \wire[d]{b} & \gate{X}\wire[d]{b} & \wire[d]{b} & \wire[d]{b} & \gate{H} & \wire[u]{c} & \gate{H} & \meter{} & \setwiretype{c} & & & \rstick{$x_4$} \\
        \lstick{$\ket{0}$} & \wire[d]{b} & \wire[d]{b} &\gate{X}\wire[d]{b} & \wire[d]{b} & \wire[d]{b} & \wire[d]{b} & \wire[u]{c} & \gate{H}\wire[u]{c} & & \meter{} & \setwiretype{c} & & & \rstick{$x_5$} \\
        \lstick{$\ket{0}$} & \wire[d]{b} & \wire[d]{b} & \wire[d]{b} & \gate{X}\wire[d]{b} & \wire[d]{b} & \wire[d]{b} & \gate{H}\wire[u]{c} & \wire[u]{c} & & \meter{} & \setwiretype{c} & & & \rstick{$x_6$} \\
        & \wire[d]{b} & \wire[d]{b} & \wire[d]{b} & \wire[d]{b} & \meter{} & \wire[d]{b}\setwiretype{c} & \phase{}\wire[u]{c} & \wire[u]{c} & & & & & & \rstick{$x_7$} \\
        & \wire[d]{b} & \wire[d]{b} & \wire[d]{b} & \wire[d]{b} & & \meter{} & \setwiretype{c} & \phase{}\wire[u]{c} & & & & & & \rstick{$x_8$} \\
        & \wire[d]{b} & \wire[d]{b} & \meter{} & \wire[d]{b} \setwiretype{c} & & & & & & & & & & \rstick{$x_9$} \\
        & \wire[d]{b} & \wire[d]{b} & & \meter{} & \setwiretype{c} & & & & & & & & & \rstick{$x_{10}$} \\
        & \meter{} & \wire[d]{b} \setwiretype{c} & & & & & & & & & & & & \rstick{$x_{11}$} \\
        & & \meter{} & \setwiretype{c} & & & & & & & & & & & \rstick{$x_{12}$}
    \end{quantikz}
\end{align*}
This is shown in \href{https://algassert.com/quirk#circuit=%7B%22cols%22%3A%5B%5B%22Y%5Et%22%2C%22X%5Et%22%5D%2C%5B1%2C1%2C%22X%22%2C%22%E2%80%A2%22%2C%22%E2%80%A2%22%5D%2C%5B1%2C1%2C1%2C%22zpar%22%2C1%2C%22zpar%22%2C1%2C1%2C1%2C1%2C1%2C1%2C1%2C%22X%22%5D%2C%5B1%2C1%2C1%2C1%2C%22zpar%22%2C1%2C%22zpar%22%2C1%2C1%2C1%2C1%2C1%2C1%2C1%2C%22X%22%5D%2C%5B1%2C1%2C%22%E2%80%A6%22%5D%2C%5B1%2C1%2C%22xpar%22%2C1%2C1%2C%22xpar%22%2C1%2C%22xpar%22%2C1%2C1%2C1%2C%22X%22%5D%2C%5B1%2C1%2C%22xpar%22%2C1%2C1%2C1%2C%22xpar%22%2C1%2C%22xpar%22%2C1%2C1%2C1%2C%22X%22%5D%2C%5B1%2C1%2C%22%E2%80%A6%22%5D%2C%5B%22zpar%22%2C1%2C1%2C%22zpar%22%2C1%2C1%2C1%2C1%2C1%2C%22X%22%5D%2C%5B1%2C%22zpar%22%2C1%2C1%2C%22zpar%22%2C1%2C1%2C1%2C1%2C1%2C%22X%22%5D%2C%5B1%2C1%2C1%2C%22H%22%2C%22H%22%5D%2C%5B1%2C1%2C1%2C%22Measure%22%2C%22Measure%22%2C1%2C1%2C1%2C1%2C%22Measure%22%2C%22Measure%22%2C%22Measure%22%2C%22Measure%22%2C%22Measure%22%2C%22Measure%22%5D%2C%5B1%2C1%2C1%2C1%2C1%2C1%2C%22H%22%2C1%2C%22H%22%2C%22%E2%80%A2%22%5D%2C%5B1%2C1%2C1%2C1%2C1%2C%22H%22%2C1%2C%22H%22%2C1%2C1%2C%22%E2%80%A2%22%5D%2C%5B1%2C1%2C1%2C1%2C1%2C%22H%22%2C%22H%22%5D%2C%5B1%2C1%2C1%2C1%2C1%2C%22Measure%22%2C%22Measure%22%2C%22Measure%22%2C%22Measure%22%5D%2C%5B%22Z%22%2C1%2C1%2C%22%E2%80%A2%22%5D%2C%5B1%2C%22Z%22%2C1%2C1%2C%22%E2%80%A2%22%5D%2C%5B%22Z%22%2C1%2C1%2C1%2C1%2C1%2C1%2C1%2C1%2C1%2C%22%E2%80%A2%22%2C%22%E2%80%A2%22%5D%2C%5B1%2C%22Z%22%2C1%2C1%2C1%2C1%2C1%2C1%2C1%2C%22%E2%80%A2%22%2C1%2C1%2C%22%E2%80%A2%22%5D%2C%5B1%2C1%2C%22X%22%2C1%2C1%2C1%2C1%2C1%2C1%2C1%2C%22%E2%80%A2%22%2C1%2C1%2C%22%E2%80%A2%22%5D%2C%5B1%2C1%2C%22X%22%2C1%2C1%2C1%2C1%2C1%2C1%2C%22%E2%80%A2%22%2C1%2C1%2C1%2C1%2C%22%E2%80%A2%22%5D%2C%5B%22Z%22%2C1%2C1%2C1%2C1%2C1%2C1%2C1%2C1%2C1%2C%22%E2%97%A6%22%5D%2C%5B1%2C1%2C%22X%22%2C1%2C1%2C%22%E2%80%A2%22%2C1%2C1%2C1%2C1%2C%22%E2%80%A2%22%5D%2C%5B%22Z%22%2C1%2C1%2C1%2C1%2C%22%E2%97%A6%22%2C1%2C1%2C1%2C1%2C%22%E2%97%A6%22%5D%2C%5B1%2C1%2C%22X%22%2C1%2C1%2C1%2C%22%E2%80%A2%22%2C1%2C1%2C%22%E2%80%A2%22%5D%2C%5B1%2C%22Z%22%2C1%2C1%2C1%2C1%2C%22%E2%80%A2%22%2C1%2C1%2C%22%E2%97%A6%22%5D%2C%5B1%2C1%2C%22X%22%2C1%2C1%2C1%2C1%2C%22%E2%80%A2%22%2C1%2C1%2C%22%E2%97%A6%22%5D%2C%5B%22Z%22%2C1%2C1%2C1%2C1%2C1%2C1%2C%22%E2%80%A2%22%2C1%2C1%2C%22%E2%80%A2%22%5D%2C%5B%22%E2%80%A6%22%2C1%2C%22X%22%2C1%2C1%2C1%2C1%2C1%2C%22%E2%80%A2%22%2C%22%E2%97%A6%22%5D%2C%5B1%2C%22Z%22%2C1%2C1%2C1%2C1%2C1%2C1%2C%22%E2%80%A2%22%2C%22%E2%80%A2%22%5D%2C%5B1%2C1%2C%22X%22%2C1%2C1%2C1%2C1%2C1%2C1%2C%22%E2%80%A2%22%2C%22%E2%80%A2%22%5D%2C%5B%22%E2%80%A2%22%2C%22%E2%80%A2%22%2C%22X%22%5D%2C%5B%22Y%5E-t%22%2C%22X%5E-t%22%5D%5D%2C%22init%22%3A%5B0%2C0%2C0%2C%22%2B%22%2C%22%2B%22%2C%22%2B%22%2C%22%2B%22%5D%7D}{Quirk}.
\end{lemma}
\begin{proof}
As before, we consider the corrections needed from the WAND gate. Note that there are only $\ell = 2$ types of fanouts required: either $ \oplus i$ or $\oplus j$. Na\"ively, one might hope then that this requires just $2$ delayed-choice consumptive \gatefanout gates. However, recall that the consumptive fanout can effect a $Z$-type error on consumed qubits; thus, we first need to make a Z-type copy, then implement the delayed-choice gates. Thus, there are $2 + 2=4$ overhead qubits, for a state of total size $7$.
\end{proof}

The Pauli correction on the first qubit is,
\[Z^{x_2\oplus x_8x_9 \oplus x_3(x_8 \oplus 1) \oplus x_5x_8} = Z^{x_2 \oplus x_3 \oplus x_8(x_3 \oplus x_5 \oplus x_9)}.\]
The Pauli correction on the second qubit is,
\[Z^{x_1 \oplus x_7x_{10} \oplus x_4(x_7 \oplus 1) \oplus x_6x_7} = Z^{x_1 \oplus x_4 \oplus x_7(x_4\oplus x_6 \oplus x_{10})}.\]
The Pauli correction on the third qubit is,
\[X^{x_8x_{11} \oplus x_7x_{12} \oplus x_3x_8 \oplus x_4x_7 \oplus x_5(x_8 \oplus 1)\oplus x_6(x_7\oplus 1) \oplus x_7x_8} = X^{x_7(x_4\oplus x_6 \oplus x_8 \oplus x_{12}) \oplus x_8(x_3\oplus x_5 \oplus x_{11}) \oplus x_5\oplus x_6}.\]

The contributive depth and reaction depth of the \gatefanout-based \texttt{AutoAND} are both also 1.

\subsection{Boolean logic in constant quantum and reaction depth}\label{ss:boolean_logic_constant_depth}

We now turn our attention to techniques for implementing arbitrary Boolean computation. We begin by defining ``wandering monomial'' states, which play an important role in the constructions of this section. 

\begin{define}[Wandering Monomial States]
    For any bit string $x \in \{0,1\}^n$ we say a state is a wandering monomial state of $x$, denoted $\ket{\mathtt{wandering\_mon(x)}}$ if it is a $2^n$ qubit state of the form: 
    \begin{align*}
        \ket{ \mathtt{wandering\_mon}(x)} = \bigotimes_{S \subseteq [n]} \ket{x_S \oplus p_S(x)}
    \end{align*}
    where $x_S := \prod_{i \in S} x_i$ and each $p_S(x)$ is a polynomial of degree $d < |S|$ supported on the bits $\{x_i\}_{i \in S}$. (So this construction associates a qubit with each of the $2^n$ possible subsets of $[n]$.) We call the $p_S(x)$ the polynomial corrections of the wandering monomial state. 
\end{define}

In all examples in this paper, we construct wandering monomial states with polynomial corrections which are determined by measurement outcomes on ancilla qubits consumed during the construction, but we do not demand this in the definition.

We now state the main result of this section, which shows it is possible to compute an $n$-bit function $f(x)$ in our circuit model with constant reaction depth and quantum depth. However, this construction requires exponentially many ancilla which are left ``dirty'' after the computation. 

\begin{theorem}
\label{thm:arbitrary_boolean}
    Let $f : \{0,1\}^n \rightarrow \{0,1\}$ be an arbitrary Boolean function. Then there is a circuit  which consists only of homogeneous PPMs acting on an n-qubit input state $\ket{\psi} = \sum_x \alpha_x \ket{x}$ along with $2^n$ $\ket{CCX}$ states, $\mathcal{O}(2^n)$ ancilla qubits in the $\ket{+}$ state, and $\mathcal{O}(2^{n})$ ancilla qubits in the $\ket{0}$ state which effects the transformation:
    \begin{align}
        \sum_x \alpha_x\ket{x} \otimes\ket{CCX}^{\otimes \, 2^n} \otimes \ket{0}^{\otimes \, \mathcal{O}(2^{n})} \otimes \ket{+}^{\otimes \, \mathcal{O}(2^n)} \rightarrow \sum_x \alpha_x \ket{x} \ket{f(x)} \ket{ \mathtt{wandering\_mon}(x)}
    \end{align}
    This circuit has constant quantum depth and reaction depth, and the polynomial corrections of the wandering monomial state are determined by measurements made on ancilla qubits in the circuit.
\end{theorem}

\begin{corollary}
    \label{cor:arbitrary_vector_boolean}
    Let $f : \{0,1\}^n \rightarrow \{0,1\}^m$ be an arbitrary vectorial Boolean function. Then there is a circuit  which consists only of homogeneous PPMs acting on an n-qubit input state $\ket{\psi} = \sum_x \alpha_x \ket{x}$ along with $2^n$ $\ket{CCX}$ states, $\mathcal{O}(m2^n)$ ancilla qubits in the $\ket{+}$ state, and $\mathcal{O}(m2^{n})$ ancilla qubits in the $\ket{0}$ state which effects the transformation:
    \begin{align}
        \sum_x \alpha_x\ket{x} \otimes\ket{CCX}^{\otimes \, 2^n} \otimes \ket{0}^{\otimes \, \mathcal{O}(m2^{n})} \otimes \ket{+}^{\otimes \, \mathcal{O}(m2^n)} \rightarrow \sum_x \alpha_x \ket{x} \ket{f(x)} \ket{ \mathtt{wandering\_mon}(x)}
    \end{align}
    This circuit has constant quantum depth and reaction depth, and the polynomial corrections of the wandering monomial state are determined by measurements made on ancilla qubits in the circuit.
\end{corollary}

Allowing for heterogeneous PPMs, the $\ket{ \mathtt{wandering\_mon}(x)}$ states left on ancilla qubits in the above two constructions can also be uncomputed in logarithmic depth, while only incurring Pauli corrections on the state $\ket{x}$.\footnote{Heterogeneous PPMs are a substitute for $CZ$s, which we do not assume can be done directly. See~\Cref{ssec:MBU} for a longer discussion of this issue.} Formalizing this observation gives a circuit which implements arbitrary Boolean computation with constant contributive depth, and without leaving any ancilla dirty.

\begin{corollary}
    \label{cor:arbitrary_vector_boolean_contributive}
    Let $f : \{0,1\}^n \rightarrow \{0,1\}^m$ be an arbitrary vectorial Boolean function. Then there is a circuit consisting of homogeneous and heterogeneous PPMs acting on an n-qubit input state $\ket{\psi} = \sum_x \alpha_x \ket{x}$ along with $2^n$ $\ket{CCX}$ states, $\mathcal{O}(m2^n)$ ancilla qubits in the $\ket{+}$ state, and $\mathcal{O}(m2^{n})$ ancilla qubits in the $\ket{0}$ state which effects the transformation:
    \begin{align}
        \sum_x \alpha_x\ket{x} \otimes\ket{CCX}^{\otimes \, 2^n} \otimes \ket{0}^{\otimes \, \mathcal{O}(m2^{n})} \otimes \ket{+}^{\otimes \, \mathcal{O}(m2^n)} \rightarrow \sum_x \alpha_x \ket{x} \ket{f(x)} 
    \end{align}
    This circuit has quantum depth and reaction depth $O(\log(n))$, and constant contributive depth.
\end{corollary}

Before giving the full proof of~\Cref{thm:arbitrary_boolean,cor:arbitrary_vector_boolean,cor:arbitrary_vector_boolean_contributive} we will prove the following helpful lemma. 

\begin{lemma}
\label{lem:wan_mon_construction}
     There is a circuit  which consists only of homogeneous PPMs acting on an n-qubit input state $\ket{\psi} = \sum_x \alpha_x \ket{x}$ along with $2^n$ $\ket{CCX}$ states and $O(2^n)$ ancilla qubits in the $\ket{0}$ state which effects the transformation:
    \begin{align}
        \sum_x \alpha_x\ket{x} \otimes\ket{CCX}^{\otimes \, 2^n} \otimes \ket{0}^{\otimes \, \mathcal{O}(2^n)}  \rightarrow \sum_x \alpha_x \ket{x}  \ket{ \mathtt{wandering\_mon}(x)}
    \end{align}
    This circuit has constant quantum depth and reaction depth, and the polynomial corrections of the wandering monomial state are determined by measurements made on ancilla qubits in the circuit.
\end{lemma}

\begin{proof}
    We first recall the $\gateand$ gate construction given in~\Cref{fig:standard_AND}. Ignoring the Pauli on the target qubit, this circuit implements the wandering-\gateand operation on unmeasured qubits: 
    \begin{align*}
        \ket{r_1}\ket{r_2}\ket{CCX} \rightarrow \ket{r_1}\ket{r_2} \ket{r_1r_2\oplus m_2r_1\oplus m_1 r_2 \oplus m_1m_2}.
    \end{align*}
    We make two further important observations. First, the output qubit storing the target of the wandering-\gateand operation is not acted on during this circuit, so a circuit consisting of a sequence of these wandering-\gateand circuits with outputs of one circuit used as inputs for the next has constant quantum depth and reaction depth. Second, when $r_1$ and $r_2$ are themselves wandering monomials with disjoint support, i.e. $r_1 = x_S \oplus p_S$, $r_2 = x_Q \oplus p_Q$ and $S \cap Q = \emptyset$, then we have: 
    \begin{align*}
        r_1r_2\oplus m_2r_1\oplus m_1 r_2 \oplus m_1m_2 &= (x_S \oplus p_S)(x_Q \oplus p_Q) \oplus m_2(x_S \oplus p_S) \oplus m_1 (x_Q \oplus p_Q) \oplus m_1m_2 \\
        &= x_Sx_Q \oplus \big( x_S p_Q  \oplus p_Sx_Q \oplus p_Sp_Q \oplus m_2(x_S \oplus p_S) \oplus m_1 (x_Q \oplus p_Q) \oplus m_1m_2\big) \\
        &= x_{(S \cup Q)} \oplus p_{(S \cup Q)},
    \end{align*} 
    where on the last line we have expressed that the final sum can also be viewed as a wandering monomial on the subset of qubits indexed by $S \cup Q$. This shows the output bit of the wandering-\gateand applied to two wandering monomials with disjoint support can itself be viewed as a wandering monomial on the union of those supports.

    These two observations form the basis of our circuit construction. We construct each wandering monomial in the state $\ket{\mathtt{wandering\_mon}(x)}$ by applying a wandering-\gateand to two wandering monomial states of lower degree. Our second observation shows that every wandering monomial can be constructed in this way. Our first shows that the quantum and reaction depths of the resulting circuit are bounded by the number of times any particular wandering monomial is used as an \textit{input} to a wandering-\gateand operation. To keep this quantity low we use \gatefanout gates (as described in~\Cref{lem:xtype_ppm}) to copy each wandering monomial state, ensuring that any given copy is used at most once as input to a wandering-\gateand gate. 
    
    There are a number of patterns we could choose for constructing higher degree monomials from lower degree ones, with many requiring similar resources. For the asymptotic analysis, it suffices to note that each degree $d > 1$ wandering monomial is constructed from a wandering-\gateand of two lower degree wandering monomials, so we will need at most $2(2^{n})$ ancilla to act as targets of fanout operations and subsequent inputs to wandering-\gateand gates, and at most $2^n$ different fanout operations total (corresponding to each of the different wandering monomials which could be fanned out). This circuit needs at most $O(2^n)$ ancilla qubits.

    After the $\ket{\mathtt{wandering\_mon}(x)}$ state is constructed, all ancilla qubits not used in the construction can be measured out in the $\ket{\pm}$ basis, and phase errors can be corrected by Pauli $Z$ operations applied to the $\ket{\mathtt{wandering\_mon}(x)}$ state, adding at most one additional layer of quantum depth.
    
\end{proof}

Now we move on to the proofs of~\Cref{thm:arbitrary_boolean,cor:arbitrary_vector_boolean}.

\begin{proof}[Proof (\Cref{thm:arbitrary_boolean})]
To begin, we write the function $f(x)$ in algebraic normal form (as an XOR of monomials): 
\begin{align*}
f(x) = \bigoplus_{M \in \mathcal{M}_f} x_M 
\end{align*}
where each $M \in \mathcal{M}_f$ is a subset of indices $M \subseteq [n]$ and the set $\mathcal{M}_f$ is determined by the function $f$ (with the convention that $x_\emptyset = 1$). Next, we consider some monomial state $\ket{\mathtt{wandering\_mon}(x)}$ and for any $S \subseteq [n]$ let 
$
    \tilde{x}_S = x_S + p_S
$
be the wandering monomial stored in the state in place of the $x_S$ monomial. Then, since the $p_S$ corrections are all of lower degree than $x_S$ (by definition of a wandering monomial state) we can also write
\begin{align*}
f(x) = \bigoplus_{M \in \tilde{\mathcal{M}}_f} \tilde{x}_M 
\end{align*}
for a set of indices $\tilde{\mathcal{M}}_f$ obtained by beginning with the highest degree monomials appearing in the sum and iteratively replacing each $x_M$ with $\tilde{x}_M \oplus p_M$. Since each monomial in $p_M$ is of lower degree than $x_M$ this procedure will eventually terminate.

These observations show we can effect the desired transformation by first preparing a wandering monomial state, then writing the parity of a subset of wandering monomials in a fresh ancilla qubit. \Cref{lem:wan_mon_construction} shows the first operation is possible with constant quantum and reaction depth, and the parity calculation can be performed using an out-of-place XOR, which adds at most one layer of reaction depth by~\Cref{lem:ztype_ppm}. Implementing this XOR directly gives a procedure with constant reaction and quantum depth for any circuit \emph{run} (see~\Cref{def:reaction_depth_run}). However, the subset of wandering monomials used as input in the parity operation is controlled by the measurement outcomes observed when preparing the wandering monomial state, and there are doubly exponentially many subsets which may be selected. Thus, the circuit as described has doubly exponential quantum depth. 

To avoid this, we introduce an additional $2^n$ ancilla in the $\ket{+}$ and $\ket{0}$ states and use $2^n$ delayed choice fanout operations (\Cref{lem:delayed-fanout}), each controlled on a different wandering monomial and targeting a fresh ancilla which are used to copy only the wandering monomials whose parity is required to compute $f(x)$, then apply a fixed XOR operation on all $2^n$ ancilla to compute the desired function. This circuit now effects the desired transformation with constant quantum and reaction depth, and $\mathcal{O}(2^n)$ ancilla used as overhead. 
\end{proof}

\begin{proof}[Proof (\Cref{cor:arbitrary_vector_boolean})]
The argument proceeds identically to~\Cref{thm:arbitrary_boolean} except that we use fanout operations to make $m$ copies of the wandering monomial state after it has been constructed. Then we compute all bits of $f(x)$ in parallel using a different wandering monomial state for each bit. Finally, we can measure out all but one wandering monomial state, and correct any phase errors using controlled Pauli operations on the one remaining wandering monomial state, producing an output of the claimed form. 
\end{proof}

\begin{proof}[Proof (\Cref{cor:arbitrary_vector_boolean_contributive})]

We first show that it is possible to perform the uncomputation
\begin{align*}
    \sum_x \alpha_x\ket{x}\ket{ \mathtt{wandering\_mon}(x)} \rightarrow \sum_x\alpha_x\ket{x}
\end{align*}\
with $O(\log(n))$ quantum and reaction depth using heterogeneous PPMs. There are a variety of ways to do this, but one straightforward approach is to first measure all qubits in the wandering monomial state corresponding to monomials of degree $> n/2$ in the $\ket{\pm}$ basis. Each $\ket{-}$ outcome observed contributes a phase $(-1)^{x_S \, \oplus \, p_S}$ to the remaining state ($\ket{+}$ outcomes contribute no phase), and these phase terms can be corrected by a collection of $CZ$ operations on the unmeasured qubits, since the degree of each measured $x_S$ and $p_S$ is at most twice the degree of the unmeasured monomials. In~\Cref{ssec:uncomputation} we review one technique for implementing $CZ$ gates, for now we just note this collection of $CZ$ corrections can be implemented in constant quantum and reaction depth by using fanout gates if necessary to parallelize multiple $CZ$ operations on the same qubit. Repeating this procedure $\lceil\log(n)\rceil$ times, each time measuring out the unmeasured wandering monomials with degree $> n'/2$ for $n' = n, n/2,n/4,\cdots$ implements the desired operation. 

Now to prove the main result, we consider first implementing the circuit as described in~\Cref{cor:arbitrary_vector_boolean}, making one additional copy of $\ket{x}$ onto fresh ancilla qubits, using the procedure described above and the copy of the $\ket{x}$ qubits to uncompute the $\ket{\mathtt{wandering\_mon}(x)}$ state,  and then measuring out the $\ket{x}$ copy in the $\{\pm\}$ basis and correcting phase errors by Pauli $Z$ operations on the input $\ket{x}$ state. This circuit has logarithmic quantum depth by the above analysis. Moreover when this circuit is compared to the one described in~\Cref{cor:arbitrary_vector_boolean}, the only new operations which touch the circuit's output qubits are the $\ket{x}$ copying operations, which occur before the uncomputation step, and the final Pauli $Z$ corrections which do not increase the circuit's contributive depth. So this circuit also has constant contributive depth, as desired.
\end{proof}

Finally, we remark that the heterogeneous PPMs used in the proof of~\Cref{cor:arbitrary_vector_boolean_contributive} are only used in the uncomputation step and do not add to the contributive depth of the circuit. Thus, even in an architecture where heterogeneous PPMs are considerably slower than homogeneous PPMs, the techniques above can allow for fast Boolean computation. 

\section{Examples of fault-tolerant Boolean logic}\label{sec:compilation_strategy}

Of course, using exponential space and, in particular, an exponential number of $\toffstate$ states to minimize the depth of coherent Boolean logic is unlikely to be viable for large Boolean circuits. However, small subcircuits can be compiled into efficient implementations, and then composed together. The main idea is that, rather than converting large Boolean circuits into unitary quantum circuits (e.g. Ref.~\cite{gidney2018halving}) and then converting these into PPMs, one should translate the Boolean circuits directly into PPMs, replacing $\gateand$ and $\gatexor$ gates with their PPM implementations. 
We give a series of examples in this section. For this, we rely on the techniques based on delayed-choice PPMs described in \Cref{sec:boolean_results}. There are several other optimizations which we use on a case-by-case basis.

\subsection{Measurement-based uncomputation}
\label{ssec:MBU}
In earlier sections, we either constructed operations which did not require measurement-based uncomputation (such as the \gateand gate), or operations where we left qubits dirty (\cref{ss:boolean_logic_constant_depth}).

To perform useful computation, such as adders and multipliers, we will have to `uncompute' qubits, so that they may be reused again in the larger quantum algorithm. For example, when adding two $\ell$-bit numbers $a$ and $b$, we would commonly like to effect the \textit{in-place} transformation $(a, b) \mapsto (a, a+b)$, not the \textit{out-of-place} transformation $(a, b) \mapsto (a, b, a+b)$. This is because we would like to reset the qubits containing $b$ to the $\ket{0}^{\otimes \, \ell}$. In principle one could uncompute the qubits containing $b$ by applying (or simulating) reversible gates, but this is typically inefficient.

\begin{define}[Measurement-based uncomputation]
   Let $R$, $A$ and $B$ be registers of qubits. Let $\ket{\psi} = \sum_x \ket{\phi_x}_R \ket{x}_A \ket{g(x)}_B$. Measurement-based uncomputation is the procedure whereby $B$ is measured out and corrections are applied to the other registers such that the remaining state is $\sum_x \ket{\phi_x}_R\ket{x}_A$.
\end{define}
An example of measurement-based uncomputation is the uncomputation of an \gateand gate using a measurement and a controlled $CZ$ gate~\cite[Fig.~3]{gidney2018halving}, which replaces the `UMA' gate of Cuccaro et al.~\cite{cuccaro2004new}.

In general, even though the \textit{computation} of a Boolean function can be performed using only our restricted gateset, the \textit{uncomputation} of dirty qubits cannot without exorbitant cost (for example, using additional $\toffstate$ states). Thus, for uncomputation, we permit ourselves to use heterogeneous (i.e. non-homogeneous) PPMs, which we consider `slower' than the homogeneous PPMs. 

\subsection{Ripple-carry adder}\label{sec:ripple_carry}
We now compile the quantum ripple-carry adder~\cite{cuccaro2004new,gidney2018halving}, which has linear depth, gate count and linear \gateand count. Our implementation is a variation on Ref.~\cite[App.~A.2]{gidney2025how}. 

\subsubsection{Auto-corrected bitwise addition}

For addition mod $2^\ell$, let $a$ and $b$ be $\ell$-bit inputs and let $c_k$ be the carry into position $k$. Then, letting $s = a+b$ and $c_0 = 0$,
\[s_k = a_k \oplus b_k \oplus c_k, \qquad c_{k+1} = a_k\cdot b_k \oplus (a_k\oplus b_k)\cdot c_k\]
Therefore, by substitution and factorizing, for $0 \leq k < \ell - 1$,
\[s_{k+1} = ((a_k\oplus s_k)\cdot(b_k\oplus s_k))\oplus (a_k \oplus b_k \oplus s_k \oplus a_{k+1} \oplus b_{k+1}).\]
$s_{k+1}$ is the output bit, which is used in the next bit addition circuit for $s_{k+2}$. Each $b_k$ is measured out after it is no longer needed, resulting in an overall map:
\[(a,b) \mapsto (a, s = a+b \mod 2^\ell).\]
For non-modular addition, set $a_\ell = b_\ell = 0$ and retain $s_\ell = c_\ell$. (This is called the carry-out bit.)
Viewed as a Boolean function, the sum function for a given position contains only one \gateand gate, and three \gatexor gates (with unbounded fan-in).

\begin{lemma}\label{lem:adder_circuit}
    The following circuit $\mathcal{C}$ implements 
    \[s_{k+1} = ((a_k\oplus s_k)\cdot(b_k\oplus s_k))\oplus (a_k \oplus b_k \oplus s_k \oplus a_{k+1} \oplus b_{k+1})\]
    up to Pauli corrections.

\begin{align*}
    \scalebox{0.8}{\begin{quantikz}[wire types = {q,q,q,q,q,q,q,q,q,q,q,q,q,n,n,n,n,n,n,n}, row sep={0.9cm,between origins}]
    \lstick{$a_k$}                            & & & & & \gate{Z}\wire[d]{b} & \gate{Z}\wire[d]{b} & & & & & & & & \rstick{$a_k$} \\
    \lstick{$b_k$}                            & & & & & \gate{Z}\wire[d]{b} & \wire[d]{b} & \gate{Z}\wire[d]{b} & & & & & & & \rstick{$b_k$} \\
    \lstick{$s_k$}                            & & & & & \gate{Z}\wire[d]{b} & \gate{Z}\wire[d]{b} & \gate{Z}\wire[d]{b} & & & & & & & \rstick{$s_k$} \\
    \lstick{$a_{k+1}$}    & & & & & \gate{Z}\wire[d]{b} & \wire[d]{b} & \wire[d]{b} & & & & & & & \rstick{$a_{k+1}$} \\
    \lstick{$b_{k+1}$}    & & & & & \gate{Z}\wire[d]{b} & \wire[d]{b} & \wire[d]{b} & & & & & & & \rstick{$b_{k+1}$} \\
    \lstick[3]{$\toffstate$}    & & \gate{Z}\wire[d]{b} & & & \wire[d]{b} & \wire[d]{b} & \gate{Z}\wire[d]{b} & \gate{H} & \meter{} & \setwiretype{c} & & & & \rstick{$x_1$}\\
                                        & \gate{Z}\wire[d]{b} & \wire[d]{b} & & & \wire[d]{b} & \gate{Z}\wire[d]{b} & \wire[d]{b} &  \gate{H} & \meter{} & \setwiretype{c} & & & & \rstick{$x_2$}\\
                                        & \wire[d]{b} & \wire[d]{b} & \gate{X}\wire[d]{b} & \gate{X}\wire[d]{b} & \gate{Z}\wire[d]{b} & \wire[d]{b} & \wire[d]{b} & \gate{H} & \meter{} & \setwiretype{c} & & & & \rstick{$x_{3}$} \\
        \lstick{$\ket{+}$} & \gate{Z}\wire[d]{b} & \wire[d]{b} & \gate{X}\wire[d]{b} & \wire[d]{b} & \wire[d]{b} & \wire[d]{b} & \wire[d]{b} & & \gate{H} & \gate{H} & \meter{} & \setwiretype{c} & & \rstick{$x_4$} \\
        \lstick{$\ket{+}$} & \wire[d]{b} & \gate{Z}\wire[d]{b} & \wire[d]{b} & \gate{X}\wire[d]{b} & \wire[d]{b} & \wire[d]{b} & \wire[d]{b} & \gate{H} & \wire[u]{c} & \gate{H} & \meter{} & \setwiretype{c} & & \rstick{$x_5$} \\
        \lstick{$\ket{0}$} & \wire[d]{b} & \wire[d]{b} &\gate{X}\wire[d]{b} & \wire[d]{b} & \wire[d]{b} & \wire[d]{b} & \wire[d]{b} & \wire[u]{c} & \gate{H}\wire[u]{c} & & \meter{} & \setwiretype{c} & & \rstick{$x_6$} \\
        \lstick{$\ket{0}$} & \wire[d]{b} & \wire[d]{b} & \wire[d]{b} & \gate{X}\wire[d]{b} & \wire[d]{b} & \wire[d]{b} & \wire[d]{b} & \gate{H}\wire[u]{c} & \wire[u]{c} & & \meter{} & \setwiretype{c} & & \rstick{$x_7$} \\
        \lstick{$\ket{+}$} & \wire[d]{b} & \wire[d]{b} & \wire[d]{b} & \wire[d]{b} & \gate{Z}\wire[d]{b} & \wire[d]{b} & \wire[d]{b} &  \wire[u]{c} & \wire[u]{c} & & & & & \rstick{$s_{k+1}$}\\
        & \wire[d]{b} & \wire[d]{b} & \wire[d]{b} & \wire[d]{b} & \wire[d]{b} & \meter{} & \wire[d]{b}\setwiretype{c} & \phase{}\wire[u]{c} & \wire[u]{c} & & & & & \rstick{$x_8$} \\
        & \wire[d]{b} & \wire[d]{b} & \wire[d]{b} & \wire[d]{b} & \wire[d]{b} & & \meter{} & \setwiretype{c} & \phase{}\wire[u]{c} & & & & & \rstick{$x_9$} \\
        & \wire[d]{b} & \wire[d]{b} & \meter{} & \wire[d]{b} \setwiretype{c} & \wire[d]{b} & & & & & & & & & \rstick{$x_{10}$} \\
        & \wire[d]{b} & \wire[d]{b} & & \meter{} & \wire[d]{b}\setwiretype{c} & & & & & & & & & \rstick{$x_{11}$} \\
        & \meter{} & \wire[d]{b} \setwiretype{c} & & & \wire[d]{b} & & & & & & & & & \rstick{$x_{12}$} \\
        & & \meter{} & \setwiretype{c} & & \wire[d]{b} & & & & & & & & & \rstick{$x_{13}$} \\
        & & & & & \meter{} & \setwiretype{c} & & & & & & & & \rstick{$x_{14}$}
    \end{quantikz}}
\end{align*}
    
\end{lemma}
\proof
First, take the \gatefanout-based \gateautoand from~\cref{lem:fanout_autoand} and apply it to $(a_k \oplus s_k)$ and $(b_k\oplus s_k)$. The XORs can be merged into the $Z$-type PPMs which perform the injection; this can be seen by the fact that the output of, say, $(a_k \oplus s_k)$ is used only in the $Z$ PPM for the \gateautoand, so the output can be subsumed into the PPM doing the injection. Second, the XOR $\cdots \oplus (a_k \oplus b_k \oplus s_k \oplus a_{k+1} \oplus b_{k+1})$ is performed on the output of the \gateautoand using the out-of-place XOR from~\cref{lem:ztype_ppm}. The obsolete qubit of the initial $\toffstate$ state is then measured out in the $X$ basis.
\endproof

$\mathcal{C}$ uses 8 ancilla qubits, has quantum depth 4 and contributive quantum depth 3. In an $\ell$-bit ripple-carry adder, $\mathcal{C}$ will be performed $\ell$ times, each output being the input to the next. However, the $Z$-type PPMs for the $k+1$th copy of $\mathcal{C}$ do not need to be performed after the $Z$-type PPMs for the $k$th copy of $\mathcal{C}$. They can be performed before, as they commute, and so one can acquire the following circuit for the computation of an $\ell$-bit ripple carry adder:

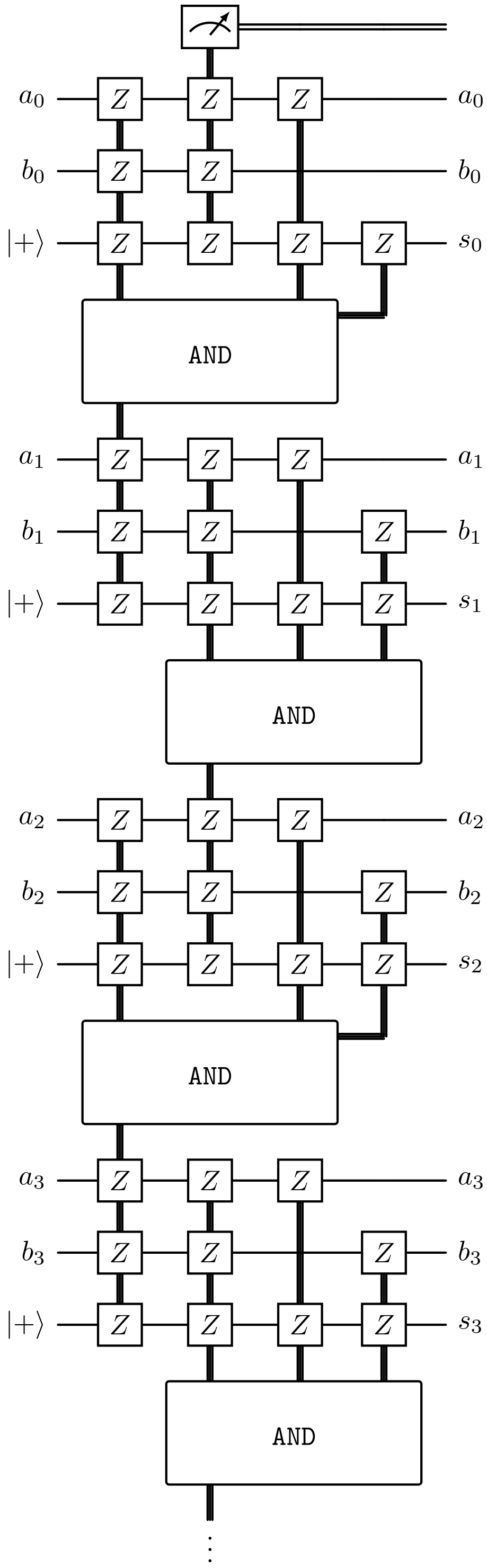
\begin{align*}
    \begin{quantikz}[wire types = {n,q,q,q,n,n,q,q,q,n,n,q,q,q,n,n,q,q,q,n,n,n}, row sep={0.9cm,between origins}]
        & & \meter{}\wire[d]{b} & \setwiretype{c} & & \\
        \lstick{$a_0$} & \gate{Z}\wire[d]{b} & \gate{Z}\wire[d]{b} & \gate{Z}\wire[d]{b} & & \rstick{$a_0$} \\
        \lstick{$b_0$} & \gate{Z}\wire[d]{b} & \gate{Z}\wire[d]{b} & \wire[d]{b} & & \rstick{$b_0$} \\
        \lstick{$\ket{+}$} & \gate{Z}\wire[d]{b} & \gate{Z} & \gate{Z}\wire[d]{b} & \gate{Z}\wire[d]{b} & \rstick{$s_0$}\\
        & \mygate & & & \setwiretype{b} & \setwiretype{n}\\
        & & & & & \\
        \lstick{$a_1$} & \gate{Z}\wire[u]{b} & \gate{Z}\wire[d]{b} & \gate{Z}\wire[d]{b} & & \rstick{$a_1$} \\
        \lstick{$b_1$} & \gate{Z}\wire[u]{b} & \gate{Z}\wire[d]{b} & \wire[d]{b} & \gate{Z}\wire[d]{b} & \rstick{$b_1$} \\
        \lstick{$\ket{+}$} & \gate{Z}\wire[u]{b} & \gate{Z}\wire[d]{b} & \gate{Z}\wire[d]{b} & \gate{Z}\wire[d]{b} & \rstick{$s_1$} \\
        & & \mygate & & & \\
        & & & & & \\
        \lstick{$a_2$} & \gate{Z}\wire[d]{b} & \gate{Z}\wire[u]{b} & \gate{Z}\wire[d]{b} & & \rstick{$a_2$} \\
        \lstick{$b_2$} & \gate{Z}\wire[d]{b} & \gate{Z}\wire[u]{b} & \wire[d]{b} & \gate{Z}\wire[d]{b} & \rstick{$b_2$} \\
        \lstick{$\ket{+}$} & \gate{Z}\wire[d]{b} & \gate{Z}\wire[u]{b} & \gate{Z}\wire[d]{b} & \gate{Z}\wire[d]{b} & \rstick{$s_2$} \\
        & \mygate & & & \setwiretype{b} & \setwiretype{n}\\
        & & & & & \\
        \lstick{$a_3$} & \gate{Z}\wire[u]{b} & \gate{Z}\wire[d]{b} & \gate{Z}\wire[d]{b} & & \rstick{$a_3$} \\
        \lstick{$b_3$} & \gate{Z}\wire[u]{b} & \gate{Z}\wire[d]{b} & \wire[d]{b} & \gate{Z}\wire[d]{b} & \rstick{$b_3$} \\
        \lstick{$\ket{+}$} & \gate{Z}\wire[u]{b} & \gate{Z}\wire[d]{b} & \gate{Z}\wire[d]{b} & \gate{Z}\wire[d]{b} & \rstick{$s_3$} \\
        & & \mygate & & & \\
        & & \wire[d]{b} & & & \\
        & & \vdots & & &
    \end{quantikz}
\end{align*}

Each \gateand box includes a $\toffstate$ state and 4 other ancilla qubits, all of which are measured out as per~\cref{lem:adder_circuit}. That is, in the circuit shown in~\cref{lem:adder_circuit}, the qubits which output classical bits $x_1,x_2,\cdots,x_7$ are contained in an \gateand box. The contributive quantum depth of the $\ell$-bit adder is therefore only 4. The additional space cost of the adder, other than $a$ and $b$, is therefore $8\ell$.

The reaction depth of the $\ell$-bit adder is still $\ell$ (before uncomputation): the $X$ Pauli corrections from~\cref{lem:fanout_autoand} in a given copy of $\mathcal{C}$ anticommute with the \gateparity gate from~\cref{lem:ztype_ppm} which leads to the next \gateautoand, and change the bases the next ancillae must be measured in. As a consequence, the quantum depth of $\ell$-bit adder is $\Theta(\ell)$. However, all these bases can be chosen \textit{after} the quantum computer has moved onto the next subroutine on $a$, $b$ and $s$, which is why the contributive depth is constant: as with a single \gateautoand, the quantum depth has been shifted from the main qubits onto the ancillae.

\subsubsection{Uncomputation of the adder}
\label{ssec:uncomputation}

\begin{lemma}[Delayed-$CZ$]\cite{litinski2022active}\label{lem:delayed_cz}
    The following circuit implements a delayed-choice $CZ$ gate, up to Pauli corrections.
    
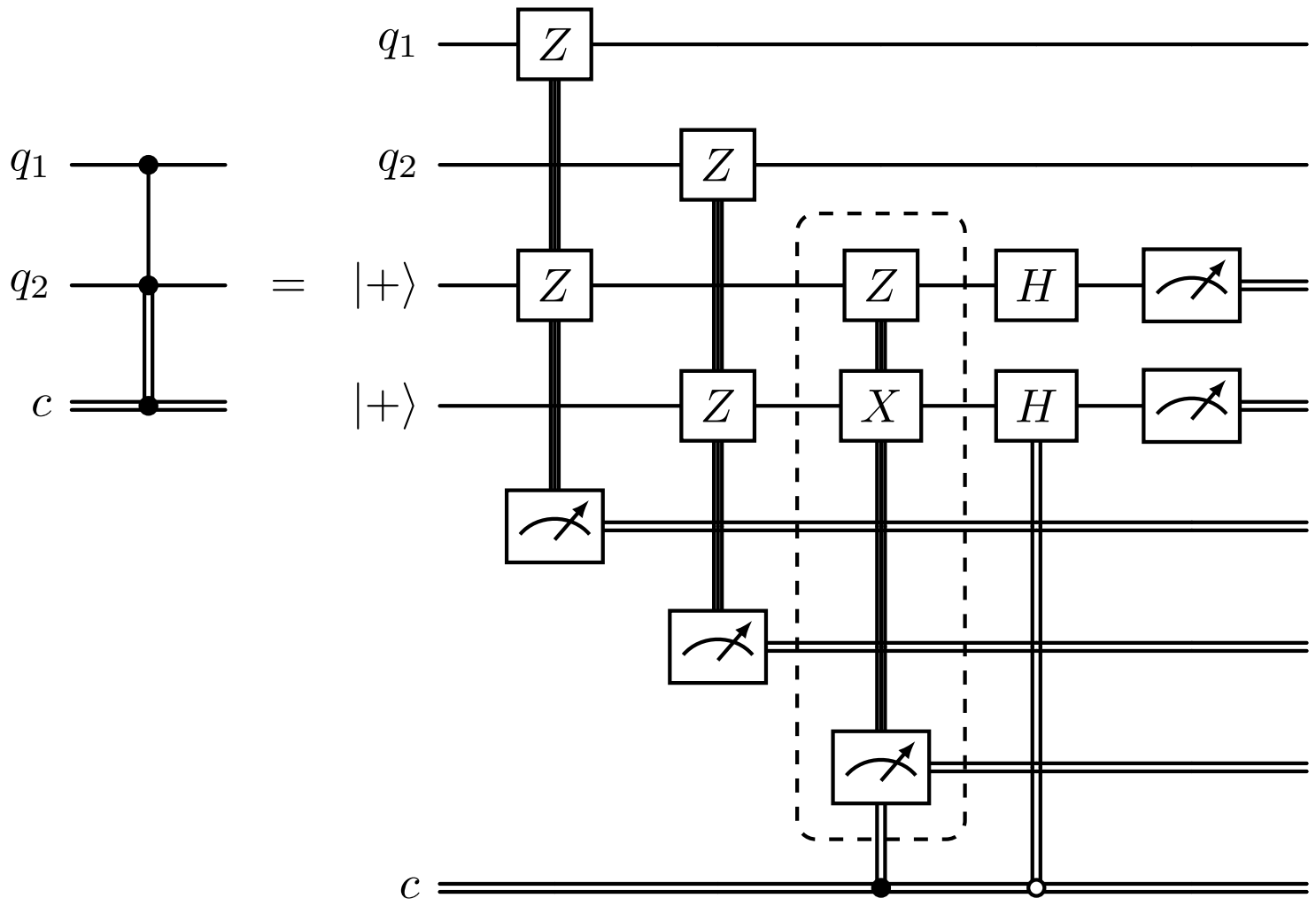
\begin{align*}
    \begin{quantikz}[wire types = {q,q,c}, row sep={0.9cm,between origins}]
    \lstick{$q_1$} & \ctrl{1} & \\
    \lstick{$q_2$} & \ctrl{0} & \\
    \lstick{$c$} & \ctrl{0}\wire[u]{c} &
    \end{quantikz} 
    \,\,&= \,\,
    \begin{quantikz}[wire types = {q,q,q,q,n,n,n,c}, row sep={0.9cm,between origins}, align equals at=3]
        \lstick{$q_1$} & \gate{Z}\wire[d]{b} & & & & & \\
        \lstick{$q_2$} & \wire[d]{b} & \gate{Z}\wire[d]{b} & & & & \\
        \lstick{$\ket{+}$} & \gate{Z}\wire[d]{b} & \wire[d]{b} & \gate{Z}\gategroup[5,steps=1,style={dashed, rounded corners}]{}\wire[d]{b} & \gate{H} & \meter{} & \setwiretype{c}\\
        \lstick{$\ket{+}$} & \wire[d]{b} & \gate{Z}\wire[d]{b} & \gate{X}\wire[d]{b} & \gate{H} & \meter{} & \setwiretype{c}\\
        & \meter{} & \wire[d]{b}\setwiretype{c} & \wire[d]{b} & \wire[u]{c} & & \\
        & & \meter{} & \wire[d]{b}\setwiretype{c} & \wire[u]{c} & & \\
        & & & \meter{} & \setwiretype{c}\wire[u]{c} & & \\
        \lstick{$c$} & & & \ctrl{0}\wire[u]{c} & \octrl{0}\wire[u]{c} & &
    \end{quantikz}
    \end{align*}
\end{lemma}
\proof
See Appendix~\ref{app:proofs}.
\endproof

We can, when measuring a given $b_k$, effect the correction by modifying~\cref{lem:adder_circuit} slightly:

\begin{align*}
    \scalebox{0.8}{\begin{quantikz}[wire types = {q,q,q,q,q,q,q,n,q,q,q,q,q,q,q,q,n,n,n,n,n,n,n,n,n}, row sep={0.9cm,between origins}]
    \lstick{$a_k$}                            & & & & & \gate{Z}\wire[d]{b} & \gate{Z}\wire[d]{b} & & & & & & & & \rstick{$a_k$} \\
    \lstick{$b_k$}                            & & & & & \gate{Z}\wire[d]{b} & \wire[d]{b} & \gate{Z}\wire[d]{b} & & & & & & & \rstick{$b_k$} \\
    \lstick{$s_k$}                            & & & & & \gate{Z}\wire[d]{b} & \gate{Z}\wire[d]{b} & \gate{Z}\wire[d]{b} & & & & & & & \rstick{$s_k$} \\
    \lstick{$a_{k+1}$}    & & & & & \gate{Z}\wire[d]{b} & \wire[d]{b} & \wire[d]{b} & & & & & & & \rstick{$a_{k+1}$} \\
    \lstick{$b_{k+1}$}    & & & & & \gate{Z}\wire[d]{b} & \wire[d]{b} & \wire[d]{b} & \gate{H} & \meter{} & \ctrl{0}\wire[d]{c}\setwiretype{c} & \octrl{0}\wire[d]{c} & & & \\
    \lstick{$\ket{+}$} & & & & & \wire[d]{b} & \gate{Z}\wire[d]{b} & \wire[d]{b} & \gate{Z}\wire[d]{b} & & \gate{X}\wire[d]{b} & \gate{H} & \meter{} & \setwiretype{c} & \rstick{$c_1$} \\
    \lstick{$\ket{+}$} & & & & & \wire[d]{b} & \wire[d]{b} & \gate{Z}\wire[d]{b} & \wire[d]{b} & \gate{Z}\wire[d]{b} & \gate{Z}\wire[d]{b} & \gate{H} & \meter{} & \setwiretype{c} & \rstick{$c_2$} \\
    & & & & & \wire[d]{b} & \wire[d]{b} & \wire[d]{b} & \wire[d]{b} & \wire[d]{b} & \meter{} & \setwiretype{c} & & & \rstick{$c_3$} \\ 
    \lstick[3]{$\toffstate$} & & \gate{Z}\wire[d]{b} & & & \wire[d]{b} & \wire[d]{b} & \wire[d]{b} & \wire[d]{b} & \gate{Z}\wire[d]{b} & \gate{H} & \meter{} & \setwiretype{c} & & \rstick{$x_1$}\\
                                        & \gate{Z}\wire[d]{b} & \wire[d]{b} & & & \wire[d]{b} &  \wire[d]{b} & \wire[d]{b} & \gate{Z}\wire[d]{b} & \wire[d]{b} &  \gate{H} & \meter{} & \setwiretype{c} & & \rstick{$x_2$}\\
                                        & \wire[d]{b} & \wire[d]{b} & \gate{X}\wire[d]{b} & \gate{X}\wire[d]{b} & \gate{Z}\wire[d]{b} & \wire[d]{b} &  \wire[d]{b} & \wire[d]{b} & \wire[d]{b} & \gate{H} & \meter{} & \setwiretype{c} & & \rstick{$x_{3}$} \\
        \lstick{$\ket{+}$} & \gate{Z}\wire[d]{b} & \wire[d]{b} & \gate{X}\wire[d]{b} & \wire[d]{b} & \wire[d]{b} &  \wire[d]{b} & \wire[d]{b} & \wire[d]{b} & \wire[d]{b} & & \gate{H} & \gate{H} & \meter{} & \setwiretype{c} \rstick{$x_4$} \\
        \lstick{$\ket{+}$} & \wire[d]{b} & \gate{Z}\wire[d]{b} & \wire[d]{b} & \gate{X}\wire[d]{b} & \wire[d]{b} & \wire[d]{b} & \wire[d]{b} & \wire[d]{b} & \wire[d]{b} & \gate{H} & \wire[u]{c} & \gate{H} & \meter{} & \setwiretype{c} \rstick{$x_5$} \\
        \lstick{$\ket{0}$} & \wire[d]{b} & \wire[d]{b} & \gate{X}\wire[d]{b} & \wire[d]{b} & \wire[d]{b} &  \wire[d]{b} & \wire[d]{b} & \wire[d]{b} & \wire[d]{b} & \wire[u]{c} & \gate{H}\wire[u]{c} & & \meter{} & \setwiretype{c} \rstick{$x_6$} \\
        \lstick{$\ket{0}$} & \wire[d]{b} & \wire[d]{b} & \wire[d]{b} & \gate{X}\wire[d]{b} & \wire[d]{b} &  \wire[d]{b} & \wire[d]{b} & \wire[d]{b} & \wire[d]{b} & \gate{H}\wire[u]{c} & \wire[u]{c} & & \meter{} & \setwiretype{c} \rstick{$x_7$} \\
        \lstick{$\ket{+}$} & \wire[d]{b} & \wire[d]{b} & \wire[d]{b} & \wire[d]{b} & \gate{Z}\wire[d]{b} &  \wire[d]{b} & \wire[d]{b} & \wire[d]{b} & \wire[d]{b} &  \wire[u]{c} & \wire[u]{c} & & & \rstick{$s_{k+1}$}\\
        & \wire[d]{b} & \wire[d]{b} & \wire[d]{b} & \wire[d]{b} & \wire[d]{b} & \wire[d]{b} & \wire[d]{b} & \meter{} & \wire[d]{b}\setwiretype{c} & \phase{}\wire[u]{c} & \wire[u]{c} & & & \rstick{$x_8$} \\
        & \wire[d]{b} & \wire[d]{b} & \wire[d]{b} & \wire[d]{b} & \wire[d]{b} & \wire[d]{b} & \wire[d]{b} & & \meter{} & \setwiretype{c} & \phase{}\wire[u]{c} & & & \rstick{$x_9$} \\
        & \wire[d]{b} & \wire[d]{b} & \meter{} & \wire[d]{b} \setwiretype{c} & \wire[d]{b} & \wire[d]{b} & \wire[d]{b} & & & & & & & \rstick{$x_{10}$} \\
        & \wire[d]{b} & \wire[d]{b} & & \meter{} & \wire[d]{b}\setwiretype{c} & \wire[d]{b} & \wire[d]{b} & & & & & & & \rstick{$x_{11}$} \\
        & \meter{} & \wire[d]{b} \setwiretype{c} & & & \wire[d]{b} & \wire[d]{b} & \wire[d]{b} & & & & & & & \rstick{$x_{12}$} \\
        & & \meter{} & \setwiretype{c} & & \wire[d]{b} &  \wire[d]{b} & \wire[d]{b} & & & & & & & \rstick{$x_{13}$} \\
        & & & & & \meter{} & \wire[d]{b}\setwiretype{c} & \wire[d]{b} & & & & & & & \rstick{$x_{14}$} \\
        & & & & & & \meter{} & \wire[d]{b}\setwiretype{c} & & & & & & & \rstick{$x_{15}$} \\
        & & & & & & & \meter{} & \setwiretype{c} & & & & & & \rstick{$x_{16}$}
    \end{quantikz}}
\end{align*}

As a consequence, the uncomputation can be performed by a chain of delayed-choice $CZ$s:

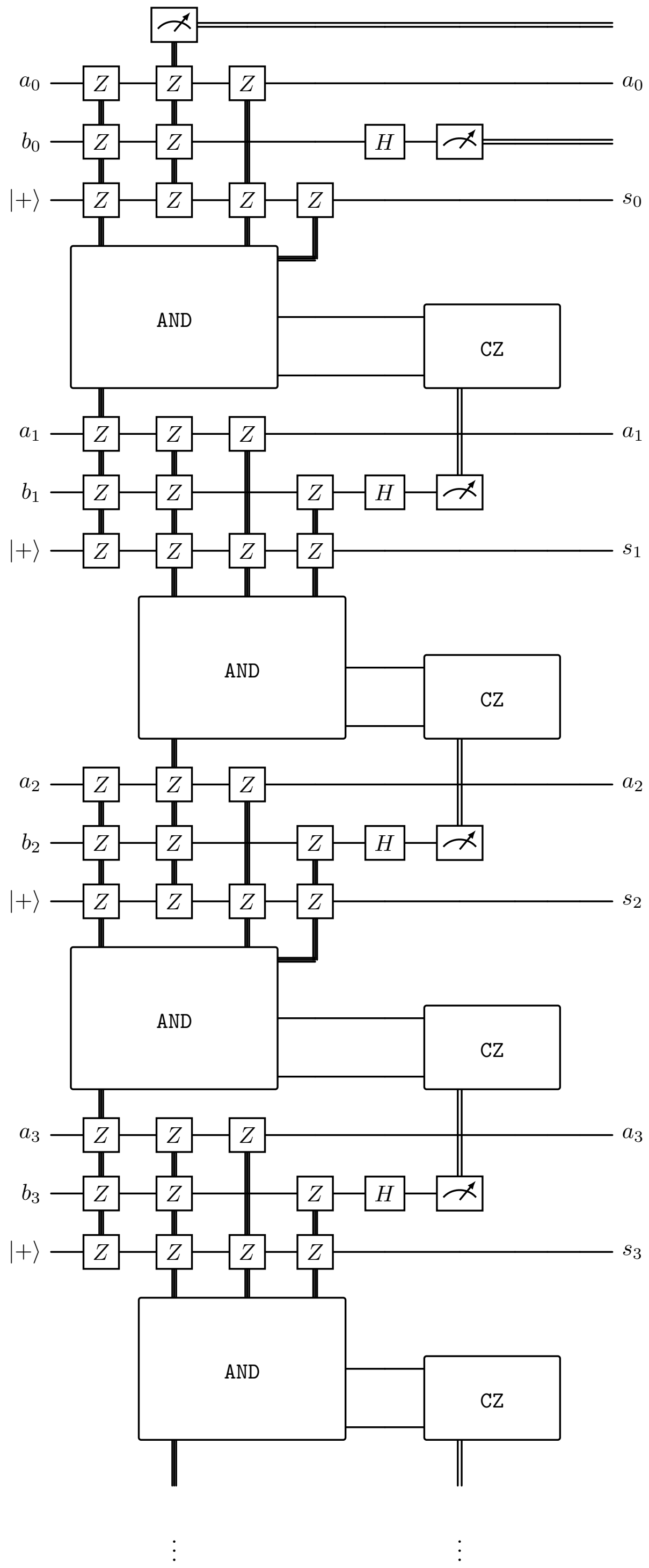
\begin{align*}
    \scalebox{0.8}{\begin{quantikz}[wire types = {n,q,q,q,n,n,n,q,q,q,n,n,n,q,q,q,n,n,n,q,q,q,n,n,n,n,n,n}, row sep={0.9cm,between origins}]
        & & \meter{}\wire[d]{b} & \setwiretype{c} & & & & & & & \\
        \lstick{$a_0$} & \gate{Z}\wire[d]{b} & \gate{Z}\wire[d]{b} & \gate{Z}\wire[d]{b} & & &  & & & & \rstick{$a_0$} \\
        \lstick{$b_0$} & \gate{Z}\wire[d]{b} & \gate{Z}\wire[d]{b} & \wire[d]{b} & & \gate{H} & \meter{} & \setwiretype{c} & & &\\
        \lstick{$\ket{+}$} & \gate{Z}\wire[d]{b} & \gate{Z} & \gate{Z}\wire[d]{b} & \gate{Z}\wire[d]{b} & & & & & & \rstick{$s_0$}\\
        & \andtwo & & & \setwiretype{b} & \setwiretype{n} & & & & & \\
        & & & & \setwiretype{q} & & \quantikzcz & \setwiretype{n} & & & \\
        & \wire[u]{b} & & & \setwiretype{q} & & &  \setwiretype{n} & & & \\
        \lstick{$a_1$} & \gate{Z}\wire[u]{b} & \gate{Z}\wire[d]{b} & \gate{Z}\wire[d]{b} & & & \wire[u]{c} & & & & \rstick{$a_1$} \\
        \lstick{$b_1$} & \gate{Z}\wire[u]{b} & \gate{Z}\wire[d]{b} & \wire[d]{b} & \gate{Z}\wire[d]{b} & \gate{H} & \meter{}\wire[u]{c} & \setwiretype{n} & & & \\
        \lstick{$\ket{+}$} & \gate{Z}\wire[u]{b} & \gate{Z}\wire[d]{b} & \gate{Z}\wire[d]{b} & \gate{Z}\wire[d]{b} & & & & & & \rstick{$s_1$} \\
        & & \andtwo & & & & & & & & \\
        & & & & \setwiretype{q} & & \quantikzcz & \setwiretype{n} & & & \\
        & & \wire[u]{b} & & \setwiretype{q} & & & \setwiretype{n} & & & \\
        \lstick{$a_2$} & \gate{Z}\wire[d]{b} & \gate{Z}\wire[u]{b} & \gate{Z}\wire[d]{b} & & & \wire[u]{c} & & & & \rstick{$a_2$} \\
        \lstick{$b_2$} & \gate{Z}\wire[d]{b} & \gate{Z}\wire[u]{b} & \wire[d]{b} & \gate{Z}\wire[d]{b} & \gate{H} & \meter{}\wire[u]{c} & \setwiretype{n} & & & \\
        \lstick{$\ket{+}$} & \gate{Z}\wire[d]{b} & \gate{Z}\wire[u]{b} & \gate{Z}\wire[d]{b} & \gate{Z}\wire[d]{b} & & & & & & \rstick{$s_2$} \\
        & \andtwo & & & \setwiretype{b} & \setwiretype{n} & & & & & \\
        & & & & \setwiretype{q} & & \quantikzcz & \setwiretype{n} & & & \\
        & \wire[u]{b} & & & \setwiretype{q} & & & \setwiretype{n} & & & \\
        \lstick{$a_3$} & \gate{Z}\wire[u]{b} & \gate{Z}\wire[d]{b} & \gate{Z}\wire[d]{b} & & & \wire[u]{c} & & & & \rstick{$a_3$} \\
        \lstick{$b_3$} & \gate{Z}\wire[u]{b} & \gate{Z}\wire[d]{b} & \wire[d]{b} & \gate{Z}\wire[d]{b} & \gate{H} & \meter{}\wire[u]{c} & \setwiretype{n} & & & \\
        \lstick{$\ket{+}$} & \gate{Z}\wire[u]{b} & \gate{Z}\wire[d]{b} & \gate{Z}\wire[d]{b} & \gate{Z}\wire[d]{b} & & & & & & \rstick{$s_3$} \\
        & & \andtwo & & & & & & & & \\
        & & & & \setwiretype{q} & & \quantikzcz & \setwiretype{n} & & & \\
        & & & & \setwiretype{q} & & & \setwiretype{n} & & & \\
        & & \wire[u]{b} & & & & \wire[u]{c} & & & & \\
        & & \vdots & & & & \vdots & & & & 
    \end{quantikz}}
\end{align*}

The adder now uses between $0$ and $\ell$ non-homogeneous PPMs, with an average of $\ell/2$ non-homogeneous PPMs required, and has reaction depth $2\ell$.
Happily, the contributive depth of the adder circuit, including the uncomputation, is still 4. The space cost, other than the original $2\ell$ qubits for storing $a$ and $b$, is $10\ell$. We compare our adder to previous works in~\cref{tab:adder_comparison}.

\tikzexternaldisable

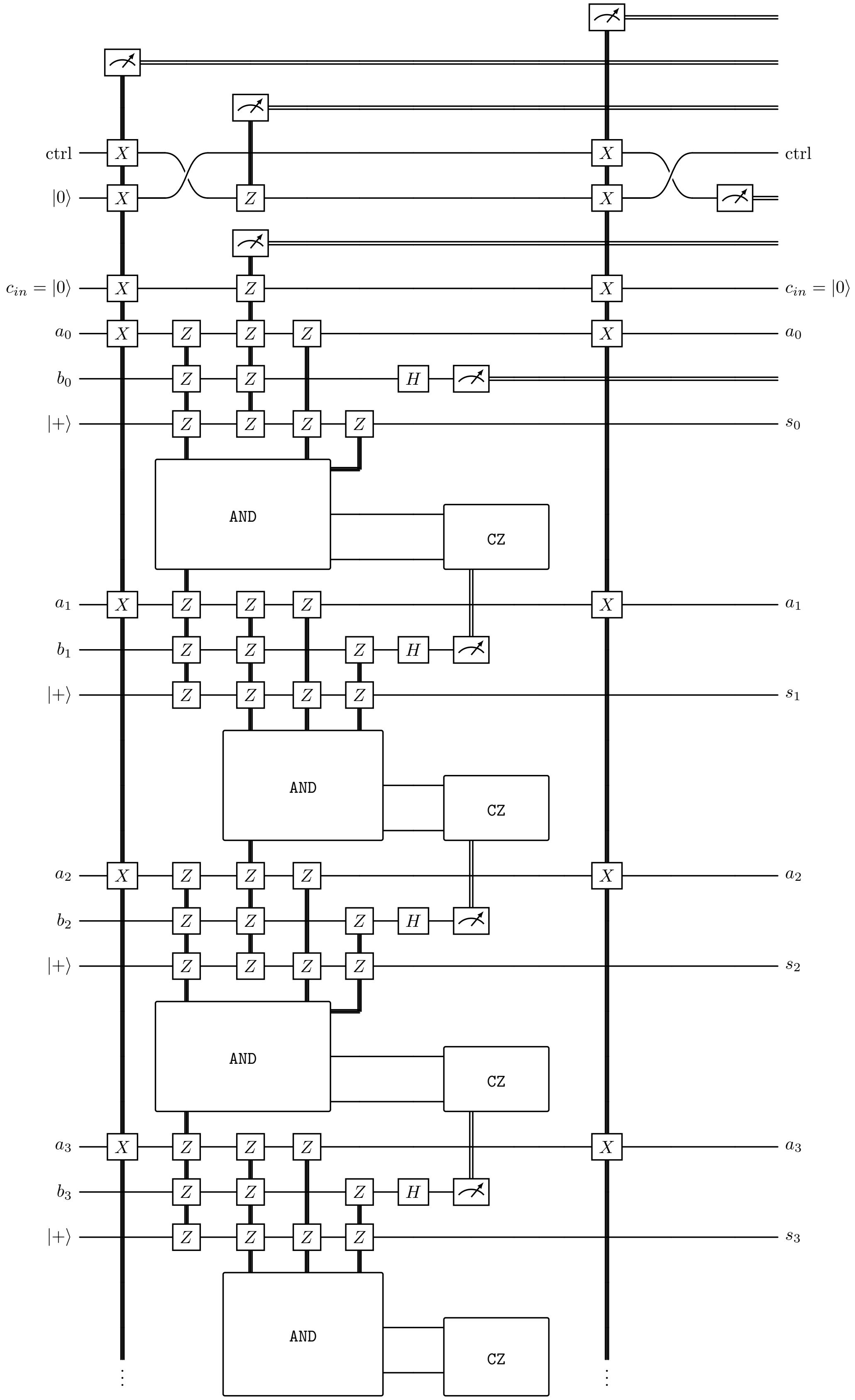
\begin{figure}
    \centering
    
    \scalebox{0.8}{\begin{quantikz}[wire types = {n,n, n,q,q,n,q,q,q,q,n,n,n,q,q,q,n,n,n,q,q,q,n,n,n,q,q,q,n,n,n}, row sep={0.9cm,between origins}]
        & & & & & & & & & & & \meter{}\wire[d]{b} & \setwiretype{c} & & \\
        & \meter{}\wire[d]{b} & \setwiretype{c} & & & & & & & & & \wire[d]{b} & & & \\
        & \wire[d]{b} & & \meter{}\wire[d]{b} & \setwiretype{c} & & & & & & & \wire[d]{b} & & & \\
        \lstick{ctrl} & \gate{X}\wire[d]{b} & \permute{2,1} & \wire[d]{b} & & & & & & & & \gate{X}\wire[d]{b} & \permute{2,1} & & \rstick{ctrl} \\
        \lstick{$\ket{0}$} & \gate{X}\wire[d]{b} & & \gate{Z} & & & & & & & & \gate{X}\wire[d]{b} & & \meter{} & \setwiretype{c} \\
        & \wire[d]{b} & & \meter{}\wire[d]{b} & \setwiretype{c} & & & & & & & \wire[d]{b} & & & \\
        \lstick{$c_{in} = \ket{0}$} & \gate{X}\wire[d]{b} & & \gate{Z}\wire[d]{b} & & & & & & & & \gate{X}\wire[d]{b} & & & \rstick{$c_{in} = \ket{0}$} \\
        \lstick{$a_0$} & \gate{X}\wire[d]{b} & \gate{Z}\wire[d]{b} & \gate{Z}\wire[d]{b} & \gate{Z}\wire[d]{b} & & & & & & & \gate{X}\wire[d]{b} & & & \rstick{$a_0$} \\
        \lstick{$b_0$} & \wire[d]{b} & \gate{Z}\wire[d]{b} & \gate{Z}\wire[d]{b} & \wire[d]{b} & & \gate{H} & \meter{} & \setwiretype{c} & & & \wire[d]{b} & & & \\
        \lstick{$\ket{+}$} & \wire[d]{b} & \gate{Z}\wire[d]{b} & \gate{Z} & \gate{Z}\wire[d]{b} & \gate{Z}\wire[d]{b} & & & & & & \wire[d]{b} & & & \rstick{$s_0$} \\
        & \wire[d]{b} & \andtwo & & & \setwiretype{b} & \setwiretype{n} & & & & & \wire[d]{b} & & & \\
        & \wire[d]{b} & & & \setwiretype{q} & & & \quantikzcz & \setwiretype{n} & & & \wire[d]{b} & & & \\
        & \wire[d]{b} & \wire[u]{b} & & \setwiretype{q} & & & & \setwiretype{n} & & & \wire[d]{b} & & & \\
        \lstick{$a_1$} & \gate{X}\wire[d]{b} & \gate{Z}\wire[u]{b} & \gate{Z}\wire[d]{b} & \gate{Z}\wire[d]{b} & & & \wire[u]{c} & & & & \gate{X}\wire[d]{b} & & & \rstick{$a_1$} \\
        \lstick{$b_1$} & \wire[d]{b} & \gate{Z}\wire[u]{b} & \gate{Z}\wire[d]{b} & \wire[d]{b} & \gate{Z}\wire[d]{b} & \gate{H} & \meter{}\wire[u]{c} & \setwiretype{n} & & & \wire[d]{b} & & & \\
        \lstick{$\ket{+}$} & \wire[d]{b} & \gate{Z}\wire[u]{b} & \gate{Z}\wire[d]{b} & \gate{Z}\wire[d]{b} & \gate{Z}\wire[d]{b} & & & & & & \wire[d]{b} & & & \rstick{$s_1$} \\
        & \wire[d]{b} & & \andtwo & & & & & & & & \wire[d]{b} & & & \\
        & \wire[d]{b} & & & \setwiretype{q} & & & \quantikzcz & \setwiretype{n} & & & \wire[d]{b} & & & \\
        & \wire[d]{b} & & & \setwiretype{q} & & & & \setwiretype{n} & & & \wire[d]{b} & & & \\
        \lstick{$a_2$} & \gate{X}\wire[d]{b} & \gate{Z}\wire[d]{b} & \gate{Z}\wire[u]{b} & \gate{Z}\wire[d]{b} & & & \wire[u]{c} & & & & \gate{X}\wire[d]{b} & & & \rstick{$a_2$} \\
        \lstick{$b_2$} & \wire[d]{b} & \gate{Z}\wire[d]{b} & \gate{Z}\wire[u]{b} & \wire[d]{b} & \gate{Z}\wire[d]{b} & \gate{H} & \meter{}\wire[u]{c} & \setwiretype{n} & & & \wire[d]{b} & & & \\
        \lstick{$\ket{+}$} & \wire[d]{b} & \gate{Z}\wire[d]{b} & \gate{Z}\wire[u]{b} & \gate{Z}\wire[d]{b} & \gate{Z}\wire[d]{b} & & & & & & \wire[d]{b} & & & \rstick{$s_2$} \\
        & \wire[d]{b} & \andtwo & & & \setwiretype{b} & \setwiretype{n} & & & & & \wire[d]{b} & & & \\
        & \wire[d]{b} & & & \setwiretype{q} & & & \quantikzcz & \setwiretype{n} & & & \wire[d]{b} & & & \\
        & \wire[d]{b} & & & \setwiretype{q} & & & & \setwiretype{n} & & & \wire[d]{b} & & & \\
        \lstick{$a_3$} & \gate{X}\wire[d]{b} & \gate{Z}\wire[u]{b} & \gate{Z}\wire[d]{b} & \gate{Z}\wire[d]{b} & & & \wire[u]{c} & & & & \gate{X}\wire[d]{b} & & & \rstick{$a_3$} \\
        \lstick{$b_3$} & \wire[d]{b} & \gate{Z}\wire[u]{b} & \gate{Z}\wire[d]{b} & \wire[d]{b} & \gate{Z}\wire[d]{b} & \gate{H} & \meter{}\wire[u]{c} & \setwiretype{n} & & & \wire[d]{b} & & & \\
        \lstick{$\ket{+}$} & \wire[d]{b} & \gate{Z}\wire[u]{b} & \gate{Z}\wire[d]{b} & \gate{Z}\wire[d]{b} & \gate{Z}\wire[d]{b} & & & & & & \wire[d]{b} & & & \rstick{$s_3$} \\
        & \wire[d]{b} & & \andtwo & & & & & & & & \wire[d]{b} & & & \\
        & \wire[d]{b} & & & \setwiretype{q} & & & \quantikzcz & \setwiretype{n} & & & \wire[d]{b} & & & \\
        & \vdots & & \vdots & \setwiretype{q} & & & \vdots & \setwiretype{n} & & & \vdots &
    \end{quantikz}}
    \caption{Controlled $\ell$-bit add-subtract circuit, performed using at most $\ell$ non-homogeneous PPMs and with contributive depth 5. In~\cite{litinski2024quantum} the ctrl wire fires if it is 0, rather than 1, but this is equivalent up to $X$ gates, which can be tracked in frame.}
    \label{fig:add_subtract}
\end{figure}

\subsection{Schoolbook multiplication}

Quantum schoolbook multiplication on $\ell$-bit integers can be performed using $\ell^2+4\ell+3$ \gateand gates~\cite{litinski2024quantum}. A core subroutine for these circuits is a controlled add-subtract, which performs an addition $(a,b)\mapsto(a,a+b)$ if the control qubit is 1, and a subtraction $(a,b)\mapsto(a,b-a)$ if the control qubit is 0. The controlled add-subtract is accomplished by applying a \gatefanout, see~\cref{lem:xtype_ppm}, from the control qubit to (a) an input carry bit, and (b) the register containing $a$ before and after the addition circuit, see~\cite[Fig.~1(f)]{litinski2024quantum}. The circuit is shown in~\cref{fig:add_subtract}.

Therefore, the controlled add-subtract subroutine can be performed with contributive depth 5. To perform the multiplication, there are $\ell$ controlled add-subtracts applied sequentially, along with 3 further adders. The contributive depth of $\ell$-bit integer multiplication using our adder circuit is therefore $5\ell+12$.

\section{Discussion}

This paper formalizes the notion of reaction depth, giving a new tool for studying the time-complexity of quantum circuits. It also develops techniques for compilation using homogeneous PPMs: a gateset adapted to the anticipated constraints of early fault tolerant architectures. Consequently, this work has connections to both theoretical and practical aspects of quantum information. We discuss these in turn.

On the theoretical side, the notion of reaction depth seems closely related to other notions such as the T-depth or magic depth of a unitary~\cite{zhang2025classical,selinger2013quantum,amy2013meet}. Circuits with low magic depth can be converted to circuits with similar reaction depth by replacing every non-Clifford operation with an appropriate magic-state injection.
Consequently, the magic depth of a unitary often provides a natural upper bound on the reaction depth of any circuit implementing that unitary. However, the converse direction is not true in general. In Appendix~\ref{app:reaction_depth} we give an example of a unitary which has a magic depth of 2, but possesses a circuit with reaction depth 1. It is unclear if the constant reaction-depth Boolean logic circuits given in~\Cref{sec:boolean_results} have a constant magic-depth analogue, without relying on a different gateset that includes e.g. small angle rotations~\cite{hoyer2005quantum,nehoran2026unitaries,valiron2014quantum,allcock2024constant,strelchuk2026power}.

The constant depth Boolean logic constructions of~\Cref{sec:boolean_results} can be compared to other results compiling coherent classical logic with constant quantum depth. Our gateset allows for \gatefanout in constant depth. It is known that \gatefanout gates, together with arbitrary single qubit operations, can implement Toffoli gates in constant depth~\cite{hoyer2005quantum,takahashi2009quantum}. Given access to constant-depth Toffoli gates and fanout, standard truth-table arguments allow for implementing arbitrary Boolean logic with exponentially many ancilla, potentially giving an alternate route to the main theorem of~\Cref{sec:boolean_results}. However, small angle single qubit rotations either have non-constant T-depth~\cite{selinger2014efficient}, or require resource states~\cite{kitaev2002classical,kim2026catalytic}. We leave a fuller comparison between this approach and the techniques of this paper to future work. 

On the practical side, we constructed relatively cheap $\gateand$ circuits, quantum adders and multipliers, making use of homogeneous PPMs and sparing use of non-homogeneous PPMs, and compiling directly from Boolean circuits rather than taking a circuitous route through unitary gates first.

\begin{mdframed}[style=aidisclosure]
\noindent\textbf{AI disclosure statement:}

\medskip

We did not use AI in the writing or mathematics of this paper, except to check for typos and to prove~\cref{ex:reaction_counterexample}. 
In fact, we anticipate that one outcome of this paper is the reverse. Current LLMs excel at problem-solving, but their abilities to perform `theory-building’~\cite{gowers2000two} appear to be more modest~\cite{jiang2026solvers}.
This work gives precise definitions and justifications for reaction depth and related concepts; we anticipate that our work will allow AI models to attack the problem of compilation~\cite{khattar2026leveraging,crichigno2026automating,gu2026discovering} while incorporating such concepts, such that the results can be verified explicitly.
\end{mdframed}

\section{Acknowledgments}
CK thanks Juan Miguel Arrazola and Danial Motlagh for sponsoring the internship at Xanadu, and Katie Chang, Chris Pattison, Nathan Wiebe, and Harry Zhou for encouraging conversations.

\newpage
\printbibliography

\appendix

\section{Circuit notation}\label{app:circuit_notation}

This appendix briefly reviews the circuit notation used in this paper, with emphasis on any non-standard notation.

\subsection{Wires}

Qubits are always denoted by single wires, while classical bits are always denoted by double wires.

\begin{figure}[h]
  \centering
  \begin{subfigure}[b]{0.45\textwidth}
    \centering
        \begin{quantikz}[wire types = {c}]
            &
        \end{quantikz}
    \caption{A classical bit.}
  \end{subfigure}
  \hfill
  \begin{subfigure}[b]{0.45\textwidth}
      \centering
        \begin{quantikz}[wire types = {q}]
            &
        \end{quantikz}
    \caption{A qubit.}
  \end{subfigure}
  \caption{Wire Types.}
\end{figure}

\noindent A variable, such as $x$, appearing to the left of a classical wire will always label the initial value of that bit. Similarly a state, such as $\ket{+}$ or $\ket{\psi}$, appearing to the left of a qubit will always label the input state of that qubit. However, a variable such as $s$ not in ket notation appearing to the left of a \textit{quantum wire} should be read as labeling the qubit, not the logical information stored in the register.  


\begin{figure}[h]
  \begin{subfigure}[b]{0.3\textwidth}
    \centering
        \begin{quantikz}[wire types = {c}]
            \lstick{$x$}  &
        \end{quantikz}
    \caption{A bit whose initial value is $x$.}
  \end{subfigure}
  \hfill
  \begin{subfigure}[b]{0.3\textwidth}
      \centering
        \begin{quantikz}[wire types = {q}]
           \lstick{$\ket{+}$}  &
        \end{quantikz}
    \caption{A qubit initially in the state~$\ket{+}$.}
  \end{subfigure}
  \hfill
    \begin{subfigure}[b]{0.3\textwidth}
      \centering
        \begin{quantikz}[wire types = {q}]
           \lstick{$s$}  &
        \end{quantikz}
    \caption{A qubit labeled $s$, whose initial state is not specified.}
  \end{subfigure}
    \caption{Input labels. }
\end{figure}

\noindent A variable $x$ appearing to the right of a classical wire indicates the final value of that bit. Occasionally a label will appear above a classical bit to represent the value of that bit at an intermediate point in the circuit. 

\begin{figure}[h]
  \centering
  \begin{subfigure}[b]{0.45\textwidth}
    \centering
        \begin{quantikz}[wire types = {c}]
            & \targ{} & \rstick{$x_1$}
        \end{quantikz}
    \caption{The output value of this bit is denoted $x_1$.}
  \end{subfigure}
  \hfill
  \begin{subfigure}[b]{0.45\textwidth}
    \centering
        \begin{quantikz}[wire types = {c}]
            & \targ{} \arrow[r,"x_2"{yshift=2pt},draw=none] & \targ{} &
        \end{quantikz}
    \caption{This bit takes value $x_2$ at the midpoint of this circuit (between the two bit flip operations).}
  \end{subfigure}
  \caption{Other wire labels.}
\end{figure}
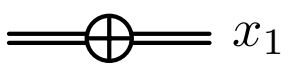
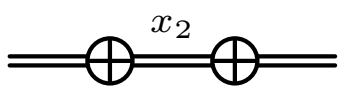

\subsection{Simple Unitaries}

Pauli $X$ operations are denoted using either $X$ gates or $\oplus$ notation. Classical bit flips are always denoted with the $\oplus$ notation.
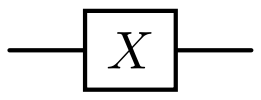
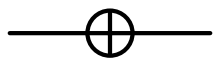
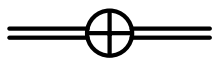
\begin{figure}[h]
  \begin{subfigure}[b]{0.3\textwidth}
      \centering
        \begin{quantikz}[wire types = {q}]
            & \gate{X} &
        \end{quantikz}
    \caption{A Pauli $X$ gate.}
  \end{subfigure}
  \hfill
  \begin{subfigure}[b]{0.3\textwidth}
      \centering
        \begin{quantikz}[wire types = {q}]
            & \targ{} &
        \end{quantikz}
    \caption{Also a Pauli $X$ gate.}
  \end{subfigure}
  \hfill
  \begin{subfigure}[b]{0.3\textwidth}
    \centering
        \begin{quantikz}[wire types = {c}]
            & \targ{} &
        \end{quantikz}
    \caption{A classical bit flip.}
  \end{subfigure}
    \caption{Quantum and Classical bit flips. }
\end{figure}
Classical and quantum swap gates are denoted using either wire crossings or connected $\times$ symbols. The latter notation is used especially in the case of controlled-swap operations.  

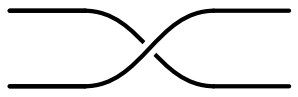
\begin{figure}[h]
  \centering
  \begin{subfigure}[b]{0.45\textwidth}
    \centering
        \begin{quantikz}[wire types = {q,q}]
            & \swap{1}  & \\
            & \targX{}  &
        \end{quantikz}
    \caption{A swap gate.}
  \end{subfigure}
  \hfill
  \begin{subfigure}[b]{0.45\textwidth}
      \centering
        \begin{quantikz}[wire types = {q,q}]
            & \permute{2,1} & \\
            &               &
        \end{quantikz}
    \caption{Also a swap gate.}
  \end{subfigure}
  \caption{Swap Gates.}
\end{figure}

All other notation for simple unitaries is standard. 

\subsection{PPMs}

Multi-qubit Pauli Product Measurements (PPMs) are indicated by Pauli gates connected to each other by a thick vertical wire, and to a ``meter'' symbol. The Pauli gates indicate the qubits and Paulis involved in the PPM, while the classical wire leaving the meter stores the outcome of the PPM. 

\begin{figure}[h]
\centering
\begin{quantikz}[wire types = {q,q,q,n}]
    &\gate{X} \wire[d]{b}   &\\
    &\wire[d]{b}            &\\
    &\gate{X} \wire[d]{b}   &\\
    &\meter{}               & \setwiretype{c} \rstick{$m$}
\end{quantikz}
\caption{A measurement of the Pauli Product $X_1X_3$ on the top three qubits, producing outcome $m$.}
\end{figure}
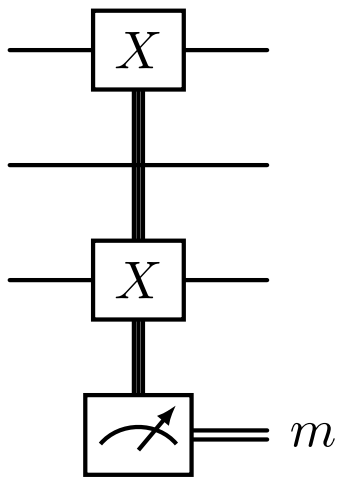

\noindent A quantum wire feeding directly into a meter always denotes a computational basis measurement, with the classical wire leaving the meter denoting the outcome of that measurement. 

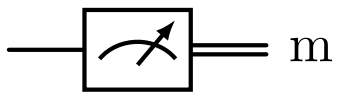
\begin{figure}[H]
\centering
\begin{quantikz}[wire types = {q}]
    &\meter{}               & \setwiretype{c} \rstick{m}
\end{quantikz}
\caption{A single qubit measured in the computational basis, producing outcome $m$.}
\end{figure}

\noindent Because of this convention, we denote measurements in the $\ket{\pm}$ basis by a (possibly classically-controlled) Hadamard gate followed by a computational basis measurement. In practice, in many fault-tolerant architectures this Hadamard gate will not be performed physically, but instead the logical operation will be implemented by an appropriate choice of measurement basis. 

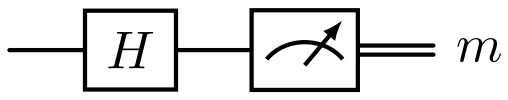
\begin{figure}[H]
\centering
\begin{quantikz}[wire types = {q}]
    & \gate{H} &\meter{}               & \setwiretype{c} \rstick{$m$}
\end{quantikz}
\caption{This circuit is equivalent to a single qubit measurement in the $\ket{\pm}$ basis, producing outcome $m$.}
\end{figure}

\subsection{Controlled Operations}

Following standard notation, controlled operations are indicated by a closed circle on the controlling qubit, with a single wire connecting that control to the controlled unitary. The same notation is used for classically controlled operations, now with a classical wire connecting the control and operation. 

\begin{figure}[H]
\begin{subfigure}[b]{0.45\textwidth}
\centering
\begin{quantikz}[wire types = {q,q}]
    & \ctrl{1}  &\\       
    & \targ{}   &  
\end{quantikz}
\caption{A CNOT gate.}
\end{subfigure}
\begin{subfigure}[b]{0.45\textwidth}
\centering
\begin{quantikz}[wire types = {c,q}]
    & \phase{}\wire[d]{c}   &\\              
    & \targ{}               & 
\end{quantikz}
\caption{A classically-controlled Pauli $X$ gate.}
\end{subfigure}
\caption{Basic controls.}
\end{figure}
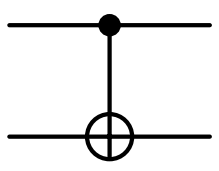
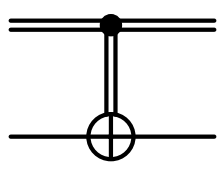

\noindent Open circles denote operations controlled on a $0$ value. These equivalently can be thought of as closed controls conjugated by bit-flip operations. 

\begin{figure}[H]
\centering
\begin{quantikz}[wire types = {q,q}, ]
    & \octrl{1}  &\\       
    & \targ{}   & 
\end{quantikz}
=
\begin{quantikz}[wire types = {q,q}, column sep={0.6cm,between origins}]
    & \targ{}   & \ctrl{1}      & \targ{}   &\\       
    &           & \targ{}       &           & 
\end{quantikz}
\caption{An open-controlled NOT gate, and an equivalent circuit written in terms of X gates and CNOT operations.}
\end{figure}
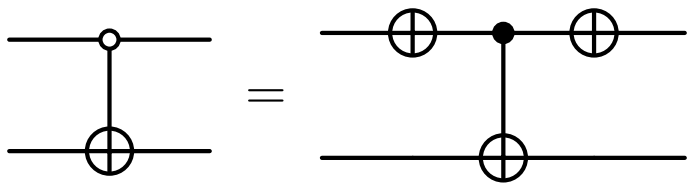

\noindent A non-standard notation used in this paper is a \textit{parity control}. These controls are denoted by a white $\oplus$ symbol inside a closed control. A gate connected to a set of such controls is understood as being controlled on the parity of the indicated bits and/or qubits, occurring only if their overall parity is odd. 

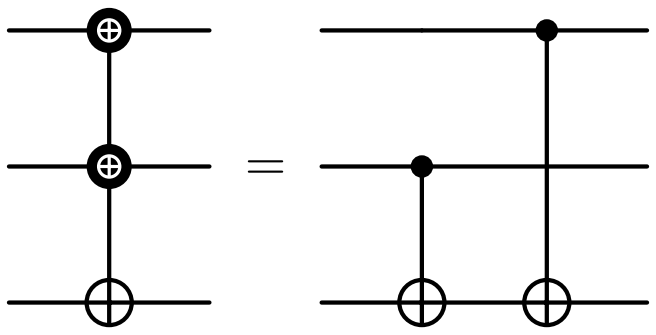
\begin{figure}[H]
\centering
\begin{quantikz}[wire types = {q,q,q}, row sep={0.9cm,between origins}]
    & \pctrl{}\wire[d]{q}   &\\ 
    & \pctrl{}\wire[d]{q}   &\\
    & \targ{}               & 
\end{quantikz}
=
\begin{quantikz}[wire types = {q,q,q}, row sep={0.9cm,between origins}]
    &           & \ctrl{2}  & \\
    & \ctrl{1}  &           & \\
    & \targ{}   & \targ{}   & 
\end{quantikz}
\caption{A parity-controlled X gate, and an equivalent operation written in terms of CNOT gates.}
\end{figure}

\noindent When multiple open or closed controls are connected to the same gate this gate is understood as being controlled on the \gateand of all the indicated controls. If the set of controls connected to a gate includes parity controls, then that gate is controlled by the parity indicated by the parity controls, \gateand-ed with all other controls. 

If multiple gates are connected to the same set of controls, then each gate is controlled as if it were individually attached to that set of controls. (Or, equivalently the product of those gates is controlled as indicated by the attached controls). 

\begin{figure}[H]
\centering
\begin{quantikz}[wire types = {q,q,c,c,c}, row sep={0.9cm,between origins}]
                    & \targ{}               &  \\
                    & \targ{}\wire[u]{c}    &  \\
    \lstick{$m_1$}  & \octrl{0}\wire[u]{c}  &  \\
    \lstick{$m_2$}  & \pctrl{}\wire[u]{c}   &  \\ 
    \lstick{$m_3$}  & \pctrl{}\wire[u]{c}   & 
\end{quantikz}
\caption{A mixture of parity controls and regular controls. This is equivalent to the operation $(X_1X_2)^{(1 \oplus m_1) (m_2 \oplus m_3)}$ being implemented on the top two qubits.}
\end{figure}
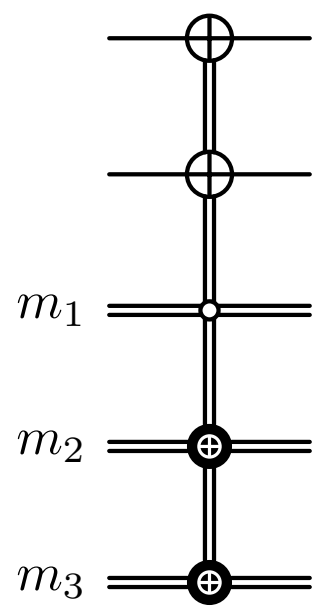

\section{Subtleties of reaction depth}\label{app:reaction_depth}

\subsection{Toffoli circuit with reduced reaction depth}

In Ref.~\cite{gidney2026eat}, Gidney describes a 4-Toffoli circuit $\mathcal{C}$, implementing a $CCCX$, which exhibits reaction depth 2, rather than the reaction depth 4 which one might expect naively. The 4-Toffoli circuit is shown below:
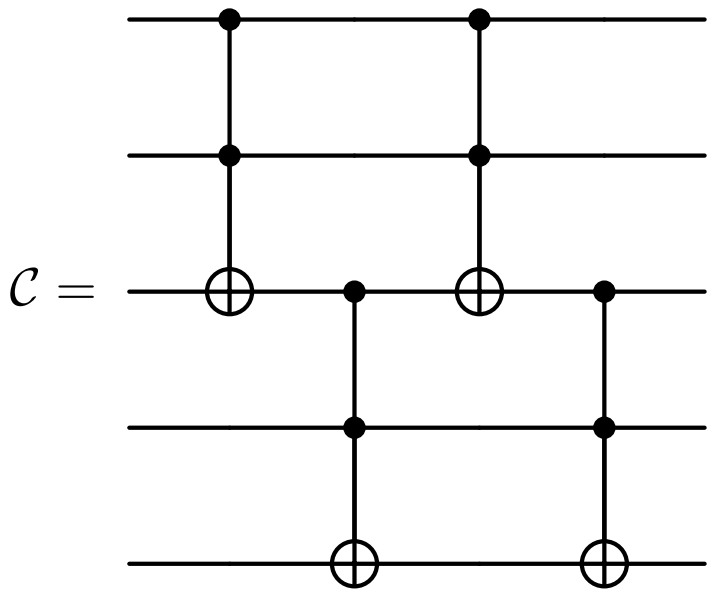
\begin{align}
    \mathcal{C} = \begin{quantikz}[wire types = {q,q,q,q,q}, row sep={0.9cm,between origins}]
                                & \ctrl{2} & & \ctrl{2} & & \\
                                & \ctrl{1} & & \ctrl{1} & & \\
                                & \targ{} & \ctrl{2} & \targ{} & \ctrl{2} & \\
                                & & \ctrl{1} & & \ctrl{1} & \\
                                & & \targ{} & & \targ{} &
        \end{quantikz}
        \label{eq:4_toffolis}
\end{align}
$\mathcal{C}$ implements a $CCCX$ on the first two and last two qubits: labeling the qubit states $\ket{a}\otimes\ket{b}\otimes\ket{c}\otimes\ket{d}\otimes\ket{e}$, the action is $\ket{a}\otimes\ket{b}\otimes\ket{c}\otimes\ket{d}\otimes\ket{e} \mapsto \ket{a}\otimes\ket{b}\otimes\ket{c}\otimes\ket{d}\otimes\ket{abd \oplus e}$. The Toffoli depth $\mathrm{Depth}_v^{CCX}(G(\mathcal{C}))$ of the circuit is 4, and so one would expect that when implementing the circuit by injection of $\toffstate$ states with Clifford corrections the reaction depth would be 4, applying each set of Clifford corrections before the next $\toffstate$ is injected.

A standard $\toffstate$ injection circuit is,
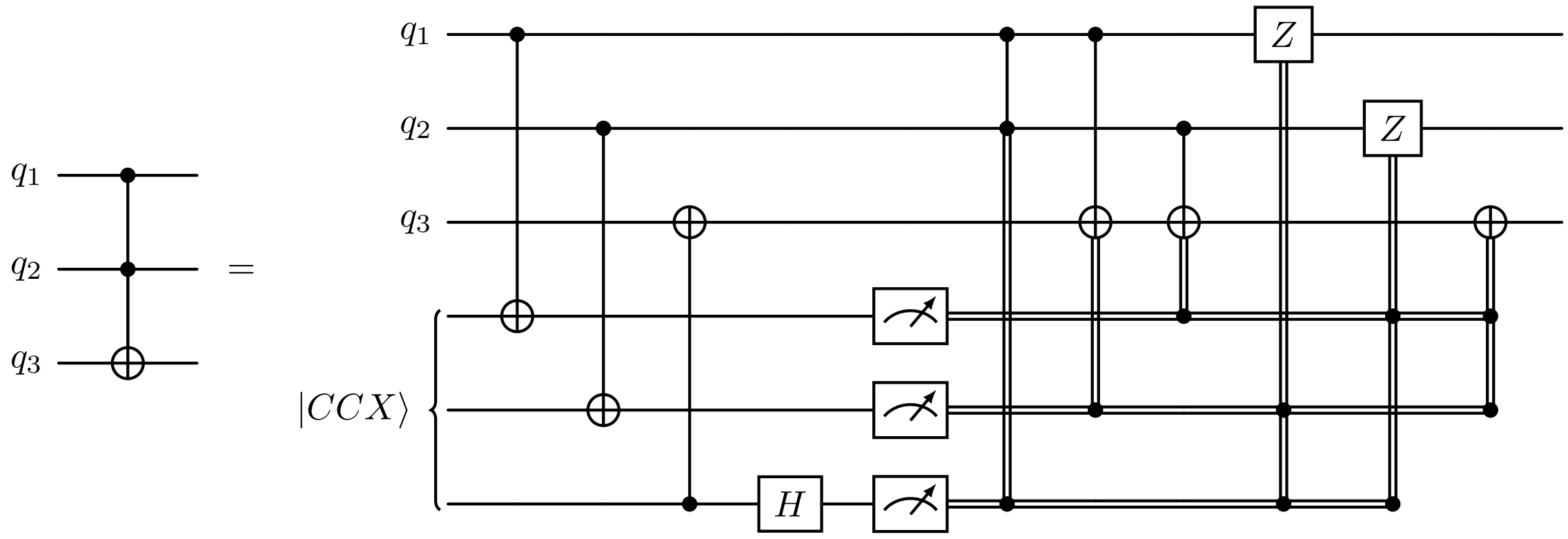
\begin{align*}
        \begin{quantikz}[wire types = {q,q,q}, row sep={0.9cm,between origins}]
        \lstick{$q_1$}          & \ctrl{2} & \\
        \lstick{$q_2$}          & \ctrl{1} & \\
        \lstick{$q_3$}          & \targ{} & \\
        \end{quantikz}
    \,\,\,=\,\,\,
    \begin{quantikz}[wire types = {q,q,q,q,q,q}, row sep={0.9cm,between origins}]
        \lstick{$q_1$}          & \ctrl{3} & & & & & \ctrl{1} & \ctrl{2} & & \gate{Z} & & & \\
        \lstick{$q_2$}          & & \ctrl{3} & & & & \ctrl{-1} & & \ctrl{1} & \wire[u]{c} & \gate{Z} & & \\
        \lstick{$q_3$}          & & & \targ{} & & & \wire[u]{c} & \targ{} & \targ{} & \wire[u]{c} & \wire[u]{c} & \targ{} & \\
        \lstick[3]{$\ket{CCX}$} & \targ{} & & & & \meter{} & \wire[u]{c}\setwiretype{c} & \wire[u]{c} & \phase{}\wire[u]{c}\setwiretype{c} & \wire[u]{c} & \phase{}\wire[u]{c} & \phase{}\wire[u]{c} & \setwiretype{n} \\
        & & \targ{} & & & \meter{} & \wire[u]{c}\setwiretype{c} & \phase{}\wire[u]{c} & & \phase{}\wire[u]{c} & \wire[u]{c} & \phase{}\wire[u]{c} & \setwiretype{n}\\
        & & & \ctrl{-3} & \gate{H} & \meter{} & \phase{}\wire[u]{c}\setwiretype{c} & & & \phase{}\wire[u]{c} & \phase{}\wire[u]{c} &  \setwiretype{n}  &
    \end{quantikz}
\end{align*}
where the last qubit of the $\toffstate$ is the `target' of the Toffoli.

\begin{claim}By commuting the Clifford corrections appropriately, the 4-Toffoli circuit $\mathcal{C}$ can be performed using the standard $\toffstate$ injection circuit with only reaction depth 2.\end{claim}

\proof
We show that there exists a gate and measurement schedule such that the measurements can be divided into two sets $\{\mathcal{M}_1, \mathcal{M}_2\}$, where $\mathcal{M}_1$ occur simultaneously and then $\mathcal{M}_2$ occur simultaneously.
Label the measurements for the first $CCX$ injection $\{1,2,3\}$ in vertical order; then label the measurements for the second $CCX$ injection $\{4,5,6\}$, and so on, up until measurement $12$. The claim is that 
\[\mathcal{M}_1 = \{1,2,5,6,7,8,11,12\};\qquad \mathcal{M}_2 = \{3,4,9,10\}.\]

The Clifford correction from measurement $1$ is $CX(q_2,q_3)$, from $2$ is $CX(q_1,q_3)$, and from $3$ is $CZ(q_1,q_2)$; continuing to the later measurements yields Table~\ref{tab:corrections_table}.

\begin{table}[]
    \centering
    \begin{tabular}{c|c}
        1 & $CX(q_2,q_3)$ \\
        2 & $CX(q_1,q_3)$ \\
        3 & $CZ(q_1,q_2)$ \\
        4 & $CX(q_4,q_5)$ \\
        5 & $CX(q_3,q_5)$ \\
        6 & $CZ(q_3,q_4)$ \\
        7 & $CX(q_2,q_3)$ \\
        8 & $CX(q_1,q_3)$ \\
        9 & $CZ(q_1,q_2)$ \\
        10 & $CX(q_4,q_5)$ \\
        11 & $CX(q_3,q_5)$ \\
        12 & $CZ(q_3,q_4)$
    \end{tabular}
    \caption{Measurements and their Clifford corrections.}
    \label{tab:corrections_table}
\end{table}

Observe that all the CNOTs performing the injections other than those with support on $q_3$ can be commuted with all Clifford and Pauli corrections. We therefore commute all these CNOTs to the start of the circuit. The measurements after those CNOTs are precisely those in $\mathcal{M}_1$, so these can be applied in the same timestep.

The corrections from $\mathcal{M}_1$ include $q_3$, and do not commute with the injection CNOTs with support on $q_3$. Therefore, all corrections from $\mathcal{M}_1$ are applied next, interspersed with the injection CNOTs with support on $q_3$. These will increase the quantum depth, but crucially only use 1 classical edge-depth.

Then measure the set $\mathcal{M}_2$. All the corrections from $\mathcal{M}_2$ are at the end of the circuit, so there are no more classical dependencies, and the reaction depth is 2.
\endproof

\begin{remark}
    The reaction depth plateaus at 2 even if the circuit is repeated $n$ times. That is, if we implement the circuit,

    
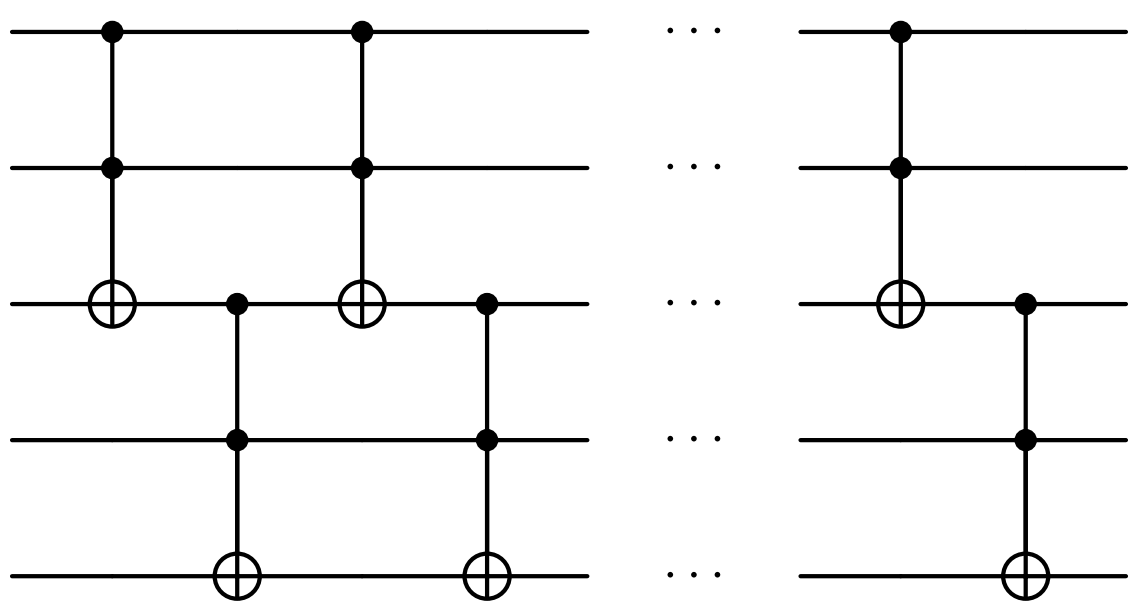
\begin{align*}
    \begin{quantikz}[wire types = {q,q,q,q,q}, row sep={0.9cm,between origins}]
                                & \ctrl{2} & & \ctrl{2} & & & \cdots\setwiretype{n} & & \ctrl{2}\setwiretype{q} & &\\
                                & \ctrl{1} & & \ctrl{1} & & & \cdots\setwiretype{n} & & \ctrl{1}\setwiretype{q} & &\\
                                & \targ{} & \ctrl{2} & \targ{} & \ctrl{2} & & \cdots\setwiretype{n} & & \targ{}\setwiretype{q} & \ctrl{2} &\\
                                & & \ctrl{1} & & \ctrl{1} & & \cdots\setwiretype{n} & & \setwiretype{q} & \ctrl{1} &\\
                                & & \targ{} & & \targ{} & & \cdots\setwiretype{n} & & \setwiretype{q} & \targ{} &
        \end{quantikz}
    \end{align*}
    
    via $\toffstate$ injection then the sets of measurements can still be divided into two sets, each of which can be performed simultaneously.
\end{remark}

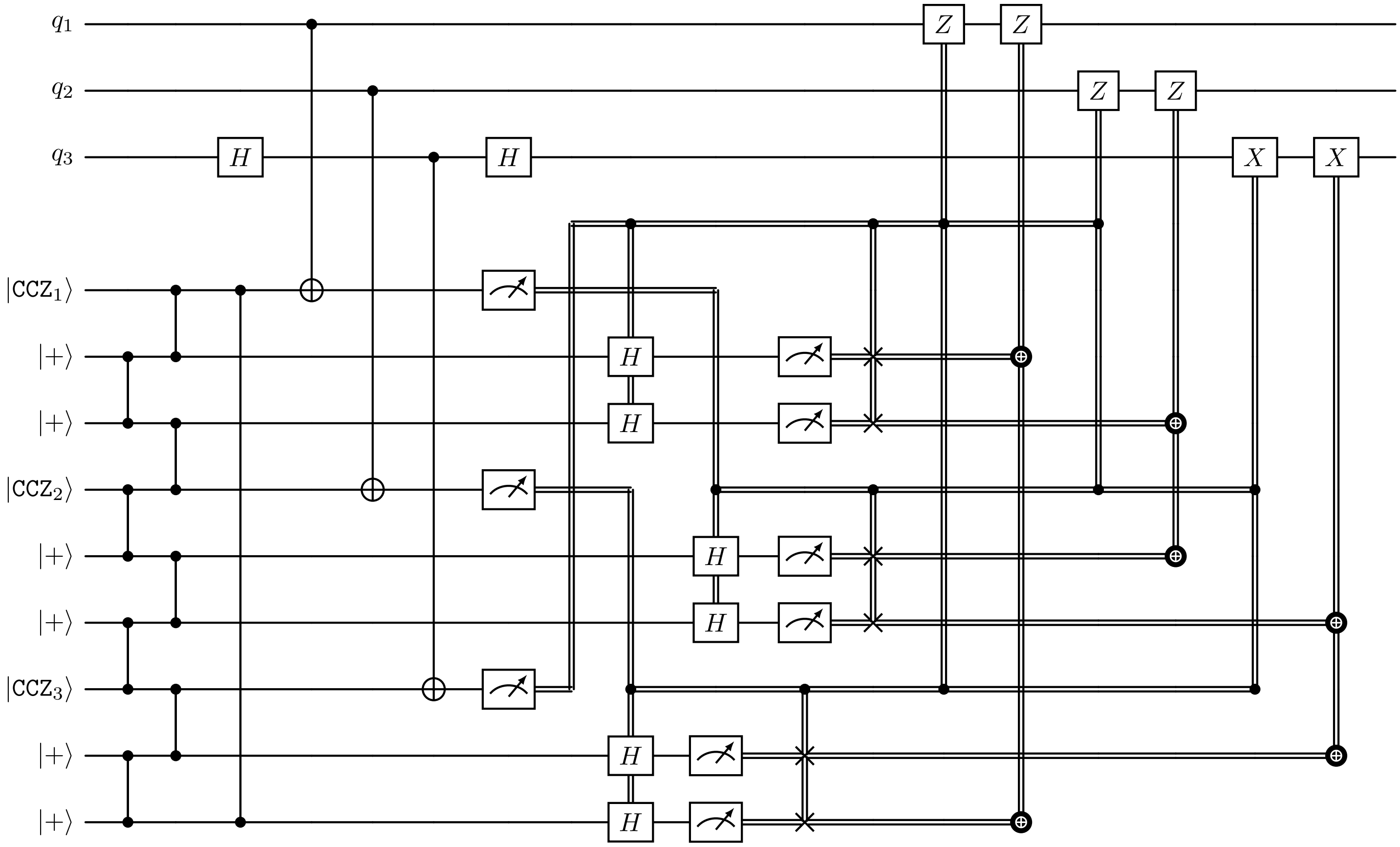
\begin{figure}[h]
    \centering
    \begin{align*}
\hspace{-10mm}\begin{quantikz}[wire types = {q,q,q,n,q,q,q,q,q,q,q,q,q}, row sep={0.9cm,between origins}]
        \lstick{$q_1$}          & & & & \ctrl{4} & & & & & & & & & \gate{Z}\wire[d]{c} & \gate{Z}\wire[d]{c} & & & & & \\
        \lstick{$q_2$}          & & & & & \ctrl{6} & & & & & & & & \wire[d]{c} & \wire[d]{c} & \gate{Z}\wire[d]{c} & \gate{Z}\wire[d]{c} & & & \\
        \lstick{$q_3$}          & & & \gate{H} & & & \ctrl{8} & \gate{H} & & & & & & \wire[d]{c} & \wire[d]{c} & \wire[d]{c} & \wire[d]{c} & \gate{X}\wire[d]{c} & \gate{X}\wire[d]{c} & \\
                                & & & & & & & & & \phase{}\wire[d]{c}\setwiretype{c} & & & \phase{}\wire[d]{c} & \phase{}\wire[d]{c} & \wire[d]{c} & \phase{}\wire[d]{c}& \wire[d]{c} \setwiretype{n} & \wire[d]{c} & \wire[d]{c} & \\
        \lstick{$\ket{\mathtt{CCZ}_1}$} & & \ctrl{1} & \ctrl{8}& \targ{} & & & \meter{} & \wire[u]{c}\setwiretype{c} & \wire[d]{c} & \wire[d]{c} &\setwiretype{n} & \wire[d]{c} & \wire[d]{c} & \wire[d]{c} & \wire[d]{c} & \wire[d]{c} & \wire[d]{c} & \wire[d]{c} & \\
        \lstick{$\ket{+}$}     & \ctrl{1} & \ctrl{-1} & & & & & & \wire[u]{c} & \gate{H}\wire[d]{c} & \wire[d]{c} & \meter{} & \swap{0}\setwiretype{c} & \wire[d]{c} &  \pctrl{0}\wire[d]{c} & \wire[d]{c}\setwiretype{n}  & \wire[d]{c} & \wire[d]{c} & \wire[d]{c} & \\
        \lstick{$\ket{+}$}      & \ctrl{-1} & \ctrl{1} & & & & & & \wire[u]{c} & \gate{H} & \wire[d]{c} & \meter{} & \swap{0}\wire[u]{c}\setwiretype{c} & \wire[d]{c} & \wire[d]{c} & \wire[d]{c} & \pctrl{0}\wire[d]{c} & \wire[d]{c}\setwiretype{n} & \wire[d]{c} & \\
        \lstick{$\ket{\mathtt{CCZ}_2}$} & \ctrl{1} & \ctrl{-1} & & & \targ{} & & \meter{} & \wire[u]{c}\setwiretype{c}& \wire[d]{c} & \phase{}\wire[d]{c}\setwiretype{n} & \setwiretype{c} & \phase{}\wire[d]{c} & \wire[d]{c} & \wire[d]{c} & \phase{} & \wire[d]{c} & \phase{}\wire[d]{c} & \wire[d]{c}\setwiretype{n} & \\
        \lstick{$\ket{+}$}      & \ctrl{-1} & \ctrl{1} & & & & & & \wire[u]{c} & \wire[d]{c} & \gate{H}\wire[d]{c} & \meter{} & \swap{0}\setwiretype{c} & \wire[d]{c} & \wire[d]{c} & & \pctrl{0}  & \wire[d]{c} \setwiretype{n} & \wire[d]{c} & \\
        \lstick{$\ket{+}$}      & \ctrl{1} & \ctrl{-1} & & & & & & \wire[u]{c} & \wire[d]{c} & \gate{H} & \meter{} & \swap{0}\wire[u]{c}\setwiretype{c} & \wire[d]{c} & \wire[d]{c} & & & \wire[d]{c} & \pctrl{0}\wire[d]{c} & \setwiretype{n}\\
        \lstick{$\ket{\mathtt{CCZ}_3}$} & \ctrl{-1} & \ctrl{1} & & & & \targ{} & \meter{} & \wire[u]{c}\setwiretype{c} & \phase{}\wire[d]{c}\setwiretype{n} & \setwiretype{c} & \phase{}\wire[d]{c} & & \phase{} & \wire[d]{c} & & & \phase{} & \wire[d]{c}\setwiretype{n} & \\
        \lstick{$\ket{+}$}      & \ctrl{1} & \ctrl{-1} & & & & & & & \gate{H}\wire[d]{c} & \meter{} & \swap{0}\setwiretype{c} & & & \wire[d]{c} & & & & \pctrl{0} & \setwiretype{n}\\
        \lstick{$\ket{+}$}     & \ctrl{-1} & & \ctrl{-8} & & & & & & \gate{H} & \meter{} & \swap{0}\wire[u]{c}\setwiretype{c} & & & \pctrl{0} & \setwiretype{n} &  & & & 
    \end{quantikz}
    \end{align*}
    \caption{Auto-corrected Toffoli circuit.}
    \label{fig:auto_toffoli}
\end{figure}

When injecting each of the Toffolis in an auto-corrected manner~\cite{gidney2019flexible}, the reaction depth is also 2. The auto-corrected $\toffstate$ injection circuit is shown in Fig.~\ref{fig:auto_toffoli}; note that it is just an AutoCCZ, with $H$ gates on pass-through qubits to convert it to a Toffoli. That the reaction depth of the auto-corrected circuit is 2 can be computed by tracking the Pauli corrections through the circuit~\cite{gidney2026eat}.

Curiously, there exists an alternative unitary implementation of the $CCCX$ which has Toffoli depth 2. Compared to~\cref{eq:4_toffolis}, this requires clean ancillae, but still returns the ancillae back to the $\ket{0}$ state afterwards:
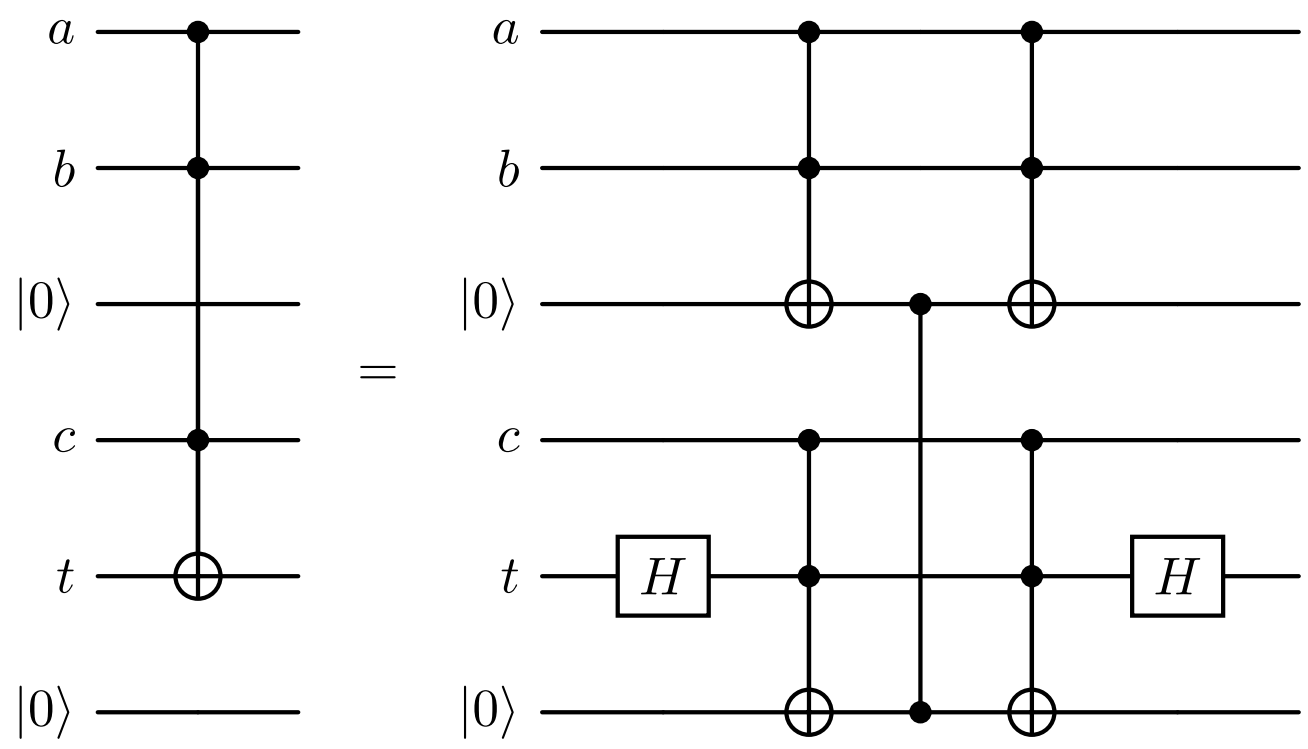
\begin{align*}
    \begin{quantikz}[wire types = {q,q,q,q,q,q}, row sep={0.9cm,between origins}]
        \lstick{$a$}                    & \ctrl{4} & \\
        \lstick{$b$}                    & \ctrl{3} & \\
        \lstick{$\ket{0}$}  & & \\
        \lstick{$c$}                   & \ctrl{1} & \\
        \lstick{$t$}                    & \targ{} & \\
        \lstick{$\ket{0}$}  & &                    
        \end{quantikz}
    \,\,\, = \,\,\,
    \begin{quantikz}[wire types = {q,q,q,q,q,q}, row sep={0.9cm,between origins}]
        \lstick{$a$}                    & & \ctrl{2} & & \ctrl{2} & & \\
        \lstick{$b$}                    & & \ctrl{1} & & \ctrl{1} & & \\
        \lstick{$\ket{0}$}  & & \targ{} & \ctrl{3} & \targ{} & & \\
        \lstick{$c$}                    & & \ctrl{2} & & \ctrl{2} & & \\
        \lstick{$t$}                    & \gate{H} & \ctrl{1} & & \ctrl{1} & \gate{H} & \\
        \lstick{$\ket{0}$}  & & \targ{} & \ctrl{0} & \targ{} & &
        \end{quantikz}
\end{align*}
\proof
Ignore the Hadamards for now. After the first layer of $CCX$ gates the state is 
\[\ket{a}\ket{b}\ket{ab}\ket{c}\ket{t}\ket{ct}.\]
Then the $CZ$ gate maps to,
\[(-1)^{abct}\ket{a}\ket{b}\ket{ab}\ket{c}\ket{t}\ket{ct}.\]
Applying the second layer of $CCX$ gates uncomputes the ancilla, leaving a circuit equivalent to a $CCCZ$, and then conjugating by the Hadamards maps to,
\[\ket{a}\ket{b}\ket{0}\ket{c}\ket{abc\oplus t}\ket{0}.\]
\endproof

The above example motivates the following lemma.

\begin{lemma}
Let $U$ be a unitary. Let $G$ be the gateset consisting of (a) arbitrary Clifford gates, (b) clean ancillae in $\ket{0}$ states and (c) magic gates $M$ (e.g. $CCX$) in the 3rd level of the Clifford hierarchy. Let $A$ be a circuit which is an exact synthesis of $U$ in the gateset $G$, returning all ancillae to their original clean $\ket{0}$ states, such that the $M$-depth $\textrm{Depth}_v^{M}(A)$ is optimal, i.e. there is no equivalent circuit $A'$ in the same gateset which has lower $M$-depth.

Let $G'$ be the gateset consisting of (a) arbitrary Clifford gates, (b) $\ket{0}$ ancillae in Pauli states, (c) magic states $M \ket{\psi}$, where $\ket{\psi}$ is a chosen Pauli state (e.g. $\ket{CCX} = CCX\ket{+}\ket{+}\ket{0}$) and (d) $Z$-basis measurements. Let $B$ be a circuit which is an exact synthesis of $U$ in the gateset $G'$, such that the reaction depth $r(B)$ is optimal, i.e. there is no equivalent circuit $B'$ in the same gateset which has lower reaction depth.

Then $r(B) \leq \textrm{Depth}_v^{M}(A)$.
\end{lemma}

\proof
Given $A$ in gateset $G$, there exists a circuit $B'$ in $G'$ such that $r(B') \leq \textrm{Depth}_v^{M}(A)$: take $A$ and replace each instance of $M$ with an injection of $M\ket{\psi}$ that implements $M$ up to corrections. As each $M$ is in the 3rd level of the Clifford hierarchy, the corrections to the injections are all Clifford (and so do not require further injection to perform). Then, for each layer of $M$ injections, correct the injections before the next layer of $M$ injections. This measurement schedule will always yield a number of rounds of classically controlled corrections at most $\textrm{Depth}_v^{M}(A)$.
\endproof

However, the converse is not true. That is, there exists a unitary $U$ for which the reaction depth is strictly lower than the magic depth. 

\begin{example}\label{ex:reaction_counterexample}
    Consider the unitary $U = CS$, a controlled-$S$ gate. Let $M = CCX$ and $M\ket{\psi} = \toffstate$. Then $\textrm{Depth}_v^{M}(A) = 2$, but $r(B) = 1$.
\end{example}
\proof
First we give a circuit $A$ which implements $U$ using the gateset $G$, then prove it has optimal $CCX$-depth.

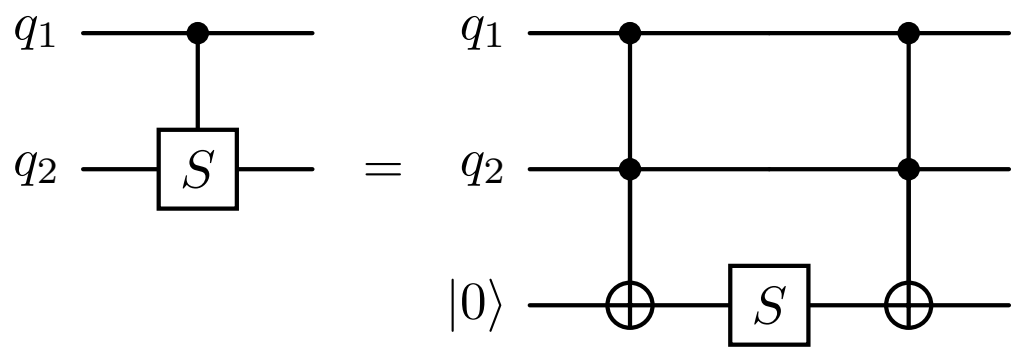
\begin{align*}
    \begin{quantikz}[wire types = {q,q,n}, row sep={0.9cm,between origins}]
            \lstick{$q_1$}          & \ctrl{1} & \\
            \lstick{$q_2$}          & \gate{S} & \\
            & &
            \end{quantikz}
    \,\,=\,
    \begin{quantikz}[wire types = {q,q,q}, row sep={0.9cm,between origins}]
    \lstick{$q_1$}                            & \ctrl{2} & & \ctrl{2} & \\
    \lstick{$q_2$}                            & \ctrl{1} & & \ctrl{1} & \\
    \lstick{$\ket{0}$}                        & \targ{} & \gate{S} & \targ{} &
        \end{quantikz}
\end{align*}
This circuit is correct because $CS\ket{a}\ket{b} = i^{ab}\ket{a}\ket{b}$. That it has optimal $CCX$-depth can be observed by the following. If there exists an implementation with lower $CCX$-depth, then it must have the structure $C\circ D\circ C'$, where $C$ and $C'$ are Clifford circuits (after some potential $\ket{0}$ state initialization) and $D$ is a tensor product of $CCZ$s, as $CCX$ is related to $CCZ$ by Hadamards, which can be included in $C$ and $C'$.

For any Clifford with symplectic matrix $\kappa$ (discarding Pauli terms, which only change signs~\cite{aaronson2004improved}), define $q_\kappa(x) = \omega(x, \kappa x) \mod 2$, where $\omega$ is the symplectic form and $x$ is a Pauli (product). For a Clifford $V$ with symplectic matrix $\sigma$, $q_{\sigma\kappa\sigma^{-1}}(\sigma x) = q_\kappa(x)$ since $\sigma$ preserves $\omega$. Hence $q_\kappa = 0$ remains true after conjugating the Clifford $\kappa$ by any Clifford.

For any Pauli $P$, observe that $D \circ P\circ D^\dagger$ is composed only of $CZ$ and Pauli gates. $q_{D \circ P\circ D^\dagger}(x) = 0$: letting $x = (u,v)$, where $u$ holds the $X$ bits and $v$ the $Z$ bits, a layer of $CZ$ gates acts as $(u, v) \mapsto (u, v+Au)$ where $A$ is symmetric with zero diagonal, so $q = u^\top Au = 0 \mod 2$. However, $CS_{ab} \circ X_a \circ CS^\dagger_{ab} = X_a\circ S_b\circ CZ_{ab}$, and $q_{S_b\circ CZ_{ab}}(x)$ is nonzero in general. Indeed, letting $x= (u,v)$, $q_{S_b\circ CZ_{ab}}(x) = u_b$.
Therefore, $CCX$-depth 1 is insufficient to synthesize a $CS$ gate, and $A$ has optimal $CCX$-depth.
Now we give a circuit $B$ using the gateset $G'$.

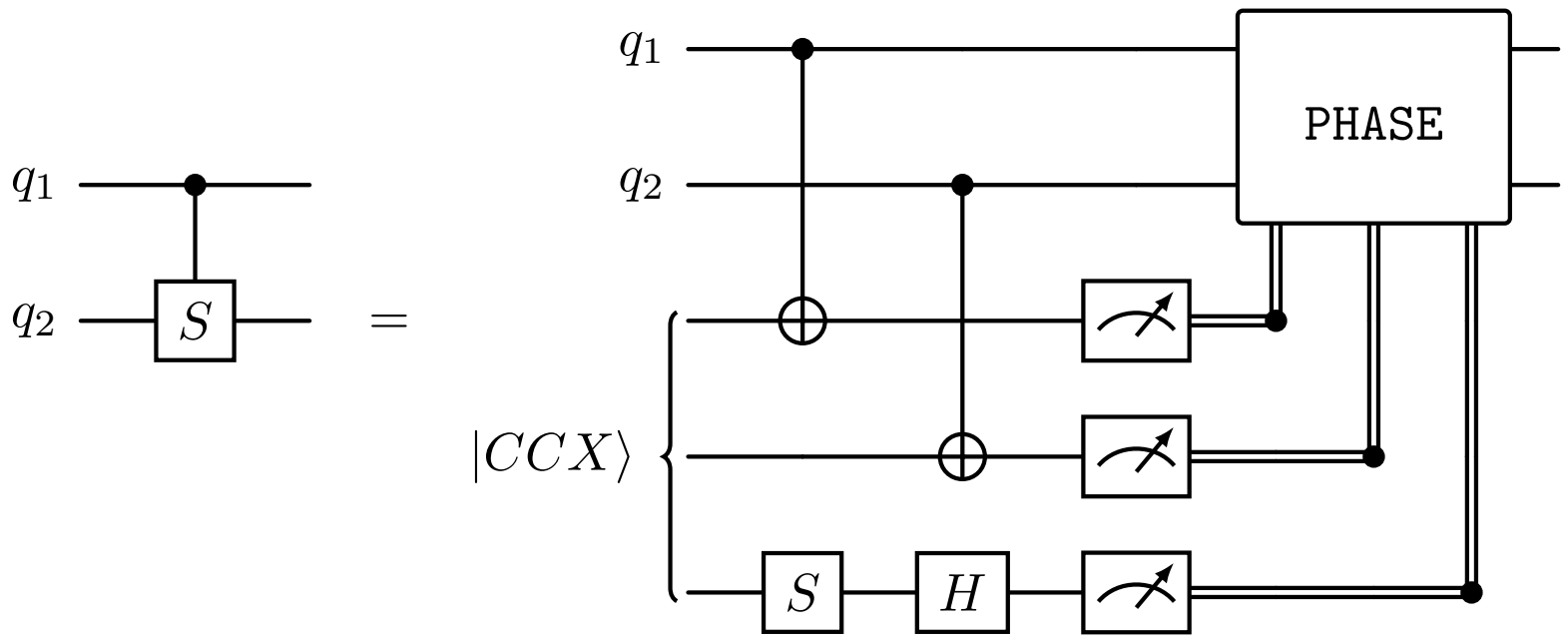
\begin{align*}
\begin{quantikz}[wire types = {q,q,n}, row sep={0.9cm,between origins}]
            \lstick{$q_1$}          & \ctrl{1} & \\
            \lstick{$q_2$}          & \gate{S} & \\
            & &
            \end{quantikz}
    \,\,\,=\,\,\,
\begin{quantikz}[wire types = {q,q,q,q,q}, row sep={0.9cm,between origins}]
        \lstick{$q_1$}          & \ctrl{2} & & & \phasepoly & & & \\
        \lstick{$q_2$}          & & \ctrl{2} & & & & & \\
        \lstick[3]{$\ket{CCX}$} & \targ{} & & \meter{} & \ctrl{0}\wire[u]{c}\setwiretype{c} & \wire[u]{c}\setwiretype{n} & \wire[u]{c} & \\
        & & \targ{} & \meter{} & \setwiretype{c} & \ctrl{0}\wire[u]{c} & \wire[u]{c}\setwiretype{n} & \\
        & \gate{S} & \gate{H} & \meter{} & \setwiretype{c} & & \ctrl{0}\wire[u]{c} & \setwiretype{n}
    \end{quantikz}
\end{align*}
Where $\texttt{PHASE}$ is a Clifford phase polynomial.

Before applying the $S$ gate to the third qubit of the $\toffstate$, $B$ holds a wandering \gateand gate (\cref{lem:gate_and}), i.e. $g = (a \oplus p_1)(b\oplus p_2)$, where $p_1$ and $p_2$ are the first two measurement outcomes. Then applying $S$ and the $X$-basis measurement yields the phase $i^{g}(-1)^{mg}$, where $m$ is the outcome of the $X$-basis measurement. The correction needed to acquire $i^{ab}$ is therefore a Clifford phase polynomial.

$B$ must use at least one $\toffstate$, as $CS$ is non-Clifford, so it must incur a non-Pauli Clifford correction and have reaction depth at least 1. Therefore $B$ has optimal reaction depth.
\endproof

This example works because of measurement-based uncomputation. That is, if we allowed the $\ket{0}$ ancilla qubit in $A$ to be reset using measurement-based uncomputation, rather than a Toffoli gate, then it would have $CCX$-depth 1. However, because measurements are non-unitary this trick cannot be used for $A$. We are not aware of any examples of unitaries where the reaction depth is strictly lower than the magic depth which are not due to measurement-based uncomputation. This is an interesting open question.

\subsection{Measurement cascades}

An intentional consequence of Definition~\ref{def:circuit_run_dag} is that, if there is a set of dependent adaptive-basis measurements occurring sequentially, the circuit run where they all output 0 still has high reaction depth.

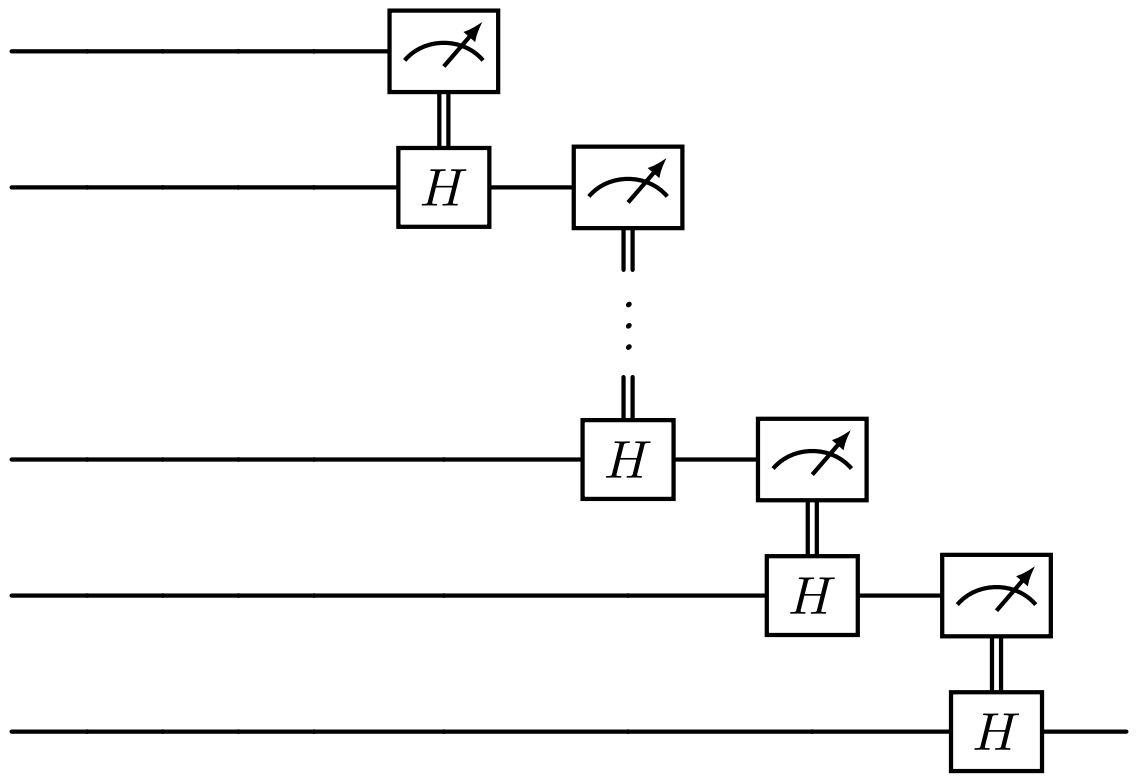
\begin{example}
    \begin{align*}
    \begin{quantikz}[wire types = {q,q,n,q,q,q}, row sep={0.9cm,between origins}, align equals at=3]
                                & & & & & \meter{}\wire[d]{c} & \setwiretype{n}& & & \\
                                & & & & & \gate{H} & \meter{}\wire[d]{c} & \setwiretype{n}& & \\
                                & & & & & & \raisebox{0.5em}{$\vdots$}\wire[d]{c} & & & \\
                                & & & & & & \gate{H} & \meter{}\wire[d]{c} & \setwiretype{n} & \\
                                & & & & & & & \gate{H} & \meter{}\wire[d]{c} & \setwiretype{n}\\
                                & & & & & & & & \gate{H} &
        \end{quantikz}
\end{align*}
Given $n$ qubits, the circuit will have reaction depth $n-1$. Moreover, even when the output of each measurement is 0, the corresponding circuit run still has reaction depth $n-1$. Explicitly, the DAG would be,
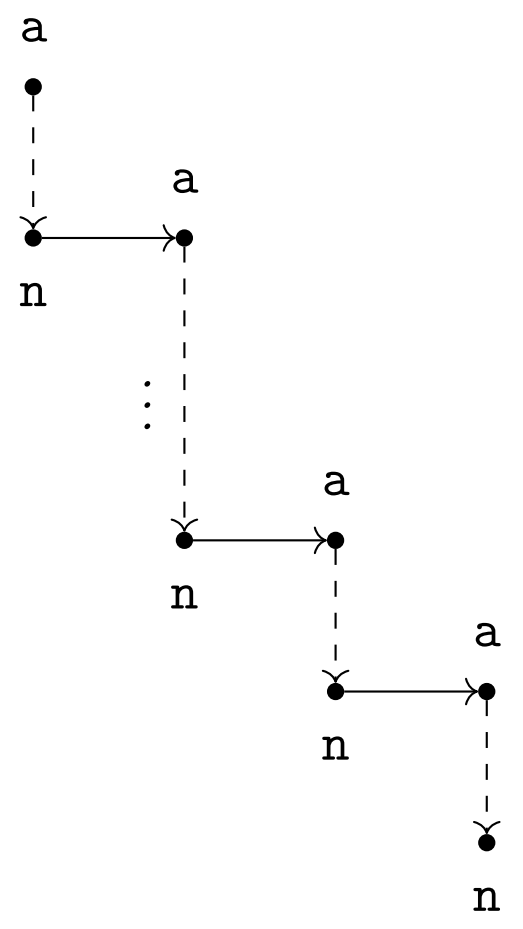
\[\begin{tikzpicture}
	\begin{pgfonlayer}{nodelayer}
		\node [style=boundary vertex] (0) at (-2, 5) {};
		\node [style=none] (16) at (-2, 5.75) {$\mathtt{a}$};
		\node [style=boundary vertex] (17) at (-2, 3) {};
		\node [style=boundary vertex] (18) at (0, 3) {};
		\node [style=boundary vertex] (19) at (0, -1) {};
		\node [style=boundary vertex] (20) at (2, -1) {};
		\node [style=boundary vertex] (21) at (2, -3) {};
		\node [style=boundary vertex] (22) at (4, -3) {};
		\node [style=none] (23) at (-0.5, 1) {$\vdots$};
		\node [style=boundary vertex] (24) at (4, -5) {};
		\node [style=none] (25) at (0, 3.75) {$\mathtt{a}$};
		\node [style=none] (26) at (2, -0.25) {$\mathtt{a}$};
		\node [style=none] (27) at (4, -2.25) {$\mathtt{a}$};
		\node [style=none] (28) at (-2, 2.25) {$\mathtt{n}$};
		\node [style=none] (29) at (0, -1.75) {$\mathtt{n}$};
		\node [style=none] (30) at (2, -3.75) {$\mathtt{n}$};
		\node [style=none] (31) at (4, -5.75) {$\mathtt{n}$};
	\end{pgfonlayer}
	\begin{pgfonlayer}{edgelayer}
		\draw [style=dashed arrow] (18) to (19);
		\draw [style=dashed arrow] (0) to (17);
		\draw [style=dashed arrow] (20) to (21);
		\draw [style=dashed arrow] (22) to (24);
		\draw [style=arrow] (17) to (18);
		\draw [style=arrow] (19) to (20);
		\draw [style=arrow] (21) to (22);
	\end{pgfonlayer}
\end{tikzpicture}\]
Even though the Hadamards are not applied, the classical wires to them are still present, and so the longest classical path through the DAG has classical edge-length $n-1$. The quantum depth is also $n-1$.

The operational interpretation is that even though the Hadamards do not need to be applied, it is not \textit{known} that they are unnecessary until the measurement results have each been decoded, and so the set of measurements cannot all be applied in parallel.
\end{example}

\subsection{On-the-fly compilation}

Another intentional consequence of Definition~\ref{def:circuit_run_dag} is that it disallows general, automatic on-the-fly recompilation of quantum gates based on previous measurement outcomes. For example, consider the following circuit,

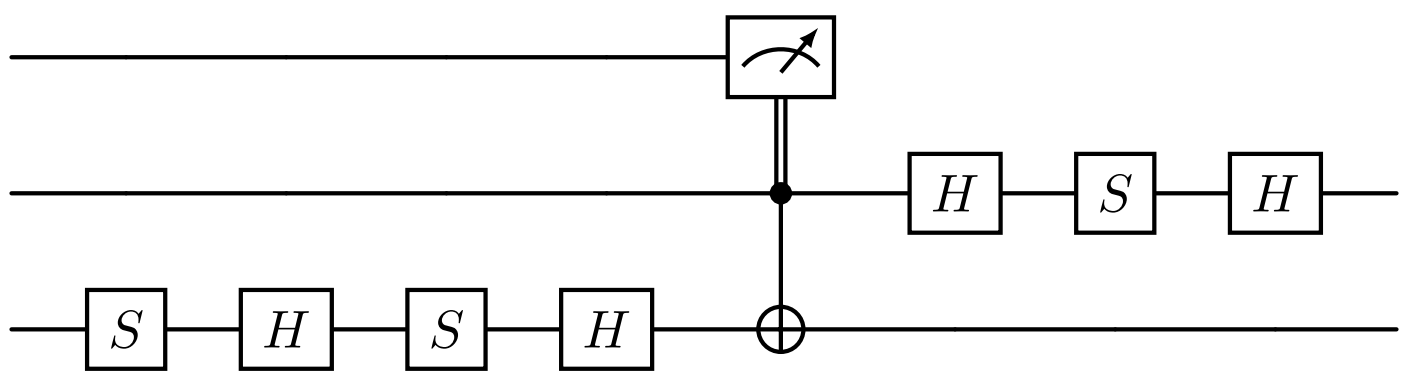
\begin{align*}
    \begin{quantikz}[wire types = {q,q,q}, row sep={0.9cm,between origins}]
                                & & & & & \meter{}\wire[d]{c} & \setwiretype{n}& & & \\
                                & & & & & \ctrl{1} & \gate{H} & \gate{S} & \gate{H} & \\
                                & \gate{S} & \gate{H} & \gate{S} & \gate{H} & \targ{} & & & &
        \end{quantikz}
\end{align*}
If the measurement returns 0, then the classically controlled CNOT is not performed. The DAG for the circuit run will in this case be,
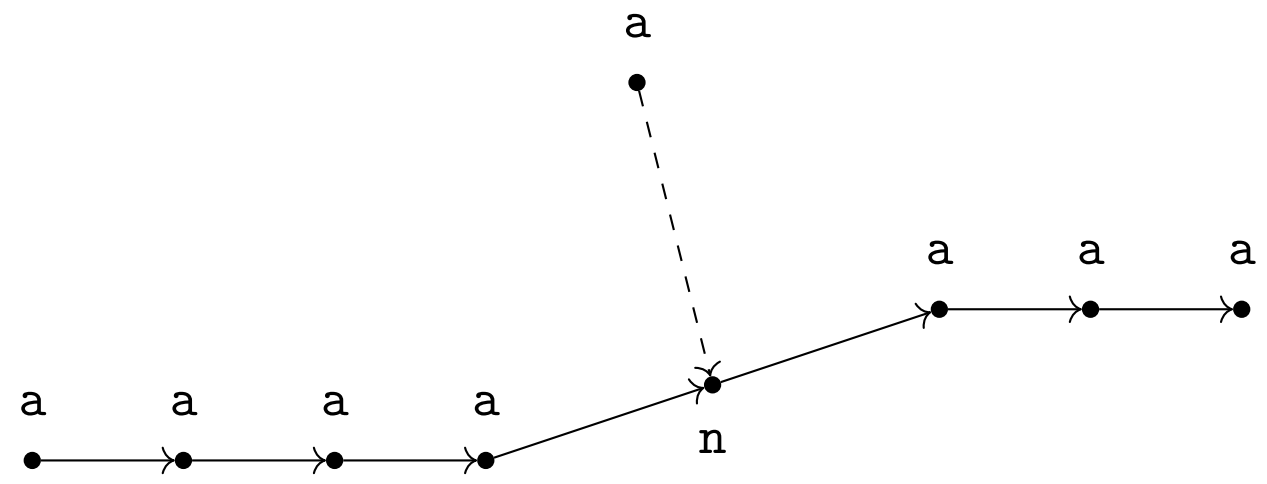
\[\begin{tikzpicture}
	\begin{pgfonlayer}{nodelayer}
		\node [style=boundary vertex] (0) at (-2, 5) {};
		\node [style=boundary vertex] (1) at (-1, 1) {};
		\node [style=boundary vertex] (2) at (2, 2) {};
		\node [style=boundary vertex] (3) at (4, 2) {};
		\node [style=boundary vertex] (4) at (6, 2) {};
		\node [style=boundary vertex] (5) at (-4, 0) {};
		\node [style=boundary vertex] (6) at (-6, 0) {};
		\node [style=boundary vertex] (7) at (-8, 0) {};
		\node [style=boundary vertex] (8) at (-10, 0) {};
		\node [style=none] (9) at (-10, 0.75) {$\mathtt{a}$};
		\node [style=none] (10) at (-8, 0.75) {$\mathtt{a}$};
		\node [style=none] (11) at (-6, 0.75) {$\mathtt{a}$};
		\node [style=none] (12) at (-4, 0.75) {$\mathtt{a}$};
		\node [style=none] (13) at (2, 2.75) {$\mathtt{a}$};
		\node [style=none] (14) at (4, 2.75) {$\mathtt{a}$};
		\node [style=none] (15) at (6, 2.75) {$\mathtt{a}$};
		\node [style=none] (16) at (-2, 5.75) {$\mathtt{a}$};
		\node [style=none] (17) at (-1, 0.25) {$\mathtt{n}$};
	\end{pgfonlayer}
	\begin{pgfonlayer}{edgelayer}
		\draw [style=dashed arrow] (0) to (1);
		\draw [style=arrow] (1) to (2);
		\draw [style=arrow] (2) to (3);
		\draw [style=arrow] (3) to (4);
		\draw [style=arrow] (5) to (1);
		\draw [style=arrow] (6) to (5);
		\draw [style=arrow] (7) to (6);
		\draw [style=arrow] (8) to (7);
	\end{pgfonlayer}
\end{tikzpicture}\]
In particular, while the CNOT is not applied, is labeled $\mathtt{n}$ and does not contribute to the quantum circuit depth, it still allows for a path in the DAG which has 7 $\mathtt{a}$-labeled vertices in it, and thus have quantum depth 7.

The operational interpretation is that while the CNOT does not need to be applied, the device does not know this far enough in advance to recompile and parallelize the $S$ and $H$ gates on different qubits. If the measurement occurs far enough in advance of the CNOT and other single-qubit gates, then the circuit could plausibly be recompiled on-the-fly. As is, this scenario can still be captured by our definitions, but it would require the classical control to branch between two entire subcircuits: one in which the CNOT is performed and the other where it is not and the single-qubit gates are hence parallelized.

\section{ZX-calculus proofs}\label{app:proofs}

To follow the proofs below, some elementary knowledge of the ZX-calculus~\cite{coecke2011interacting} is required.

Proof of Lemma~\ref{lem:ztype_ppm}:
\proof
\[\begin{tikzpicture}
	\begin{pgfonlayer}{nodelayer}
		\node [style=none] (0) at (-5, 4) {};
		\node [style=none] (1) at (-5, 2) {};
		\node [style=none] (2) at (-5, -2) {};
		\node [style=none] (3) at (6, 4) {};
		\node [style=none] (4) at (6, 2) {};
		\node [style=none] (5) at (6, -2) {};
		\node [style={white_dot}] (6) at (-3, 4) {};
		\node [style={white_dot}] (7) at (-3, 2) {};
		\node [style={white_dot}] (8) at (-3, -2) {};
		\node [style={grey_dot}] (9) at (0, 0) {};
		\node [style=none] (10) at (0.5, -0.5) {$m\pi$};
		\node [style={white_dot}] (11) at (-5, -4) {};
		\node [style=none] (12) at (6, -4) {};
		\node [style={white_dot}] (13) at (-3, -4) {};
		\node [style={grey_dot}] (14) at (4, -4) {};
		\node [style=none] (15) at (4.5, -3.5) {$m\pi$};
		\node [style=none] (16) at (-3, 0) {$\vdots$};
	\end{pgfonlayer}
	\begin{pgfonlayer}{edgelayer}
		\draw (0.center) to (3.center);
		\draw (4.center) to (1.center);
		\draw (2.center) to (5.center);
		\draw (9) to (8);
		\draw (7) to (9);
		\draw (9) to (6);
		\draw (12.center) to (11);
		\draw (13) to (9);
	\end{pgfonlayer}
\end{tikzpicture} \ \ = \ \ \begin{tikzpicture}
	\begin{pgfonlayer}{nodelayer}
		\node [style=none] (0) at (-5, 4) {};
		\node [style=none] (1) at (-5, 2) {};
		\node [style=none] (2) at (-5, -2) {};
		\node [style=none] (3) at (6, 4) {};
		\node [style=none] (4) at (6, 2) {};
		\node [style=none] (5) at (6, -2) {};
		\node [style={white_dot}] (6) at (-3, 4) {};
		\node [style={white_dot}] (7) at (-3, 2) {};
		\node [style={white_dot}] (8) at (-3, -2) {};
		\node [style={grey_dot}] (9) at (0, 0) {};
		\node [style=none] (10) at (0.75, 0.25) {$m\pi$};
		\node [style=none] (12) at (6, -4) {};
		\node [style={grey_dot}] (14) at (4, -4) {};
		\node [style=none] (15) at (4.5, -3.5) {$m\pi$};
		\node [style=none] (16) at (-3, 0) {$\vdots$};
	\end{pgfonlayer}
	\begin{pgfonlayer}{edgelayer}
		\draw (0.center) to (3.center);
		\draw (4.center) to (1.center);
		\draw (2.center) to (5.center);
		\draw (9) to (8);
		\draw (7) to (9);
		\draw (9) to (6);
		\draw (9) to (14);
		\draw (14) to (12.center);
	\end{pgfonlayer}
\end{tikzpicture} \]
\vspace{10mm}
\[= \begin{tikzpicture}
	\begin{pgfonlayer}{nodelayer}
		\node [style=none] (0) at (-5.25, 4) {};
		\node [style=none] (1) at (-5.25, 2) {};
		\node [style=none] (2) at (-5.25, -2) {};
		\node [style=none] (3) at (5.75, 4) {};
		\node [style=none] (4) at (5.75, 2) {};
		\node [style=none] (5) at (5.75, -2) {};
		\node [style={white_dot}] (6) at (-3.25, 4) {};
		\node [style={white_dot}] (7) at (-3.25, 2) {};
		\node [style={white_dot}] (8) at (-3.25, -2) {};
		\node [style={grey_dot}] (9) at (-0.25, -4) {};
		\node [style=none] (12) at (5.75, -4) {};
		\node [style=none] (13) at (-3.25, 0) {$\vdots$};
	\end{pgfonlayer}
	\begin{pgfonlayer}{edgelayer}
		\draw (0.center) to (3.center);
		\draw (4.center) to (1.center);
		\draw (2.center) to (5.center);
		\draw (9) to (8);
		\draw (7) to (9);
		\draw (9) to (6);
		\draw (9) to (12.center);
	\end{pgfonlayer}
\end{tikzpicture} \]
\endproof

Proof of Lemma~\ref{lem:xtype_ppm}:
\proof
\[\begin{tikzpicture}
	\begin{pgfonlayer}{nodelayer}
		\node [style=none] (0) at (-7.75, 4) {};
		\node [style=none] (1) at (8.25, 4) {};
		\node [style=none] (2) at (-7.75, 1.25) {};
		\node [style=none] (3) at (8.25, 1.25) {};
		\node [style=none] (4) at (-7.75, -2.75) {};
		\node [style=none] (5) at (-1.75, -2.75) {};
		\node [style={grey_dot}] (6) at (-5.75, 4) {};
		\node [style={grey_dot}] (7) at (-5.75, 1.25) {};
		\node [style={grey_dot}] (8) at (-5.75, -2.75) {};
		\node [style={grey_dot}] (9) at (-1.75, -2.75) {};
		\node [style=none] (10) at (-1.25, -2.25) {$\ell \pi$};
		\node [style={grey_dot}] (11) at (-7.75, -4.75) {};
		\node [style={grey_dot}] (12) at (-5.75, -4.75) {};
		\node [style={white_dot}] (13) at (-3.75, -0.75) {};
		\node [style=none] (14) at (8.25, -4.75) {};
		\node [style=none] (15) at (-5.75, 2.5) {$\vdots$};
		\node [style=none] (16) at (-3.25, -0.25) {$m\pi$};
		\node [style={grey_dot}] (17) at (6.25, -4.75) {};
		\node [style={grey_dot}] (18) at (6.25, 1.25) {};
		\node [style={grey_dot}] (19) at (6.25, 4) {};
		\node [style=none] (20) at (6.75, 4.5) {$\ell\pi$};
		\node [style=none] (21) at (6.75, 1.75) {$\ell\pi$};
		\node [style=none] (22) at (6.75, -4.25) {$\ell\pi$};
		\node [style={white_dot}] (23) at (4.25, -4.75) {};
		\node [style=none] (24) at (4.75, -4.25) {$m\pi$};
	\end{pgfonlayer}
	\begin{pgfonlayer}{edgelayer}
		\draw (0.center) to (1.center);
		\draw (3.center) to (2.center);
		\draw (4.center) to (5.center);
		\draw (13) to (8);
		\draw (11) to (12);
		\draw (14.center) to (12);
		\draw (12) to (13);
		\draw (13) to (7);
		\draw (6) to (13);
	\end{pgfonlayer}
\end{tikzpicture} \ \ = \ \ \begin{tikzpicture}
	\begin{pgfonlayer}{nodelayer}
		\node [style=none] (0) at (-8, 4) {};
		\node [style=none] (1) at (8, 4) {};
		\node [style=none] (2) at (-8, 1) {};
		\node [style=none] (3) at (8, 1) {};
		\node [style=none] (4) at (-8, -3) {};
		\node [style=none] (5) at (-2, -3) {};
		\node [style={grey_dot}] (6) at (-6, 4) {};
		\node [style={grey_dot}] (7) at (-6, 1) {};
		\node [style={grey_dot}] (8) at (-6, -3) {};
		\node [style={grey_dot}] (9) at (-2, -3) {};
		\node [style=none] (10) at (-1.5, -2.5) {$\ell \pi$};
		\node [style={white_dot}] (13) at (-4, -1) {};
		\node [style=none] (14) at (8, -5) {};
		\node [style=none] (15) at (-6, 2.5) {$\vdots$};
		\node [style=none] (16) at (-3.5, -0.5) {$m\pi$};
		\node [style={grey_dot}] (17) at (6, -5) {};
		\node [style={grey_dot}] (18) at (6, 1) {};
		\node [style={grey_dot}] (19) at (6, 4) {};
		\node [style=none] (20) at (6.5, 4.5) {$\ell\pi$};
		\node [style=none] (21) at (6.5, 1.5) {$\ell\pi$};
		\node [style=none] (22) at (6.5, -4.5) {$\ell\pi$};
		\node [style={white_dot}] (23) at (4, -5) {};
		\node [style=none] (24) at (4.5, -4.5) {$m\pi$};
	\end{pgfonlayer}
	\begin{pgfonlayer}{edgelayer}
		\draw (0.center) to (1.center);
		\draw (3.center) to (2.center);
		\draw (4.center) to (5.center);
		\draw (13) to (8);
		\draw (13) to (7);
		\draw (6) to (13);
		\draw [in=180, out=-90, looseness=1.50] (13) to (23);
		\draw (23) to (17);
		\draw (17) to (14.center);
	\end{pgfonlayer}
\end{tikzpicture} \]
\vspace{10mm}
\[= \ \ \begin{tikzpicture}
	\begin{pgfonlayer}{nodelayer}
		\node [style=none] (0) at (-8, 4) {};
		\node [style=none] (1) at (8, 4) {};
		\node [style=none] (2) at (-8, 1) {};
		\node [style=none] (3) at (8, 1) {};
		\node [style=none] (4) at (-8, -3) {};
		\node [style={grey_dot}] (6) at (-6, 4) {};
		\node [style={grey_dot}] (7) at (-6, 1) {};
		\node [style=none] (10) at (-3.25, -2) {$\ell \pi$};
		\node [style={white_dot}] (11) at (-4, -1) {};
		\node [style=none] (12) at (8, -5) {};
		\node [style=none] (13) at (-6, 2.5) {$\vdots$};
		\node [style={grey_dot}] (15) at (6, -5) {};
		\node [style={grey_dot}] (16) at (6, 1) {};
		\node [style={grey_dot}] (17) at (6, 4) {};
		\node [style=none] (18) at (6.5, 4.5) {$\ell\pi$};
		\node [style=none] (19) at (6.5, 1.5) {$\ell\pi$};
		\node [style=none] (20) at (6.5, -4.5) {$\ell\pi$};
		\node [style=none] (21) at (4, -5) {};
		\node [style={grey_dot}] (22) at (-3.95, -2.25) {};
		\node [style={grey_dot}] (23) at (-5, 0) {};
		\node [style={grey_dot}] (24) at (-4.625, 0.5) {};
		\node [style=none] (25) at (-4, 0.5) {$\ell \pi$};
		\node [style=none] (26) at (-5.75, -0.25) {$\ell \pi$};
	\end{pgfonlayer}
	\begin{pgfonlayer}{edgelayer}
		\draw (0.center) to (1.center);
		\draw (3.center) to (2.center);
		\draw (11) to (7);
		\draw (6) to (11);
		\draw (15) to (12.center);
		\draw (21.center) to (15);
		\draw [in=-90, out=180, looseness=1.50] (21.center) to (11);
		\draw (4.center) to (11);
	\end{pgfonlayer}
\end{tikzpicture} \ \ = \ \ \begin{tikzpicture}
	\begin{pgfonlayer}{nodelayer}
		\node [style=none] (0) at (-8, 4) {};
		\node [style=none] (1) at (8, 4) {};
		\node [style=none] (2) at (-8, 1) {};
		\node [style=none] (3) at (8, 1) {};
		\node [style=none] (4) at (-8, -3) {};
		\node [style={grey_dot}] (5) at (-2, 4) {};
		\node [style={grey_dot}] (6) at (-2, 1) {};
		\node [style={white_dot}] (8) at (-4, -3) {};
		\node [style=none] (9) at (8, -3) {};
		\node [style=none] (10) at (-6, 2.5) {$\vdots$};
	\end{pgfonlayer}
	\begin{pgfonlayer}{edgelayer}
		\draw (0.center) to (1.center);
		\draw (3.center) to (2.center);
		\draw (8) to (6);
		\draw (5) to (8);
		\draw (4.center) to (8);
		\draw (8) to (9.center);
	\end{pgfonlayer}
\end{tikzpicture} \]
\endproof

Proof of Lemma~\ref{lem:delayed_cz}:
\proof
We consider first the case where the classical control is 0:
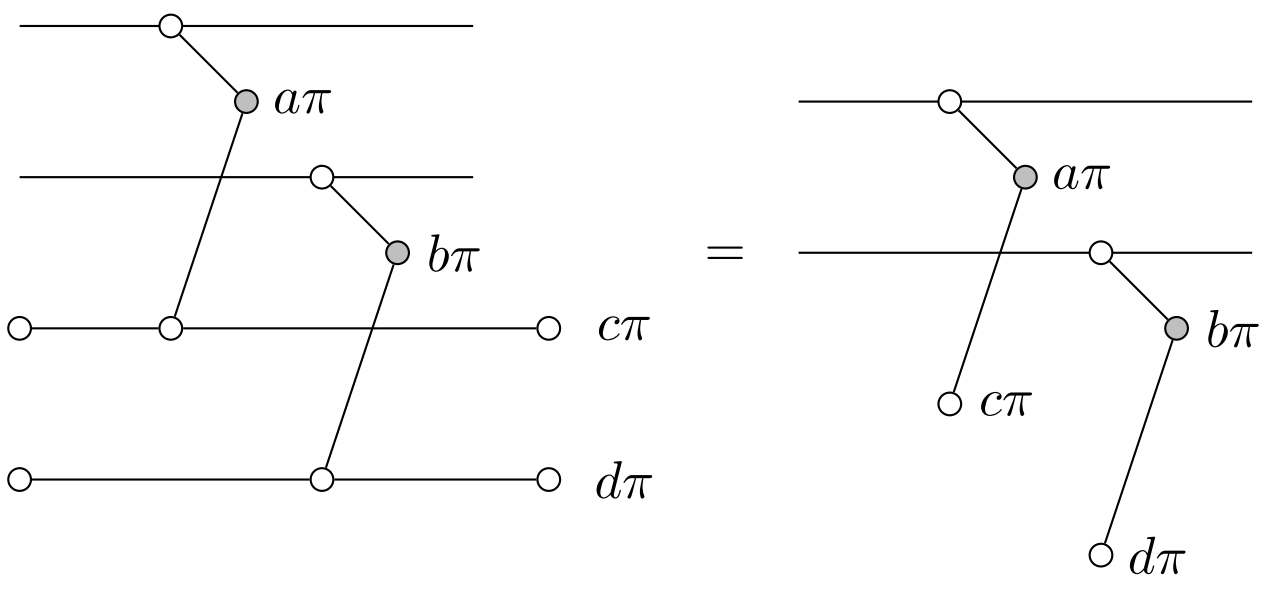
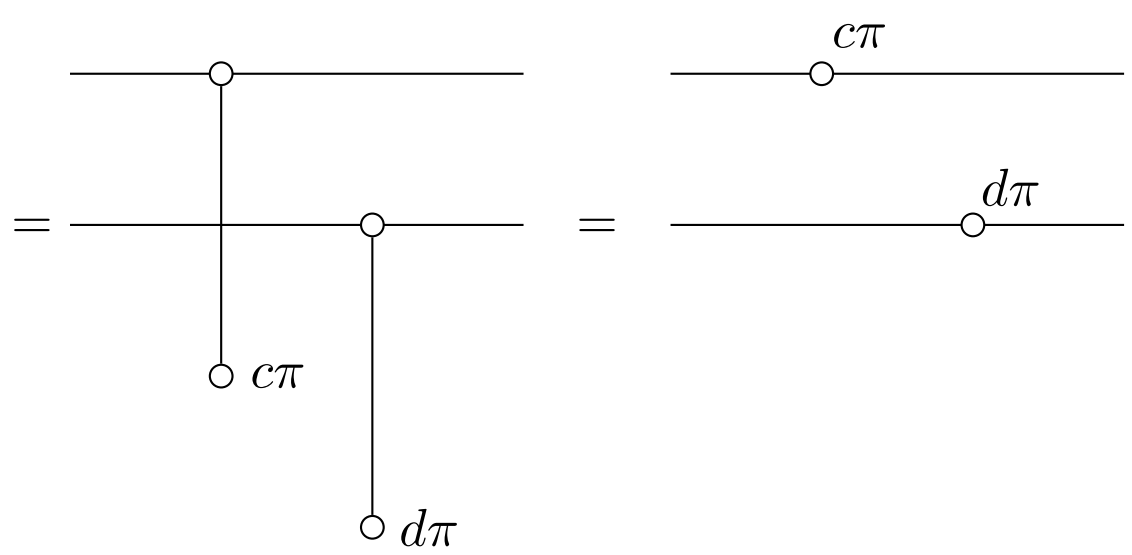
\[\begin{tikzpicture}
	\begin{pgfonlayer}{nodelayer}
		\node [style=none] (0) at (-4, 3) {};
		\node [style=none] (1) at (2, 3) {};
		\node [style=none] (2) at (-4, 1) {};
		\node [style=none] (3) at (2, 1) {};
		\node [style={white_dot}] (4) at (-2, 3) {};
		\node [style={white_dot}] (5) at (-2, -1) {};
		\node [style={grey_dot}] (6) at (-1, 2) {};
		\node [style={white_dot}] (7) at (0, 1) {};
		\node [style={white_dot}] (8) at (0, -3) {};
		\node [style={grey_dot}] (9) at (1, 0) {};
		\node [style={white_dot}] (10) at (-4, -1) {};
		\node [style={white_dot}] (11) at (-4, -3) {};
		\node [style={white_dot}] (12) at (3, -1) {};
		\node [style={white_dot}] (13) at (3, -3) {};
		\node [style=none] (14) at (-0.25, 2) {$a\pi$};
		\node [style=none] (15) at (1.75, 0) {$b\pi$};
		\node [style=none] (16) at (4, -1) {$c\pi$};
		\node [style=none] (17) at (4, -3) {$d\pi$};
	\end{pgfonlayer}
	\begin{pgfonlayer}{edgelayer}
		\draw (0.center) to (1.center);
		\draw (2.center) to (3.center);
		\draw (5) to (6);
		\draw (6) to (4);
		\draw (9) to (8);
		\draw (7) to (9);
		\draw (11) to (8);
		\draw (10) to (5);
		\draw (5) to (12);
		\draw (8) to (13);
	\end{pgfonlayer}
\end{tikzpicture} \ \ = \ \ \begin{tikzpicture}
	\begin{pgfonlayer}{nodelayer}
		\node [style=none] (0) at (-4, 2) {};
		\node [style=none] (1) at (2, 2) {};
		\node [style=none] (2) at (-4, 0) {};
		\node [style=none] (3) at (2, 0) {};
		\node [style={white_dot}] (4) at (-2, 2) {};
		\node [style={white_dot}] (5) at (-2, -2) {};
		\node [style={grey_dot}] (6) at (-1, 1) {};
		\node [style={white_dot}] (7) at (0, 0) {};
		\node [style={white_dot}] (8) at (0, -4) {};
		\node [style={grey_dot}] (9) at (1, -1) {};
		\node [style=none] (14) at (-0.25, 1) {$a\pi$};
		\node [style=none] (15) at (1.75, -1) {$b\pi$};
		\node [style=none] (16) at (-1.25, -2) {$c\pi$};
		\node [style=none] (17) at (0.75, -4) {$d\pi$};
	\end{pgfonlayer}
	\begin{pgfonlayer}{edgelayer}
		\draw (0.center) to (1.center);
		\draw (2.center) to (3.center);
		\draw (5) to (6);
		\draw (6) to (4);
		\draw (9) to (8);
		\draw (7) to (9);
	\end{pgfonlayer}
\end{tikzpicture} \]
\vspace{10mm}
\[= \begin{tikzpicture}
	\begin{pgfonlayer}{nodelayer}
		\node [style=none] (0) at (-3, 2) {};
		\node [style=none] (1) at (3, 2) {};
		\node [style=none] (2) at (-3, 0) {};
		\node [style=none] (3) at (3, 0) {};
		\node [style={white_dot}] (4) at (-1, 2) {};
		\node [style={white_dot}] (5) at (-1, -2) {};
		\node [style={white_dot}] (7) at (1, 0) {};
		\node [style={white_dot}] (8) at (1, -4) {};
		\node [style=none] (12) at (-0.25, -2) {$c\pi$};
		\node [style=none] (13) at (1.75, -4) {$d\pi$};
	\end{pgfonlayer}
	\begin{pgfonlayer}{edgelayer}
		\draw (0.center) to (1.center);
		\draw (2.center) to (3.center);
		\draw (4) to (5);
		\draw (7) to (8);
	\end{pgfonlayer}
\end{tikzpicture} \ \ = \ \ \begin{tikzpicture}
	\begin{pgfonlayer}{nodelayer}
		\node [style=none] (0) at (-3, 2) {};
		\node [style=none] (1) at (3, 2) {};
		\node [style=none] (2) at (-3, 0) {};
		\node [style=none] (3) at (3, 0) {};
		\node [style={white_dot}] (4) at (-1, 2) {};
		\node [style={white_dot}] (6) at (1, 0) {};
		\node [style=none] (8) at (-0.5, 2.5) {$c\pi$};
		\node [style=none] (9) at (1.5, 0.5) {$d\pi$};
	\end{pgfonlayer}
	\begin{pgfonlayer}{edgelayer}
		\draw (0.center) to (1.center);
		\draw (2.center) to (3.center);
	\end{pgfonlayer}
\end{tikzpicture} \]
Now consider the case where the classical control is 1:
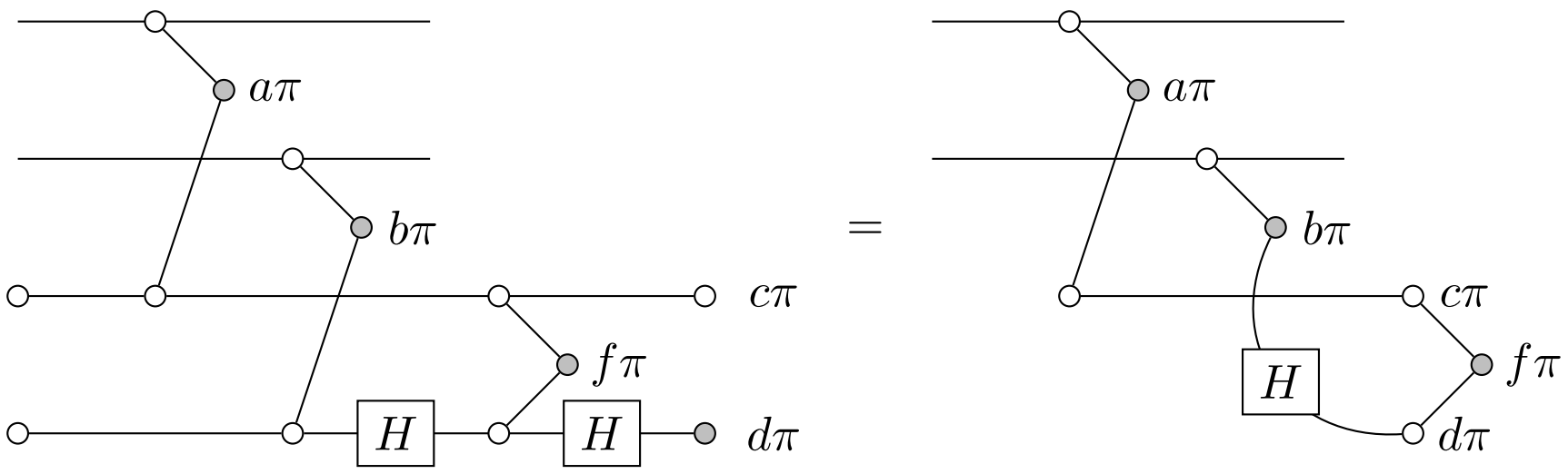
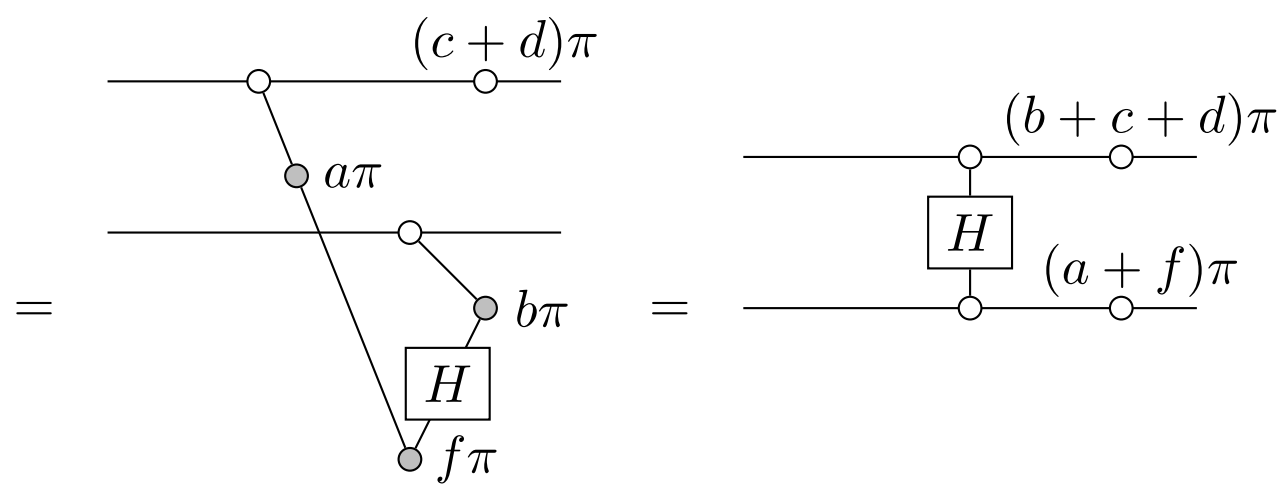
\[\begin{tikzpicture}
	\begin{pgfonlayer}{nodelayer}
		\node [style=none] (0) at (-5, 3) {};
		\node [style=none] (1) at (1, 3) {};
		\node [style=none] (2) at (-5, 1) {};
		\node [style=none] (3) at (1, 1) {};
		\node [style={white_dot}] (4) at (-3, 3) {};
		\node [style={white_dot}] (5) at (-3, -1) {};
		\node [style={grey_dot}] (6) at (-2, 2) {};
		\node [style={white_dot}] (7) at (-1, 1) {};
		\node [style={white_dot}] (8) at (-1, -3) {};
		\node [style={grey_dot}] (9) at (0, 0) {};
		\node [style={white_dot}] (10) at (-5, -1) {};
		\node [style={white_dot}] (11) at (-5, -3) {};
		\node [style={white_dot}] (12) at (5, -1) {};
		\node [style=none] (14) at (-1.25, 2) {$a\pi$};
		\node [style=none] (15) at (0.75, 0) {$b\pi$};
		\node [style=none] (16) at (6, -1) {$c\pi$};
		\node [style=none] (17) at (6, -3) {$d\pi$};
		\node [style={grey_dot}] (18) at (5, -3) {};
		\node [style={white_dot}] (19) at (2, -1) {};
		\node [style={white_dot}] (20) at (2, -3) {};
		\node [style={grey_dot}] (21) at (3, -2) {};
		\node [style=none] (22) at (3.75, -2) {$f\pi$};
		\node [style=box] (23) at (0.5, -3) {$H$};
		\node [style=box] (24) at (3.5, -3) {$H$};
	\end{pgfonlayer}
	\begin{pgfonlayer}{edgelayer}
		\draw (0.center) to (1.center);
		\draw (2.center) to (3.center);
		\draw (5) to (6);
		\draw (6) to (4);
		\draw (9) to (8);
		\draw (7) to (9);
		\draw (11) to (8);
		\draw (10) to (5);
		\draw (5) to (12);
		\draw (18) to (8);
		\draw (21) to (20);
		\draw (21) to (19);
	\end{pgfonlayer}
\end{tikzpicture} \ \ = \ \ \begin{tikzpicture}
	\begin{pgfonlayer}{nodelayer}
		\node [style=none] (0) at (-5, 3) {};
		\node [style=none] (1) at (1, 3) {};
		\node [style=none] (2) at (-5, 1) {};
		\node [style=none] (3) at (1, 1) {};
		\node [style={white_dot}] (4) at (-3, 3) {};
		\node [style={white_dot}] (5) at (-3, -1) {};
		\node [style={grey_dot}] (6) at (-2, 2) {};
		\node [style={white_dot}] (7) at (-1, 1) {};
		\node [style={grey_dot}] (9) at (0, 0) {};
		\node [style=none] (13) at (-1.25, 2) {$a\pi$};
		\node [style=none] (14) at (0.75, 0) {$b\pi$};
		\node [style=none] (15) at (2.75, -1) {$c\pi$};
		\node [style=none] (16) at (2.75, -3) {$d\pi$};
		\node [style={white_dot}] (18) at (2, -1) {};
		\node [style={white_dot}] (19) at (2, -3) {};
		\node [style={grey_dot}] (20) at (3, -2) {};
		\node [style=none] (21) at (3.75, -2) {$f\pi$};
		\node [style=box] (22) at (0.075, -2.25) {$H$};
	\end{pgfonlayer}
	\begin{pgfonlayer}{edgelayer}
		\draw (0.center) to (1.center);
		\draw (2.center) to (3.center);
		\draw (5) to (6);
		\draw (6) to (4);
		\draw (7) to (9);
		\draw (20) to (19);
		\draw (20) to (18);
		\draw (18) to (5);
		\draw [bend right=60, looseness=1.25] (9) to (19);
	\end{pgfonlayer}
\end{tikzpicture} \]
\vspace{10mm}
\[= \ \ \begin{tikzpicture}
	\begin{pgfonlayer}{nodelayer}
		\node [style=none] (0) at (-4, 3) {};
		\node [style=none] (1) at (2, 3) {};
		\node [style=none] (2) at (-4, 1) {};
		\node [style=none] (3) at (2, 1) {};
		\node [style={white_dot}] (4) at (-2, 3) {};
		\node [style={grey_dot}] (6) at (-1.5, 1.75) {};
		\node [style={white_dot}] (7) at (0, 1) {};
		\node [style={grey_dot}] (8) at (1, 0) {};
		\node [style=none] (9) at (-0.75, 1.75) {$a\pi$};
		\node [style=none] (10) at (1.75, 0) {$b\pi$};
		\node [style=none] (11) at (1.25, 3.5) {$(c+d)\pi$};
		\node [style={grey_dot}] (15) at (0, -2) {};
		\node [style=none] (16) at (0.75, -2) {$f\pi$};
		\node [style=box] (17) at (0.5, -1) {$H$};
		\node [style={white_dot}] (18) at (1, 3) {};
	\end{pgfonlayer}
	\begin{pgfonlayer}{edgelayer}
		\draw (0.center) to (1.center);
		\draw (2.center) to (3.center);
		\draw (6) to (4);
		\draw (7) to (8);
		\draw (8) to (15);
		\draw (6) to (15);
	\end{pgfonlayer}
\end{tikzpicture} \ \ = \ \ \begin{tikzpicture}
	\begin{pgfonlayer}{nodelayer}
		\node [style=none] (0) at (-3, 2) {};
		\node [style=none] (1) at (3, 2) {};
		\node [style=none] (2) at (-3, 0) {};
		\node [style=none] (3) at (3, 0) {};
		\node [style={white_dot}] (4) at (0, 2) {};
		\node [style={white_dot}] (6) at (0, 0) {};
		\node [style=none] (10) at (2.25, 2.5) {$(b+c+d)\pi$};
		\node [style=box] (13) at (0, 1) {$H$};
		\node [style={white_dot}] (14) at (2, 2) {};
		\node [style={white_dot}] (15) at (2, 0) {};
		\node [style=none] (16) at (2.25, 0.5) {$(a+f)\pi$};
	\end{pgfonlayer}
	\begin{pgfonlayer}{edgelayer}
		\draw (0.center) to (1.center);
		\draw (2.center) to (3.center);
		\draw (13) to (6);
		\draw (13) to (4);
	\end{pgfonlayer}
\end{tikzpicture} \]
\endproof

\section{Methods for fast measurements with quantum LDPC codes}\label{app:fast_ldpc_logic}

A common framework for viewing CSS codes, and operations upon them, is via homological algebra. CSS codes are well-known to be in bijection with chain complexes (of length 2); chain maps describe coherent maps between CSS codes, and longer chain complexes can describe detectors.

There are known methods of constructing CSS codes with favorable parameters, sparsity and girth, such as ``lifted products''~\cite{panteleev2021degenerate}. As a consequence, most practical quantum LDPC codes are CSS codes.

Two main algebraic methods of describing fast logical measurements on CSS codes are:
\begin{itemize}
    \item Chain maps, which describe CNOT circuits, and are used to define \textit{homomorphic measurements}~\cite{huang2023homomorphic}, whereby an ancilla system is initialized, CNOTs are performed to the memory code, and then the ancilla system is measured out. Homomorphic measurements generalize Shor and Steane measurements.
    \item Mapping cones, which describe code deformations. These are used to define surgeries (in particular, \textit{homological measurements}~\cite{ide2024fault}), whereby additional qubits and checks are incorporated into the code and then subsequently measured out. Homological measurements generalize lattice surgery~\cite{Horsman2012LatticeSurgery}.
\end{itemize}
In both cases, the measurements made by the method must be all-$Z$ or all-$X$-type. \footnote{In~\cite{yuan2026noncss} mapping cones are extended to the non-CSS case, but this doesn't directly yield new fast surgery methods, and there appear to be strong obstructions to performing non-CSS measurements with fast surgery~\cite{cowtan2025fast,baspin20262fast2surgery}.} Also in both cases, with some conditions the logical measurements can be performed in one syndrome cycle. Homomorphic measurements still require the ancilla code to be fault-tolerantly prepared, but this can be done separately. Fast surgery can either be done while assuming that the original code is single-shot or by using buffer rounds which are amortized into the surgery time overhead~\cite{baspin2025fast,cowtan2025fast}.

There are other logical measurement schemes~\cite{williamson2024low,cowtan2025parallel} which neither require the original code to be CSS nor require the measurement to be of all-$Z$ or all-$X$-type, but these tend to be much less efficient when measuring many logical operators; either in time or in space overhead.

\end{document}